\documentclass[10pt,prb,twocolumn,superscriptaddress,amsmath,amssymb,floatfix,preprintnumbers]{revtex4-2}
\usepackage[utf8]{inputenc}
\usepackage[T1]{fontenc}
\usepackage[pdftex]{graphicx}
\usepackage{amssymb, amsmath, bm, color, soul, tikz, xcolor, hyperref, mathtools, dcolumn, bm, float, cleveref, listings, lipsum, bbold, orcidlink, braket}
\usepackage[mathlines]{lineno}

\hypersetup{colorlinks=true,linkcolor=blue,citecolor=blue}

\definecolor{lime}{HTML}{A6CE39}
\DeclareRobustCommand{\orcidicon}{%
	\begin{tikzpicture}
	\draw[lime, fill=lime] (0,0) 
	circle [radius=0.16] 
	node[white] {{\fontfamily{qag}\selectfont \tiny ID}};
	\draw[white, fill=white] (-0.0625,0.095) 
	circle [radius=0.007];
	\end{tikzpicture}
	\hspace{-2mm}
}

\foreach \x in {A, ..., Z}{%
	\expandafter\xdef\csname orcid\x\endcsname{\noexpand\href{https://orcid.org/\csname orcidauthor\x\endcsname}{\noexpand\orcidicon}}
}

\begin{document}

\title{Energy levels and Aharonov-Bohm oscillations in twisted bilayer graphene quantum dots and rings}

\author{N. S. Bandeira}\email{nathanaell@fisica.ufc.br}
\affiliation{Departamento de F\'isica, Universidade Federal do Cear\'a, Campus do Pici, 60455-900 Fortaleza, Cear\'a, Brazil}

\author{Andrey Chaves\orcidA{}}\email{andrey@fisica.ufc.br}
\affiliation{Departamento de F\'isica, Universidade Federal do Cear\'a, Campus do Pici, 60455-900 Fortaleza, Cear\'a, Brazil}

\author{L. V. de Castro}
\affiliation{Departamento de F\'isica, Universidade Federal do Cear\'a, Campus do Pici, 60455-900 Fortaleza, Cear\'a, Brazil}

\author{R. N. Costa Filho\orcidB{}}
\affiliation{Departamento de F\'isica, Universidade Federal do Cear\'a, Campus do Pici, 60455-900 Fortaleza, Cear\'a, Brazil}

\author{M. Mirzakhani\orcidC{}}
\affiliation{Department of Physics, Yonsei University, Seoul 03722, South Korea}
\affiliation{Center for Theoretical Physics of Complex Systems, Institute for Basic Science, Daejeon, 34126, South Korea}

\author{F. M. Peeters\orcidD{}}
\affiliation{Department of Physics, University of Antwerp, Groenenborgerlaan 171, B-2020 Antwerp, Belgium}
\affiliation{Departamento de F\'isica, Universidade Federal do Cear\'a, Campus do Pici, 60455-900 Fortaleza, Cear\'a, Brazil}

\author{D. R. da Costa \orcidE{}}
\email{diego_rabelo@fisica.ufc.br}
\affiliation{Departamento de F\'isica, Universidade Federal do Cear\'a, Campus do Pici, 60455-900 Fortaleza, Cear\'a, Brazil}
\affiliation{Department of Physics, University of Antwerp, Groenenborgerlaan 171, B-2020 Antwerp, Belgium}

\begin{abstract}
We present a systematic study of the energy levels of twisted bilayer graphene (tBLG) quantum dots (QD) and rings (QR) under an external perpendicular magnetic field. The confinement structures are modeled by a circular dot-like- and ring-like-shaped site-dependent staggered potential, which prevents edge effects and leads to an energy gap between the electron and hole states. Results are obtained within the tight-binding model with interlayer hopping parameters defined by the Slater-Koster form for different interlayer twist angles $\theta$. Our findings show that, for $\theta$ around 0$^\circ$ or $60^\circ$, the energy spectra exhibit features resulting from the interplay between characteristics of the AA and AB/BA stacking orders that compose the moiré pattern of such tBLG, while the low-energy levels are shown to be nearly independent on the rotation angle for $10^\circ\lesssim \theta\lesssim 50^\circ$. In the absence of a magnetic field, the energy levels of the QR scale with its width $W$ according to a power law $W^{-\alpha}$, whose exponent $1 \lessapprox\alpha\lessapprox 2$ depends on the twist angle. Most interestingly, the lowest energy states of tBLG QRs oscillate as a function of its average radius, with the oscillation period matching half of the moiré period. In the presence of an intense magnetic field, the lowest energy levels for the tBLG QDs and QRs match almost perfectly, regardless of whether the external radius of the quantum confinement structure is smaller or on the order of the moiré period, which is due to the interplay of the trigonal warping effect and the preferential localization of the eigenstates. Our results reveal relevant information about the moiré pattern in tBLG and its role in charge particle confinement. 
\end{abstract}

\maketitle

\section{Introduction}

Recently, twisted bilayer graphene (tBLG) has been the subject of substantial theoretical and experimental works due to many fascinating physical phenomena, such as unconventional superconductivity \cite{cao2018unconventional,yankowitz2019tuning}, spin-polarized phases \cite{cao2016superlattice, lu2019superconductors}, and correlated insulator behavior \cite{cao2018correlated, lee2019theory, cao2020tunable} emerging for rotation angle between the two coupled graphene layers of order $\theta \sim 1^\circ$, named \emph{magic} twist angle \cite{wang2019properties, andrei2020graphene}. As a consequence of the twisting angle, a moiré pattern will be formed in the system. Due to the slight crystallographic offset between the two twisted graphene layers, different stacking registries inside the moiré supercell can be observed. This affects the interlayer electronic coupling, resulting in a modulation of the band edge energies that can be described by a periodic long-range interaction known as moiré potential. \cite{dos2012continuum, moon2013optical, dindorkar2023magical, wang2022intrinsic, he2021moire, PhysRevLett.99.256802, bistritzer2011moire, PhysRevB.77.045403, koshino2015electronic} Such relative rotation between the layers significantly modifies its low-energy band structure, for instance, leading to the emergence of moiré flat bands near the Fermi level, band gap changes, reduction of the band velocity of the Dirac cones, and the appearance of van Hove singularity in the far-infrared region. \cite{PhysRevB.82.121407, dos2012continuum, PhysRevB.81.165105, PhysRevLett.99.256802, PhysRevB.77.235403, bistritzer2011moire, trambly2010localization} 

Beyond twisting the graphene layers, a typical platform for controlling the physical properties of graphene-based systems is reducing their dimensionality by restricting the charge carriers' mobility via energy barriers, which consequently leads to quantized energetic states along the confined direction. Two of the extensively investigated nanostructures in BLG that have a prominent role for future technological applications are the BLG quantum dots (QDs) \cite{BQDsDiego, DiegomagQD, da2014analytical, PhysRevB.93.165410, PhysRevB.94.035415, pereira2007tunable, velasco2018visualization, ge2020visualization, ge2021imaging, PhysRevB.96.115428} and the BLG quantum rings (QRs) \cite{zarenia2010S, zarenia2009electrostatically, PhysRevB.105.115430, RASTEGARSEDEHI2022114853, zahidi2017energy, xavier2010topological}. The huge interest in these nanostructures is due to the fact that BLG QDs are a promising quantum information platform owing to their long spin decoherence times and tunability, \cite{GQDbook, harrison2016quantum, li2022recent} whereas BLG QRs are the most natural systems to investigate quantum interference phenomenon in transport properties, Aharonov–Bohm oscillations and persistent currents. \cite{PhysRev.115.485, PhysRevLett.58.2814, ford1988aharonov, fuhrer2001energy, ford1989electrostatically, VIEFERS20041, fuhrer2002energy, MANNINEN2012119, da2017electronic, da2014geometry, xavier2016electronic, PhysRevB.95.205414, araujo2022modulation, bahamon}

QDs made of AA- ($\theta=0^\circ$) and AB-stacked ($\theta=60^\circ$) BLG have been widely studied both theoretically \cite{da2014analytical, BQDsDiego, pereira2007tunable, DiegomagQD} and experimentally \cite{velasco2018visualization, ge2020visualization, ge2021imaging}. By assuming different geometries (hexagonal, triangular, and circular), edge types (zigzag, armchair, and mix edges), and stackings (AA and AB) in the presence of a perpendicular electric field to the BLG QDs, da Costa \textit{et al.} \cite{BQDsDiego} reported that the sample edges and geometries play essential roles in modifying the electronic properties of BLG QDs, and moreover, in the presence of the magnetic field they demonstrated in Ref.~\cite{DiegomagQD} that the spatial symmetry of the carrier density distributions is related to the symmetry of the confinement potential, the position of zigzag edges, and the presence or absence of interlayer inversion symmetry. Experimentally, AB-stacked BLG QDs have also been electrostatically defined using a scanning tunneling microscope \cite{velasco2018visualization, ge2020visualization, ge2021imaging}. For instance, Ge \textit{et al.} \cite{ge2020visualization} demonstrated that the confined states in a circularly symmetric tip-induced p-n junction potential profile exhibited a robust broken rotational symmetry. They attributed this lack of circular symmetry to the low energy anisotropic band as a consequence of the trigonal warping effect in BLG, thus showing the relevance of additional interlayer hoppings to model quantum-confined BLG systems.

For intermediate interlayer angles between $\theta=0^\circ$ and $\theta=60^\circ$, a number of reported studies have been carried out in the literature addressing the electronic \cite{PhysRevB.87.075433, mirzakhani2020circular, wang2021enhanced, wang2022enhanced, tilak2021flat, PhysRevB.104.235417}, optical \cite{PhysRevX.12.021055, tiutiunnyk2019opto, tepliakov2020twisted, liu2023polarizability}, magnetic \cite{luo2024tuning, mirzakhani2023magnetism}, and transport \cite{PhysRevB.101.235432} properties of tBLG QDs. Among them, irregular \cite{PhysRevB.87.075433} and well-defined shaped \cite{tiutiunnyk2019opto, tepliakov2020twisted, PhysRevX.12.021055} tBLG flakes have been explored. The former \cite{PhysRevB.87.075433} shows that a single moiré spot in a tBLG flake is sufficient for its low-energy density of states to closely resemble that of an infinite tBLG sample, implying that the low-energy physics in this system is well described as that of a moiré quantum well trapping low-energy graphene electrons. This suggests that a tBLG flake with a single moiré unit cell already leads to electron localization on the AA moiré spot, which is in agreement with the moiré quantum well picture. From an experimental point of view, Refs.~\cite{PhysRevB.104.235417, tilak2021flat} proposed strategies to obtain confined electronic states in tBLG QDs. For instance, Zhou \textit{et al.} \cite{PhysRevB.104.235417} demonstrated a general approach for fabricating stationary tBLG QDs by introducing nanoscale p-n junctions with sharp boundaries in the tBLG.

Many consequences on the electronic properties of few-layer graphene have been observed by introducing an additional inner boundary in the QD structure to form ring-like quantum confinement nanostructures, for instance, as reported for monolayer graphene QRs in Refs.~\cite{da2014geometry, xavier2016electronic, zarenia2010S, RASTEGARSEDEHI2022114853, PhysRevB.95.235427} and BLG QRs in Refs.~\cite{PhysRevB.105.115430, zarenia2010S, zarenia2009electrostatically, RASTEGARSEDEHI2022114853, zahidi2017energy, xavier2010topological}. Some of such previous theoretical studies considered electrostatically defined BLG QRs \cite{xavier2010topological, zarenia2009electrostatically} by solving the Dirac equation analytically \cite{xavier2010topological, zarenia2010S} and numerically \cite{zarenia2009electrostatically} for AB-stacked BLG. By assuming infinite-mass boundary conditions, Refs.~\cite{PhysRevB.105.115430, zarenia2010S, zahidi2017energy} explored circular BLG QRs. Zarenia \textit{et al.} \cite{zarenia2010S} applied a simplified model by freezing the motion along the radial direction for analytically deriving the energy levels of a simple zero-width-ring geometry defined in an AB-stacked BLG. Results for the energy spectrum and corresponding wave functions for realistic rings with non-zero thickness within infinite-mass boundary conditions were described in Ref.~\cite{PhysRevB.105.115430}. The authors investigated two ring systems with AB-stacking: an isolated BLG QR defined by a site-dependent staggered media, and a hybrid BLG QR sandwiched between a monolayer graphene QD and an infinite monolayer graphene region. A similar isolated BLG QR study of non-zero thickness for the AA-stacked case was carried out in Ref.~\cite{zahidi2017energy}. However, to our knowledge, there is no theoretical study on the energy spectrum of tBLG QRs or a detailed comparison with tBLG QD results aiming to address the effect of the twisting angle on the energy levels of tBLG QRs and the consequences of an additional inner boundary on the QD systems.

Within the mentioned context of moiré superlattice and quantum confinement systems, in this work, we present a systematic study of the electronic properties of tBLG QRs in both the absence and presence of an external perpendicular magnetic field. Two questions guide our work: (1) in the absence of a magnetic field, how do the energy levels of the tBLG QR scale with its width and the QR average radius? and (2) in the presence of a magnetic field, will there be any influence on the Aharonov-Bohm oscillation of regular ring structures due to the moiré pattern of the tBLG system? With these goals in mind, we organized the paper as follows: In Sec.~\ref{Sec.Model}, we present the theoretical model based on the tight-binding approach with interlayer hopping parameters defined by the Slater-Koster form. The two investigated quantum confinement structures are present: the tBLG QDs and tBLG QRs modeled by circular dot- and ring-shaped site-dependent staggered potentials. Results in the absence (Sec.~\ref{sec.B.eq.0}) and the presence (Sec.~\ref{sec.B.neq.0}) of a perpendicular magnetic field are discussed in Sec.~\ref{sec.results}. Finally, in Sec.~\ref{Sec.Conclusion}, we provide the main concluding remarks.

\begin{figure}[t]
    \centerline{\includegraphics[width = 1\linewidth]{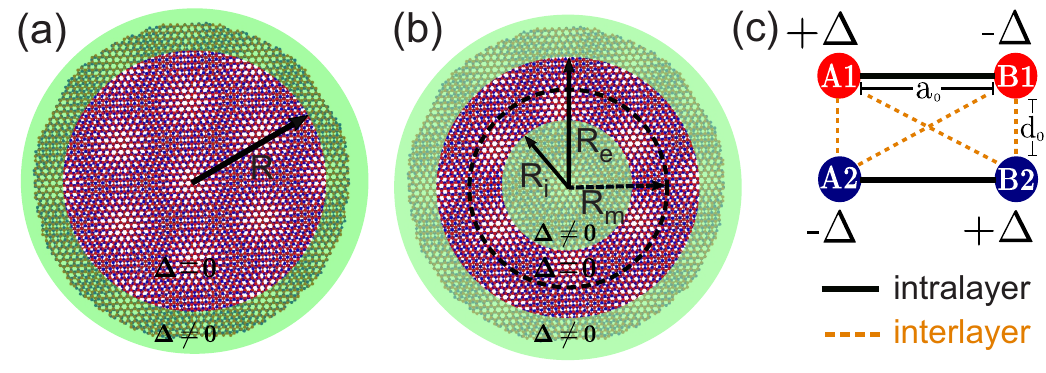}}
    \caption{(\textcolor{blue}{Color online}) Sketches of the investigated tBLG quantum confinement nanostructures: (a) QD and (b) QR with a circular shape. The structures are surrounded by a staggered potential medium, with $\Delta\neq 0$ for $r>R$ in the QD with radius $R$ and $R_i<r<R_e$ in the QR case with average radius $R_m = (R_e+R_i)/2$, as highlighted by the green shaded regions. (c) A schematic illustration depicting the assumed site-dependent mass potential at each layer. The interlayer and intralayer connections between different sublattices in tBLG are also shown. $a_0$ and $d_0$ are the interatomic and interlayer distances, respectively.} 
	\label{Fig:Sketch}
\end{figure}

\section{Theoretical Model}\label{Sec.Model}

tBLG consists of two coupled graphene layers that are twisted with respect to each other by an angle $\theta$, also viewed as two misaligned honeycomb layers vertically separated by an interlayer distance of $0.335$ nm. \cite{bistritzer2011moire, dos2012continuum, moon2013optical, koshino2015electronic} Each graphene layer is composed of carbon atoms arranged in a honeycomb structure with two sites per unit cell: $A_i$ and $B_i$, with $i=1$ and $i=2$ labeling the top and bottom layers, respectively [see Fig.~\ref{Fig:Sketch}(c)]. In AA-stacked BLG, the atoms in the upper and lower layers are all located on top of each other, \textit{i.e.} with a $\theta=0$ twist angle, whereas in an AB(BA)-stacked BLG, also known as Bernal stacking, the atoms in the $B_2$ ($A_2$) sublattices in the bottom layer are aligned with the $A_1$ ($B_1$) atoms in the top layer, respectively, which is also equivalent to a $\theta=60^\circ$ interlayer twist. \cite{mccann2013electronic} As a crystallographic consequence of a relative rotation $0^\circ<\theta<60^\circ$ between the layers, one observes the formation of a periodic moiré superlattice, which displays alternating patterns between AA and AB/BA stacked lattices. The periodic length of such moiré superlattice ($L_M$) is associated with the twist angle $\theta$ and the lattice constant of the primitive unit cells of each monolayer graphene ($a=0.246$~nm) as
\begin{equation}\label{eq.Lm}
L_M = \frac{a}{2\sin{(\theta/2)}}.    
\end{equation} 
By this inverse proportionality relation between $\theta$ and $L_M$, it is evident that the greater the twist angle, the shorter the length of the moiré period. \cite{dos2012continuum, moon2013optical, dindorkar2023magical, wang2022intrinsic, he2021moire, PhysRevLett.99.256802, bistritzer2011moire, PhysRevB.77.045403, andrei2020graphene, wang2019properties, koshino2015electronic} Here, we assumed the systems rotation reference to be the center of the unit cell of the untwisted AA-stacked BLG with the twisting $z$ axes located along the $B_1-B_2$ connection.

Despite the fact that the moiré superlattice can be defined for any $\theta$, even when the atomic lattice structure itself is incommensurate, in general, the unit cell in the lattice structure of carbon atoms in tBLG is not periodic for any twist angle $\theta$, because the periods of the two graphene layers are generally incommensurate with each layer \cite{moon2013optical, koshino2015electronic}. A commensurate structure is obtained if a certain atom in one of the layers is moved by the rotation to a position occupied by an atom of the same kind \cite{PhysRevLett.99.256802}. In other words, when the two periods match, the atomic lattice structure becomes periodic, with a finite unit cell. This happens when the lattice vector of the top layer $\mathbf{v}_1 = n\mathbf{a}_1 + m\mathbf{a}_2$ coincides with the lattice vector of the bottom layer $\mathbf{v}_2 = m\tilde{\mathbf{a}}_1 + n\tilde{\mathbf{a}}_2$, where $m$ and $n$ are integer numbers, $\mathbf{a}_1 = a(\sqrt{3/2},1/2)$ and $\mathbf{a}_2 = a(\sqrt{3/2},-1/2)$ are the Bravais lattice basis vectors of the monolayer graphene, and $\tilde{\mathbf{a}}_i = \mathcal{R}(\theta)\mathbf{a}_i$ with $i=\{1,2\}$ are the primitive lattice vectors of the rotated layer with $\mathcal{R}(\theta)$ being the $2\times 2$ rotation matrix. The lattice vectors of the unit cell of the superlattice structure are given by \cite{PhysRevB.81.161405, moon2013optical, koshino2015electronic} $\mathbf{L}_1 = \mathbf{v}_1 = \mathbf{v}_2$ and $\mathbf{L}_2 = \mathcal{R}(\theta=\pi/3)\mathbf{L}_1$. The commensurate twist angle is given in terms of the integer numbers $(n,m)$ by $\cos\left(\theta\right)=(m^2+n^2+4mn)/[2(m^2+n^2+mn)]$ and the lattice constant of the commensurate superlattice is $L=|\mathbf{L}_1|=|\mathbf{L}_2|=|m-n|L_M$, such that the lattice constant of commensurate superstructure $L$ coincides with the moiré superlattice period $L_M$ when $|m - n| = 1$. Here, we shall investigate tBLG nanostructures for different twist angles, dealing especially with the cases for $\theta = 1.08^\circ$, $\theta = 2.00^{\circ}$, $\theta = 2.65^{\circ}$, and $\theta = 5.09^\circ$ that have moiré period $L_M(\theta=1.08^{\circ}) \approx 13.05$ nm, $L_{M}(\theta=2.00^{\circ}) \approx 7.03$ nm, $L_{M}(\theta=2.65^{\circ}) \approx 5.33$ nm, and $L_M(\theta=5.09^{\circ}) \approx 2.77$ nm, respectively. Their commensurate structures obey the $|m - n| = 1$ condition and are formed with $(m,n)=(31,30)$, $(m,n) = (17,16)$, $(m,n) = (13,12)$, and $(m,n) = (7,6)$, respectively. The choice of twisting angles here for commensurate structures was made to allow us to compare limit cases of our results with those already available in the literature \cite{mirzakhani2020circular}, but our model is straightforwardly adapted for any twist angle, commensurate or not.

Two different quantum confinement nanostructures are investigated here: circular tBLG QDs and QRs, as illustrated in Figs.~\ref{Fig:Sketch}(a) and \ref{Fig:Sketch}(b), respectively. The former is characterized by a radius $R$, and the latter, by an average radius $R_m$ and a width $W$, such that $R_2 = R_m + W/2$ is the external radius and $R_1 = R_m - W/2$ the internal radius of the tBLG QR. In order to describe the charge carriers in such QDs and QRs defined in a tBLG, we use a single-orbital ($p_z$) tight-binding model with the Hamiltonian written in second quantization form as
\begin{equation}
H=\sum_{i}^{N}\left(\epsilon_{i} + M_{i}\right)c^{\dagger}_{i}c_{i} + \sum_{<i,j>}t_{i,j}(\mathbf{d}_{ij})c^{\dagger}_{i}c_{j} + H.c,
\label{eq: HTBG}
\end{equation} 
where $c_i^{\dagger}$ ($c_i$) is the creation (annihilation) operator for an electron at site $i$ with on-site energy $\epsilon_i$ and a position-dependent staggered potential term $M_{i}$ (also known as ``mass term'' due to its influence on the low energy states, which is equivalent to a mass term in Dirac-Weyl equation). \cite{berry1987neutrino, da2014geometry, xavier2016electronic, schnez2008analytic} The staggered site-dependent potential is defined in such a way that $M_{i} = \Delta$ ($M_{i} = -\Delta$) if $i$ belongs to sites in the sublattice $A$ ($B$) in one layer and $B$ ($A$) in the other one. This is illustrated in Fig.~\ref{Fig:Sketch}(c) with $M_{A1} = \Delta$ and $M_{B1} = -\Delta$ for sublattices in the top layer ($A1$ and $B1$ in red symbols) and $M_{A2} = -\Delta$ and $M_{B2} = \Delta$ for sublattices in the bottom layer ($A2$ and $B2$ in blue symbols). Such a staggered potential medium can be realized by using an appropriate nanostructured substrate for BLG, similar to monolayer graphene case deposited in SiC and hexagonal boron nitride substrates \cite{zhou2007substrate, PhysRevB.76.073103, PhysRevLett.115.136802, schnez2008analytic, giovannetti2007substrate, wang2016gaps}, and has been employed successfully to simulate electronic confinement in BLG nanostructures as reported in Refs.~\cite{da2014analytical, DiegomagQD, PhysRevB.105.115430, zarenia2010S, zahidi2017energy}. This breaks the sublattice symmetry in BLG QDs \cite{da2014analytical, DiegomagQD} and QRs \cite{PhysRevB.105.115430, zarenia2010S, zahidi2017energy}, leading to a large gap opening in their energy spectra and avoiding the influence of zero-mode edge states coming from zigzag terminations. As highlighted by the green shaded regions in Figs.~\ref{Fig:Sketch}(a) and \ref{Fig:Sketch}(b), for both confinement structures, we assume a surrounding medium in the outer boundaries with $\Delta = 2.0$ eV \cite{mirzakhani2020circular}. In the case of QRs, the internal region $r<R_1$ also has this non-zero mass-term, which defines the inner boundary of the ring.

The second term of the Hamiltonian \eqref{eq: HTBG} describes the intralayer and interlayer charge carrier motions, where $t_{i,j}(\mathbf{d}_{ij})$ corresponds to the transfer integrals associated with the hopping energies that depend on the interatomic position $\mathbf{d}_{ij} = |\mathbf{r}_{i}-\mathbf{r}_{j}|$ between the sites $i$ and $j$. To describe the interatomic coupling and knowing that the atomic orbitals have exponentially decaying tails far from their centers \cite{wang2022intrinsic, PhysRevB.77.045403, duarte2023moir}, which leads, in turn, to a hopping rate between every two sites ($i \neq j$) that decays exponentially as a function of distance $|d_{ij}|$, we assume here the transfer integrals $t_{i,j}$ to be written within the \textit{Slater-Koster} approximation \cite{moon2013optical, koshino2015electronic}. As such, they are modeled with decaying exponential functions $V_{pp\sigma}(d_{ij})$ and $V_{pp\pi}(d_{ij})$ accounting for the decomposition of the transfer integrals between the $p_z$ orbitals as $\sigma-$ and $\pi-$like bonds, such as
\begin{subequations}
\begin{align}
& t(d_{ij})=-\hspace{-0.05cm}V_{pp\pi}\hspace{-0.05cm}\left[\hspace{-0.05cm} 1 \hspace{-0.075cm}-\hspace{-0.075cm}\left(\frac{\mathbf{d_{ij}}\cdot \mathbf{e_{z}}}{d_{ij}}\hspace{-0.05cm}\right)^{2}\hspace{-0.05cm}\right] \hspace{-0.075cm}-\hspace{-0.075cm} V_{pp\sigma}\hspace{-0.05cm}\left(\hspace{-0.05cm}\frac{\mathbf{d_{ij}}\cdot \mathbf{e_{z}}}{d_{ij}}\hspace{-0.05cm}\right)^{2},\\
& V_{pp\pi}(d_{ij})=V^{0}_{pp\pi}\exp\left[-\left(\frac{d_{ij} - a_{cc}}{\delta_{0}}\right)\right],\\
& V_{pp\sigma}(d_{ij})=V^{0}_{pp\pi}\exp\left[-\left(\frac{d_{ij} - d_{0}}{\delta_{0}}\right)\right],
\end{align}
\end{subequations}
where $\delta_{0} = 0.184 a_{cc}$ is the decay length, $a_{cc}=a/\sqrt{3}$ is the carbon-carbon distance of graphene, $d_{ij} = |\mathbf{d_{ij}}|$, and $\mathbf{e_z}$ is the unit vector along the vertical $z$-direction. The intralayer and interlayer nearest-neighbor hoppings are, respectively, given by $V_{pp\pi}^{0} \approx - 2.7$ eV and $V_{pp\sigma}^{0}\approx 0.48$ eV, as to agree with band structures of the monolayer graphene and AB-stacked BLG cases. For intralayer motion, only hopping terms between the nearest neighboring lattice sites are included. For interlayer contributions, hopping energies for $d > 4a_{cc}$ are exponentially small and then shall be neglected. A similar restriction for the atomic distances has been widely employed for theoretical investigation of tBLG systems, see, e.g., Refs.~\cite{moon2013optical, koshino2015electronic, mirzakhani2023magnetism, mirzakhani2020circular}.

The effect of an external magnetic field is incorporated in the tight-binding model via the Peierls substitution, such that the hopping energy becomes $t(\mathbf{d}_{ij})\rightarrow t(\mathbf{d}_{ij}) \exp\left[i2\pi\Phi_{i,j}\right]$, with $\Phi_{i,j}=\frac{1}{\Phi_{0}}\int_{\mathbf{r}_{j}}^{\mathbf{r}_{i}} \mathbf{A}(\mathbf{r}) \cdot d\mathbf{r}$ as the Peierls phase, where $\Phi_{0}=h/e$ is a quantum of magnetic flux, and $\mathbf{A}(\mathbf{r})=(0,xB_{0},0)$ is the vector potential. We assume the Landau gauge, associated with a perpendicular magnetic field in the $\hat{z}$-direction. This results in a non-zero contribution for the hoppings only along the $y$-direction, given by $\Phi_{ij}$ = $\mbox{sgn}(y_{j}-y_{i})\frac{(x_{j} + x_{i})\Phi}{2\sqrt{3}a_{cc}\Phi_{0}}$, where $\Phi=\sqrt{3}a^{2}_{cc}b/2$ is the magnetic flux threading one carbon hexagon.

\section{Results and Discussion}\label{sec.results}

\subsection{Energy levels at zero magnetic field} \label{sec.B.eq.0}

\begin{figure}[t!]
    \centering
    \includegraphics[width=0.9\linewidth]{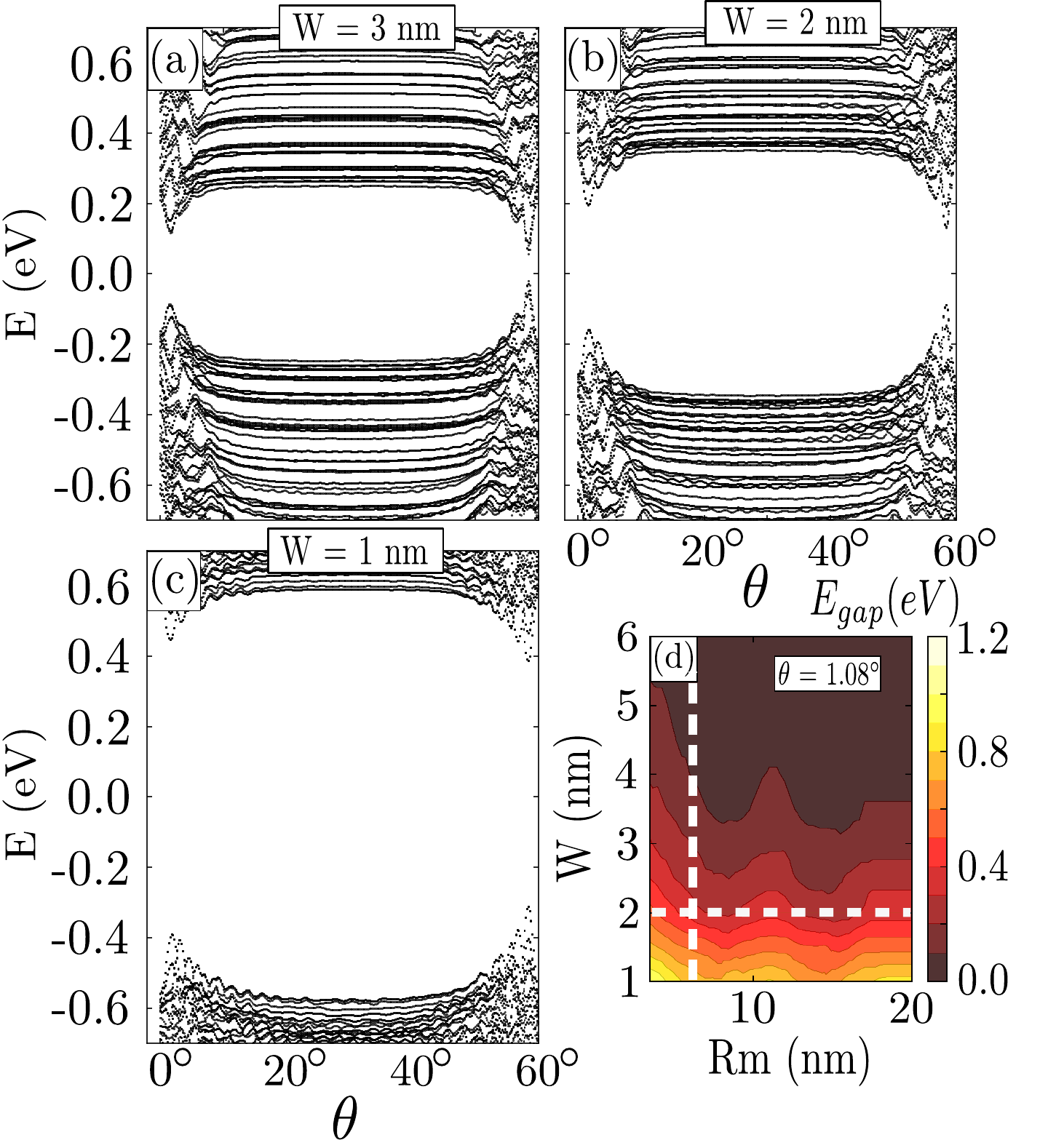}
    \caption{(\textcolor{blue}{Color online}) Energy levels of a circular tBLG QR, with average radius $R_m = 4.81$ nm, as a function of the twist angle $\theta$ for different ring width: (a) $W = 3.0$ nm, (b) $W = 2.0$ nm and (c) $W = 1.0$ nm. (d) Contour plot of the energy gap as a function of $R_{m}$ and $W$ for the magic angle $\theta = 1.08^{\circ}$. The horizontal and vertical dashed white lines in (d) emphasize the cases of a fixed ring width of $W=2$~nm (horizontal) and a fixed average radius of $R_{m} = 4.81$ nm (vertical), as shall be depicted in Fig.~\ref{ExRm}(b) and Fig.~\ref{ExW}(b) for $\theta = 1.08^{\circ}$, respectively, that the energy gap oscillates as a function of the QR average radius and the decreasing behavior of the ground state as a function of the ring width.}
	\label{ExoW}
\end{figure}

In Fig.~\ref{ExoW}, we analyze the behavior of the energy levels in a circular tBLG QR with average radius $R_{m} = 4.81$ nm as a function of the interlayer twist angle $\theta$ for different values of the ring width $W$: (a) $W=3$~nm, (b) $W=2$~nm, and (c) $W=1$~nm. Note from Figs.~\ref{ExoW}(a)-\ref{ExoW}(c) that the tBLG QRs exhibit an energy gap between the electron and hole states regardless of the twist angle. This is due to the mass-potential confinement assumed here for defining the ring-like BLG nanostructures, which also prevent edge states. \cite{mirzakhani2020circular, DiegomagQD, BQDsDiego} Indeed, edge states are absent in all results presented in this work. By comparing Figs.~\ref{ExoW}(a)--\ref{ExoW}(c) for different values of $W$, one notices that by narrowing the ring width, the energy level spacing between electron and hole states are modified, thus increasing the energy gap of the system. This is reminiscent of the simple physics of quantum confinement nanostructures, like quantum wells and QDs, where the energy levels' spacing is inversely proportional to the confinement lengthscales. \cite{harrison2016quantum} This statement of the decreasing tendency of the band gap the larger the tBLG QR width is confirmed by Fig.~\ref{ExoW}(d) for a larger range of ring width values taking the twist angle $\theta = 1.08^\circ$ as an example, chosen considering that this corresponds to the magic angle. Figure~\ref{ExoW}(d) shows a color map of the energy gap of tBLG QRs as a function of the geometric system parameters $R_m$ and $W$. To directly compare the result in Fig.~\ref{ExoW}(d) with Figs.~\ref{ExoW}(a)-\ref{ExoW}(c), a vertical white dashed line is presented, guiding the eyes to the band gap energies for $R_m = 4.81$ nm QR and showing a band gap decreasing of more than 500 meV in the range $W\in [1,6]$ nm.

Moreover, it is clear from Figs.~\ref{ExoW}(a)-\ref{ExoW}(c) that for rotation angles of $ 10^{\circ} \lesssim \theta \lesssim 50^{\circ}$, the energy levels are only weakly affected by the twist angle, \textit{i.e.} they are nearly independent of $\theta$. This is a consequence of the weak interaction between the layers behaving effectively like two decoupled graphene layers  \cite{deng2020interlayer}, verified by analyzing their few numbers of interlayer hoppings and their weak magnitudes for this twist angle range \cite{PhysRevB.81.165105, PhysRevB.87.075433, bistritzer2011moire}. On the other hand, the interlayer twist angles $\theta \lesssim 10^\circ$ and $\theta \gtrsim 50^\circ$ have a strong influence on the energy levels. This results from interference effects induced by the different regions that compose the moiré pattern for these interlayer twist angles. In the formation of moiré patterns for these twist angles, one observes well-defined AA and AB-stacked spots that consequently modify the energy spectrum \cite{moon2013optical} and the emergence of a minimum energy gap \cite{mirzakhani2020circular}. Similarly to Figs.~\ref{ExoW}(a)--\ref{ExoW}(c), by investigating tBLG QDs with a circular shape, Mirzakhani \textit{et al.} \cite{mirzakhani2020circular} observed that such minimum-energy states of the first-lowest electron energy occur at smaller angles when the dot radius increases. This is related to the fact that the smaller the twist angle, the larger the moiré period $L_M$ [see Eq.~\eqref{eq.Lm}], and consequently, a QD with a large radius is necessary to cover a well-defined AA register, behaving like a minimum in the moiré potential where the electronic wavefunctions are expected to be localized. For a tBLG QD with radius $R=4.81$ nm, Mirzakhani \textit{et al.} \cite{mirzakhani2020circular} found that such band gap minimum happens for the twist angle of $\theta = $ 1.53$^{\circ}$. Here, according to Figs.~\ref{ExoW}(a)--\ref{ExoW}(c) for tBLG QRs with a fixed average radius of $R_{m}$ = 4.81 nm, the minima-energy states close to AA ($\theta=0^\circ$) and AB ($\theta=60^\circ$) configurations were obtained roughly at the same twist angles of $\theta_{min}=1.6^\circ$ and $\theta_{min}=58.4^\circ$ for the three investigated QR widths, noting that the latter case with the larger angle close to the AB configuration corresponds to the lowest band gap energy value.

\begin{figure}[!t]
    \centering
    \includegraphics[width=0.90\linewidth]{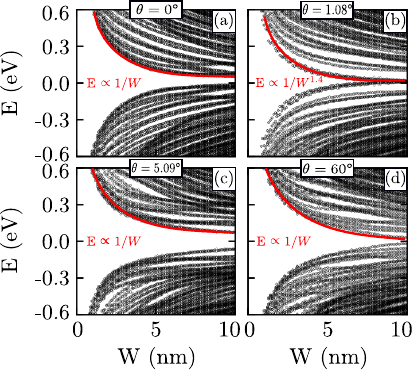}
    \caption{(\textcolor{blue}{Color online}) Energy levels of a circular tBLG QR, with average radius $R_m = 4.81$ nm, as a function of ring width $W$ for different twist angles: (a) $\theta = 0^{\circ}$, (b) $\theta = 1.08^{\circ}$, (c) $\theta = 5.09^{\circ}$ and (d) $\theta = 60^{\circ}$. Red solid curves show the power-law dependence $W^{\beta}$ that fits the lowest QR energy levels.}
    \label{ExW}
\end{figure}

To investigate deeper the effect of ring width on the energy spectrum for different interlayer twist angles, we show in Fig.~\ref{ExW} the energy levels as a function of the tBLG QR width assuming: (a) $\theta = 0^{\circ}$ (AA stacking), (b) $\theta = 1.08^{\circ}$ (magic angle), (c) $\theta = 5.09^{\circ}$, and (d) $\theta = 60^{\circ}$ (AB stacking), with a fixed average radius $R_m = 4.81$ nm. A relevant concern to be addressed here is related to the scaling law $1/W^\alpha$ of the lowest-energy levels of tBLG QRs and its tunability ($1\leq \alpha\leq 2$) by taking different twist angle $\theta$. It is known that (i) the system scalelength ($W$) dependence of semiconductor nanostructures, such as quantum wells and QDs, is such that the energy levels decay approximately as $\propto 1/W^2$ and $\propto 1/R^2$ as a function of the well width and QD radius, respectively; (ii) that the energy levels dependence on the system length of such quantum-confined structures roughly reflect the energy dispersion of the corresponding bulk system; and that (iii) the band structures of AA-stacked and AB-stacked BLGs without taking into account skewed interlayer hoppings exhibit linear [$E(k) \propto k$] and quadratic [$E(k) \propto k^2$] dispersions, and consequently, for BLG quantum confinement structures \cite{PhysRevB.94.035415, PhysRevB.93.165410, BQDsDiego, DiegomagQD, zarenia2010S, zarenia2009electrostatically, da2014analytical}, the lowest-energy levels are expected to scale as $E_{AA} \propto 1/length$ and $E_{AB} \propto 1/length^2$ for AA- and AB-stacked cases, respectively. Therefore, it is reasonable that tBLG QRs exhibit a $1/W^\alpha$ energy states tendency since the rotation between layers yields moiré patterns that are composed by coexisting of AA, AB/BA, and intermediate stacking orders that, in turn, result in changes to the band structures, such that one expects that $\alpha$ should vary from $1<\alpha<2$ for different twist angle as a consequence of the different stacking between the layers. In fact, by analyzing the lowest-energy states in Figs.~\ref{ExW}(a)-\ref{ExW}(d), one observes that the decay power $\alpha = 1$ is the same for AA- and AB-stacked tBLG QRs, as well as for $\theta=5.09^\circ$ case, while it presents a faster energy decreasing of $\alpha = 1.4$ at the magic angle, as shown in Fig.~\ref{ExW}(b). The distinct behavior at the magic angle can be understood as a consequence of the flatness of the dispersion relation close to the Fermi level. \cite{moon2013optical} This characteristic has an overall effect on the energy levels, resulting in a more abrupt close of the gap, which induces a $\alpha > 1$ parameter. \cite{mirzakhani2020circular} Based on the aforementioned statements (ii) and (iii), it was perhaps expected that the QRs with AA [Fig.~\ref{ExW}(a)] and AB [Fig.~\ref{ExW}(d)] stackings would behave differently, exhibiting a different $\alpha$ value, given that their dispersion relations in the low-energy and long-wavelength limits exhibit quite distinct behavior, namely linear and parabolic, which, in turn, would be expected to lead to $\alpha$ values close to $\alpha = 1$ and $\alpha = 2$ for the AA and AB cases, respectively. However, it is important to mention that this coefficient for the power-law is valid in the situation where only perpendicular hoppings take place, as observed for AA- and AB-stacked BLG QD in Refs.~\cite{PhysRevB.94.035415, PhysRevB.93.165410, BQDsDiego, DiegomagQD, zarenia2010S, zarenia2009electrostatically, da2014analytical}. In the situation where skewed hoppings are taken into account, as assumed here [see Sec.~\ref{Sec.Model} and Fig.~\ref{Fig:Sketch}(c)], the trigonal warping effect is relevant, slightly modifying the power-law tendency and leading to preferential localization of the probability densities of the low-energy states with three-fold symmetry, as reported for AB-stacked BLG QD \cite{ge2020visualization, velasco2018visualization, PhysRevB.101.161103, ge2021imaging} and as shall be discussed further here for tBLG QRs with different twist angles [Figs.~\ref{Fig.6} and \ref{Fig.7}]. Such slight modification in the power-law of the lowest-energy levels as a function of dot radius $R$ for infinite-mass defined AA- and AB-stacked tBLG QDs in the absence of a magnetic field, described in a similar theoretical approach as assumed here, also was reported in Ref.~\cite{mirzakhani2020circular} that showed that such energy states exhibit an $\approx 1/R$ and $\approx 1/R^{1.66}$ dependences, respectively, and in accordance with what was presented here for tBLG QRs for the magic twist angle [Fig.~\ref{ExW}(b)], the electron and hole states in tBLG QDs with $\theta=1.08^\circ$ closes rapidly when the dot radius increases by a power-law proportional to $1/R^{2.11}$.



\begin{figure}[t]
    \centering
    \includegraphics[width=\linewidth]{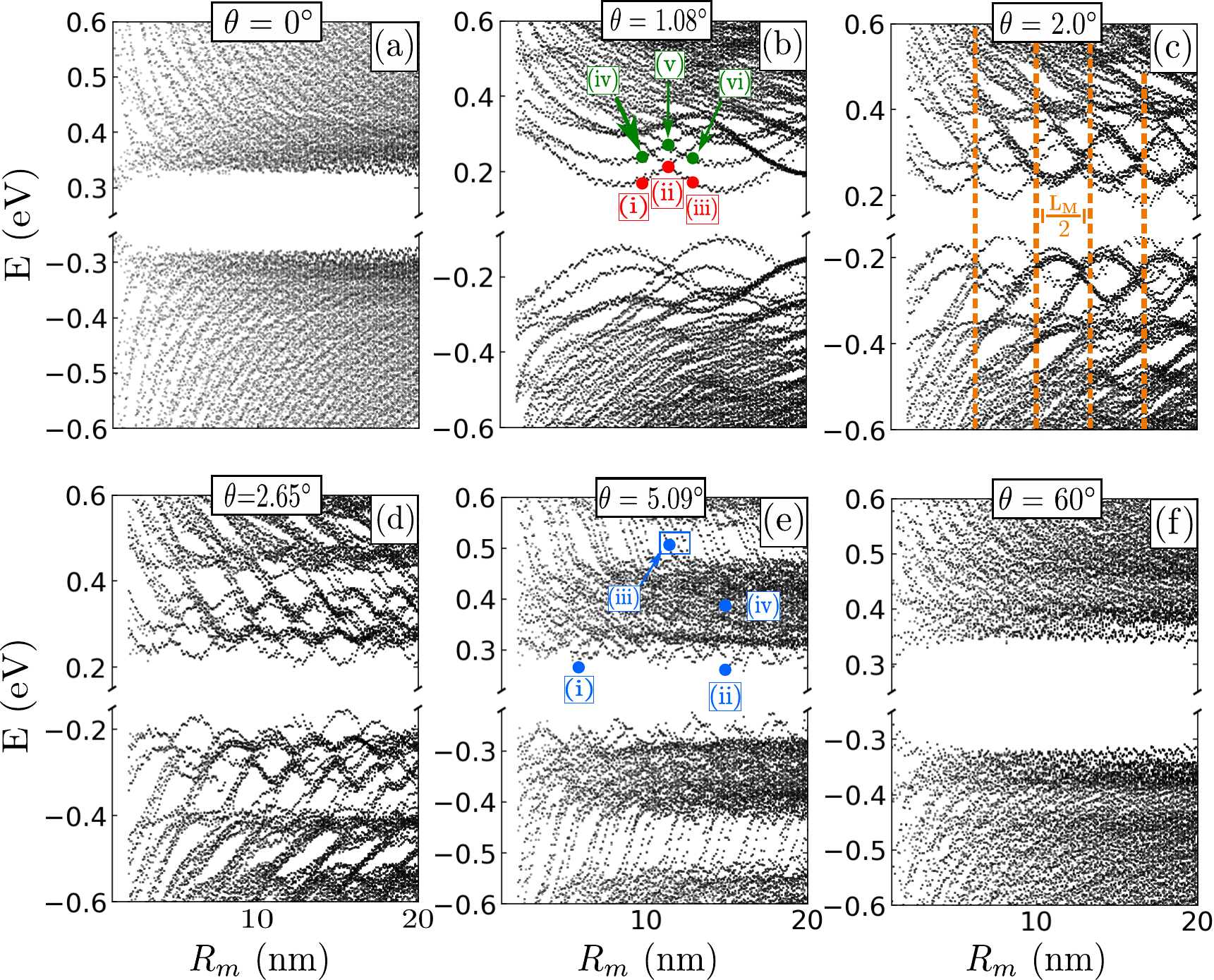}
    \caption{(\textcolor{blue}{Color online}) Energy levels of circular tBLG QRs, with ring width $W = 2.0$ nm, as a function of average radius $R_{m}$ for different twist angle $\theta$ and moiré superlattice period $L_{M}(\theta)$: (a) $\theta = 0^{\circ}$; undefined $L_{M}(0^{\circ})$, (b) $\theta = 1.08^{\circ}$; $L_{M}(1.08^{\circ}) = 13.05$ nm, (c) $\theta = 2.00^{\circ}$; $L_{M}(2.00^{\circ}) = 7.05$ nm, (d) $\theta = 2.65^{\circ}$; $L_{M}(2.65^{\circ}) = 5.32$ nm, (e) $\theta = 5.09^{\circ}$; $L_{M}(5.09^{\circ}) = 2.77$ nm, (f) $\theta = 60^{\circ}$; $L_{M}(60^{\circ}) = 0.246$ nm.}    
    \label{ExRm}
\end{figure}

Interestingly, analyzing the horizontal white dashed line in Fig.~\ref{ExoW}(d) for a fixed QR width ($W=2$~nm), one observes a periodic oscillation in the energy gap when the average radius $R_m$ increases. In addition, Fig.~\ref{ExoW}(d) shows that for large tBLG QR width, such oscillations are absent, that is due to the energy gap being already closed, as can be seen for large $W$ values in Fig.~\ref{ExW}. Such an oscillating behavior is, at first glance, surprising since one would expect to obtain oscillations of the energy levels in systems with ring topology when subjected to an external perpendicularly applied magnetic field, known as Aharonov-Bohm oscillation, as widely discussed in the literature for semiconductor QRs \cite{PhysRev.115.485, PhysRevLett.58.2814, ford1988aharonov, fuhrer2001energy, ford1989electrostatically, VIEFERS20041, fuhrer2002energy, MANNINEN2012119, da2017electronic}, monolayer graphene QRs \cite{da2014geometry, xavier2016electronic, PhysRevB.95.205414, araujo2022modulation, bahamon}, and BLG QRs \cite{PhysRevB.105.115430, zarenia2010S, zarenia2009electrostatically, RASTEGARSEDEHI2022114853, zahidi2017energy, xavier2010topological}. However, in Fig.~\ref{ExRm}, where we investigated the behavior of the energy levels as a function of $R_m$ for a fixed width $W = $ 2 nm and different rotation angles: (a) $\theta = 0^{\circ}$ (AA - stacking), (b) $\theta = 1.08^{\circ}$ (magic-angle), (c) $\theta = 2.00^{\circ}$, (d) $\theta = 2.65^{\circ}$, (e) $\theta = 5.09^{\circ}$, (f) $\theta = 60^{\circ}$ (AB - stacking), no external field is being applied and, even so, it clearly shows oscillatory behavior with the ring average radius. Moreover, by comparing Figs.~\ref{ExRm}(b)-\ref{ExRm}(e), one notices that the oscillation period ($P$) of the low-energy levels as a function of the average radius $R_m$ is strongly dependent on the twisting angle $\theta$, decreasing the larger the twist angle. In fact, a more careful analysis shows that the average radius' period is in the same order of magnitude of half wavelength of the moiré cell [Eq.~\eqref{eq.Lm}], \textit{i.e.} $P(\theta) \propto L_M(\theta)/2$, as indicated by the vertical dashed orange lines in Fig.~\ref{ExRm}(c). Given this dependence and since $L_M$ is undefined for the $\theta=0^\circ$ case [Fig.~\ref{ExRm}(a)] and very small [$L_{M}(60^{\circ}) = a = 0.246$ nm] for the $\theta=60^\circ$ case [Fig.~\ref{ExRm}(f)], these are limiting cases of indefinitely large and small oscillation periods, respectively, justifying the reason why it has a difficult visualization of clear oscillations in Figs.~\ref{ExRm}(a) and \ref{ExRm}(f). This indicates that such oscillations are associated with the presence of the moiré pattern in the crystallographic structure of the tBLG QRs since the oscillations only clearly appear for rotation angles where $L_m$ is well-defined and of the order of the ring width. This will become evident from the following discussions of the probability densities in Figs.~\ref{Fig.6} and \ref{Fig.7}. In addition, as already pointed out in the discussion of Fig.~\ref{ExoW}, note that all energy spectra in Fig.~\ref{ExRm} exhibit a gap due to the mass-potential confinement type assumed here for defining the ring-like structures and that the energy gap on the spectra of tBLG QRs shows a more significant dependence on the ring width, as shown in Fig.~\ref{ExoW}(d).

\begin{figure}[t]
    \centering
    \includegraphics[width=0.7\linewidth]{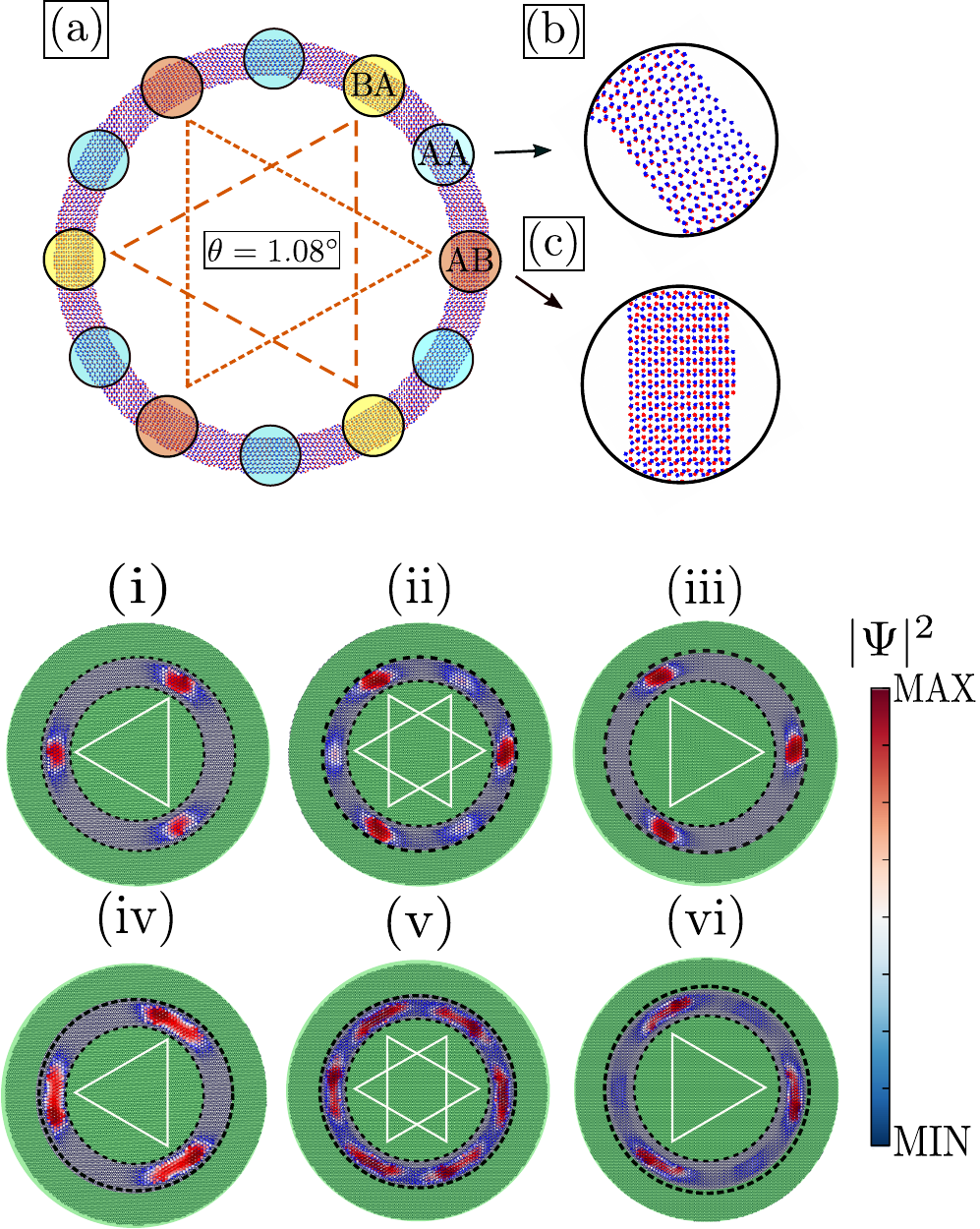}
    \caption{(\textcolor{blue}{Color online}) (a) Sketch of the regions of a $\theta=1.08^\circ$ tBLG QR emphasizing the local stacking registries BA (yellow circles), AA (cyan circles), and AB (orange circles). (b) and (c) show a magnification of the crystal lattice in the AA and AB regions to help visualization. Probability densities for the states labeled as (i)-(vi) in Fig.~\ref{ExRm}(b) are shown in panels (i)-(vi), respectively. Green-shaded regions correspond to the ones with non-zero mass potential (\textit{i.e.}, the confining barriers of the QR). The upper row shows the ground state for three values of average radius: (i) $R_m = 10.20$ nm, (ii) $R_m = 11.40$ nm, and (iii) $R_m = 12.40$ nm, for which the energies are $E = 0.179$ eV, $E = 0.211$ eV, and $E = 0.183$ eV, respectively. The bottom row shows the first excited states for radii (iv) $R_m = 10.20$ nm, (v) $R_m = 11.40$ nm, and (vi) $R_m = 12.40$ nm, with energies $E = 0.245$ eV, $E = 0.276$ eV, and $E = 0.245$ eV, respectively.}
    \label{Fig.6}
\end{figure}

\begin{figure}[t]
    \centering
    \includegraphics[width=0.7\linewidth]{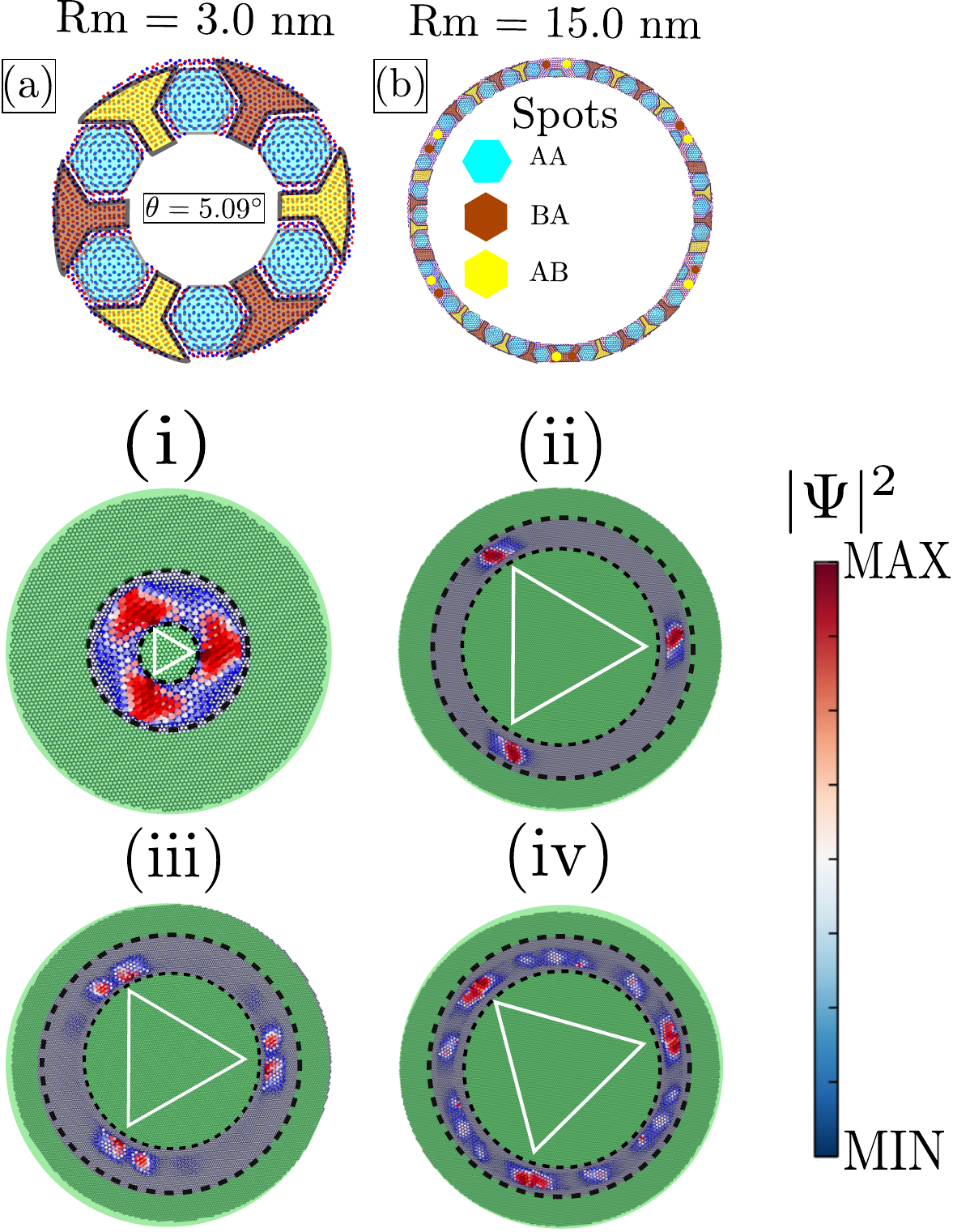}
    \caption{(\textcolor{blue}{Color online}) Sketches of the regions of a $\theta=5.09^\circ$ tBLG QR with an average radius of (a) $R_m = 3$ nm and (b) $R_m = 15$ nm, emphasizing the local stacking registries BA (brown), AA (cyan), and AB (yellow). Probability densities for the states labeled by (i)-(vi) in Fig.~\ref{ExRm}(e) for $\theta=5.09^\circ$. Green-shaded regions correspond to the ones with non-zero mass potential. (i) $E = 0.309$ eV; $R_m = 3.0$ nm, (ii) $E = 0.268$ eV; $R_m = 15.0$ nm, (iii) $E = 0.514$ eV; $R_m = 11.3$ nm, and (iv) $E = 0.392$ eV; $R_m = 15.0$ nm.}
    \label{Fig.7}
\end{figure}

In order to have a proper understanding of the physical nature of the energy levels' oscillations and their connection with the moiré pattern as the average radius $R_m$ varies, we shall examine how the probability densities $|\Psi|^2$ are distributed along the tBLG QRs. For that, we selected six [four] states in Fig.~\ref{ExRm}(b) [Fig.~\ref{ExRm}(e)] for $\theta=1.08^\circ$ [$\theta=5.09^\circ$], marked as points (i)-(vi) [(i)-(iv)] and presented their $|\Psi|^2$ distributions in panels (i)-(vi) [(i)-(iv)] of Fig.~\ref{Fig.6} [Fig.~\ref{Fig.7}], respectively. The states (i)-(iii) and (iv)-(iv) of Fig.~\ref{Fig.6} correspond to the first and second lowest-energy levels in the energy spectrum of Fig.~\ref{ExRm}(b), respectively, energetically located at the vicinity of the crossing point around the state (ii) with $R_m = 11.4$ nm and $E = 0.211$ eV. From Fig.~\ref{Fig.6}, one notices that the probability densities shown in panel (i) for $R_m = 10.20$ nm and $E = 0.179$ eV, and in panel (iii) for $R_m = 12.40$ nm and $E = 0.183$ eV exhibit a three-fold symmetry with pronounced probability density amplitudes localized close of BA and AB stacking regions of the tBLG QRs, respectively, \textit{i.e.} states (i) and (iii) are $\pi/3$-rotated with respect to each other. The preferential localization on AB and BA spots can be verified by the sketch of the BA, AA, and AB regions of the ring in Figs.~\ref{Fig.6}(a), \ref{Fig.6}(b), and \ref{Fig.6}(c). Note that the ground state in Fig.~\ref{ExRm}(b) for $\theta=1.08^\circ$ is formed by a set of two states that cross around the state (ii) that, in turn, has a mix of the probability densities of the states (i) and (iii), \textit{i.e.} the probability density in (ii) of Fig.~\ref{Fig.6} shows features both in AB and BA stacking spots with $|\Psi|^2$ having six peaks, demonstrating that the probability density distribution goes smoothly from AB [state (i)] to BA [state (iii)] preferential localization, thus leading the energy oscillation observed in Fig.~\ref{ExRm}(b) when the average radius increases by keeping the ring width fixed. A structural view of the system [see Figs.~\ref{Fig.6}(a), \ref{Fig.6}(b), and \ref{Fig.6}(c)] allows us to infer that this happens because when one varies the average radius, the area covered by the ring region defined with a fixed width passes through different regions of the moiré profile that have potential minima in the AB and BA stops, and since the moiré pattern is periodic, the preferential minima potential regions for the charge carriers are also periodically covered by the fixed-width-ring region. It leads to the energetic oscillation of the lowest energies, such that the minimum energy values correspond roughly to the situation in which the ring entirely covers the AB or BA regions, whereas the energy values that slightly increase or decrease are associated with the situations in which the ring passes between regions with other different or mixing stacks. By analogy, viewing the AB and BA spots of the moiré pattern as quantum well potentials for the charge carriers' confinement in tBLG QRs, one can easily understand that the second lowest-energy levels in Fig.~\ref{ExRm}(b) should correspond to modes that exhibit probability densities with two-peaks. By analyzing the panels (iv)-(vi) of Fig.~\ref{Fig.6} for these excited energy states around the same energetic region of the oscillation point discussed for panels (i)-(iii) in Fig.~\ref{ExRm}(b), one notices the similar preferential localization of $|\Psi|^2$ of the states (iv)-(vi) as in states (i)-(iii) at AB and BA regions, but now, for states (iv)-(vi), the spatial distribution of the confined states presents a two-peak shape at AB and BA regions, which results from its higher energy (second energy level) and two-peak nodal behavior as in standard semiconductor quantum wells. 

\begin{figure*}[t]
    \centering
    \includegraphics[width=0.8\linewidth]{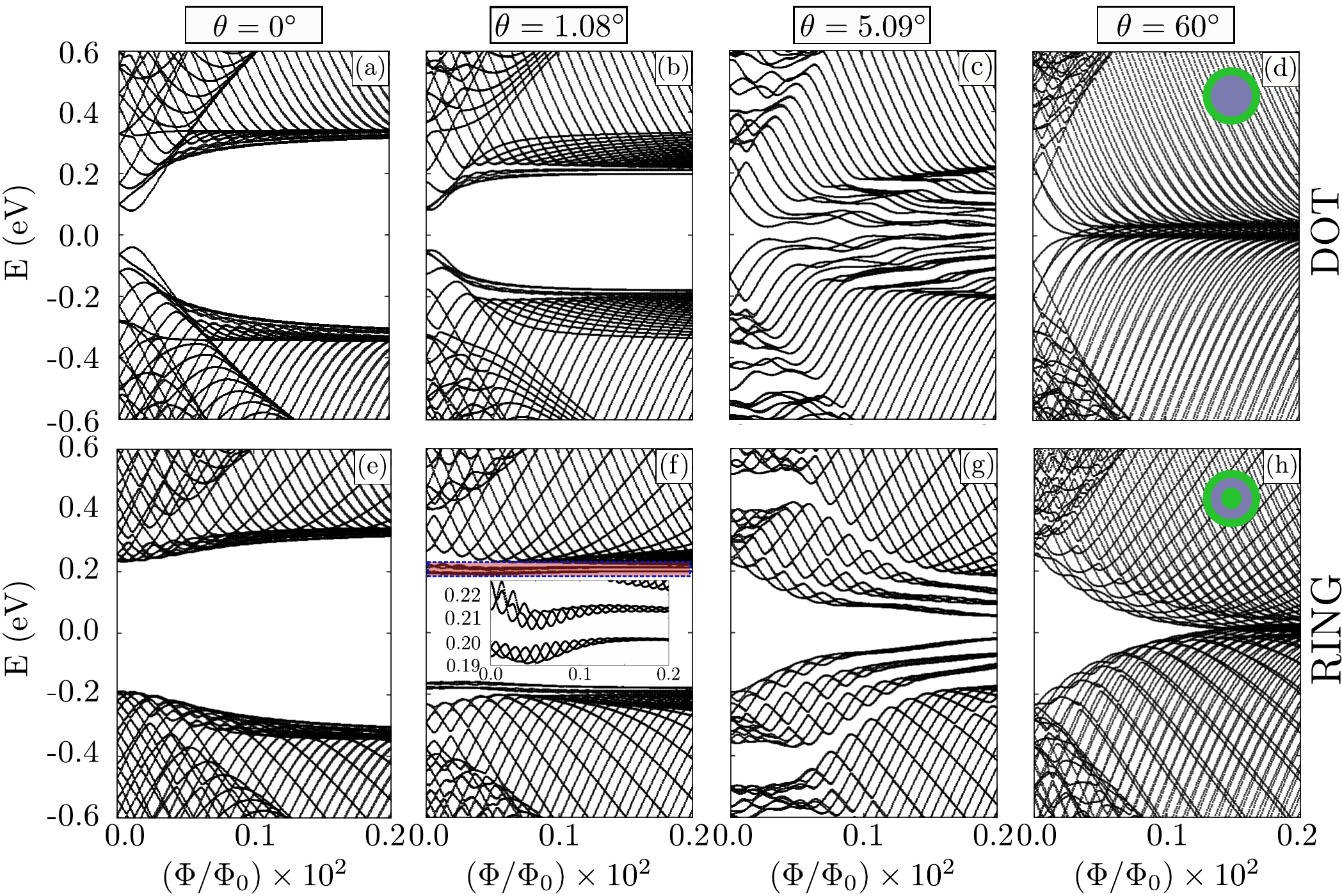}
    \caption{(\textcolor{blue}{Color online}) Energy levels of circular tBLG [top panels - (a) to (d)] QDs, with radius $R = 4.81$ nm, and [bottom panels - (e) to (h)] QRs, with outer radius $R_e = 4.81$ nm and inner radius $R_i = 2.0$ nm, as a function of a magnetic flux $\Phi$ (in units of $\Phi_{0}$) for different twist angle $\theta$: (a, e) $\theta = 0^{\circ}$, (b, f) $\theta = 1.08^{\circ}$, (c, g) $\theta = 5.09^{\circ}$, and (d, h) $\theta = 60^{\circ}$. The inset in panel (f) shows a magnification of the results highlighted by the shaded red area in the spectrum, emphasizing the low-energy levels with two subbands formed by three oscillating states.}
\label{Fig.8}
\end{figure*}

From Fig.~\ref{ExRm}, one realizes that the oscillations of the lowest energies are clearly observed up to the $\theta = 2.65^{\circ}$ twist angle in Fig.~\ref{ExRm}(d) as a consequence of the interplay between energetically favorable regions at AB and BA spots of the ring, as physically discussed for states (i)-(vi) of Fig.~\ref{ExRm}(b) for $\theta=1.08^\circ$ in Fig.~\ref{Fig.6}. For larger twist angle values, groups of energy states emerge, as shown for $\theta = 5.09^{\circ}$ in Fig.~\ref{ExRm}(e), where three groups of energy states with distinct behavior as a function of $R_m$ are observed: the first group of states is associated with the lowest energies that exhibit oscillatory behavior [see states (i) and (ii) in Fig.~\ref{ExRm}(e)]; the second group of states corresponds to highest energy levels with a decreasing tendency grouped in four states [see state (iii) in Fig.~\ref{ExRm}(e)]; and the third group of states with intermediary energies [see state (iv) in Fig.~\ref{ExRm}(e)]. The probability densities of four states of these three groups are shown in Figs.~\ref{Fig.7}(i)-\ref{Fig.7}(iv). Indeed, for such a higher twist angle, the AB, AA, and BA spots become very small along the ring perimeter, especially for larger radius $R_m$, as illustrated in Figs.~\ref{Fig.7}(a) and \ref{Fig.7}(b). This suggests a suppression of the role played by the moiré pattern in the electron confinement in the ring. In the case of a small ring radius $R_m = 3$ nm, such as the one for the state (i), the spots are still clear, and the electron is clearly confined in the AB regions. Interestingly, its wave function in these spots exhibits a trigonal warped shape as a consequence of the $\gamma_3$ term included in our model. \cite{ge2020visualization, velasco2018visualization, PhysRevB.101.161103, ge2021imaging} As we take the lowest energy state at a larger average radius, as illustrated in Fig.~\ref{Fig.7}(b) for $R_m = 15$ nm and shown in panel (iv) of Fig.~\ref{Fig.7}, the electron is still confined in these spots, although they are much narrower as compared to the $\theta = 1.08^{\circ}$ case, and accordingly, the lowest energy state still exhibits a weak oscillation in $R_m$, such as the one observed in Figs.~\ref{ExRm}(b)-\ref{ExRm}(d). Conversely, energy states in the second region, such as the one shown in panel (iii) of Fig.~\ref{Fig.7}, exhibit energies that only decrease with $R_m$ grouped in four states. These are states whose probability densities exhibit doubled peaks, which suggests they are excited states of the AB confinement regions. As for states with intermediary energy, \textit{i.e.}, in the third energy group of Fig.~\ref{ExRm}(e), the probability density illustrated in panel (iv) of Fig.~\ref{Fig.7} shows that these states are distributed along different regions of the ring with no clear dependence on $R_m$, exhibiting a mix localization with high intensity in three-fold AB/BA confinement region.

\subsection{Influence of a perpendicularly applied magnetic field}\label{sec.B.neq.0}



Under a perpendicular magnetic field, energy shifts are usually expected on the energy spectrum of QDs, \cite{harrison2016quantum} whereas Aharonov-Bohm oscillations in the energy levels are expected for QRs. \cite{PhysRev.115.485, ford1988aharonov, fuhrer2001energy, ford1989electrostatically, fuhrer2002energy, VIEFERS20041, MANNINEN2012119, da2017electronic} It is well-known that an applied magnetic field breaks the time-reversal symmetry and consequently induces in graphene-based systems the degeneracy lifting of the K/K$^\prime$ energy states. \cite{PhysRevB.92.201412, PhysRevLett.99.236809, PhysRevB.77.235406, PhysRevLett.96.136806, KANDEMIR20152120} Indeed, for low magnetic flux through the dot, QD states of tBLG shown in Figs.~\ref{Fig.8}(a)-\ref{Fig.8}(d) exhibit energy shifts that are reminiscent of the results observed for monolayer graphene QDs, \cite{TGQDs2B, GQDbook, li2022recent, lavor2020magnetic, schnez2008analytic, ZareniaQDs, PetersGQDsMagnetic, gucclu2013zero, PhysRevB.95.235427} and the Landau levels of the BLG bulk system are not meaningful at these low magnetic field amplitudes due to the important role played by the finite size of the system. On the other hand, at a high magnetic field, the magnetic length becomes smaller than the system size so that confinement effects are strongly reduced, and therefore, the convergence of excited states to the Landau levels of infinite BLG sheet is observed, matching the regions of the BLG QD spectrum with higher density of states [omitted here from the spectra of Fig.~\ref{Fig.8} for visualization reasons of the lowest energy levels]. Even though the gaps of QD and QR at zero magnetic field are of the same order of magnitude for all interlayer twist angles (including $\theta=0^{\circ}$) due to the infinite mass term assumed here to define the tBLG confinement nanostructures as discussed in Sec.~\ref{sec.B.eq.0}, as the magnetic flux increases, results for both tBLG QD and QR are qualitatively different from the untwisted ($\theta = 0^{\circ}$) case: the energy spectra for $\theta = 5.09^{\circ}$ and $\theta = 60^{\circ}$ in Figs.~\ref{Fig.8}(c), \ref{Fig.8}(d), \ref{Fig.8}(g), and \ref{Fig.8}(h) exhibit convergence to a set of states with zero energy for high magnetic flux. These energy levels approaching the zeroth Landau level at $E=0$ are the so-called quantum Hall edge states that are grouped in pairs of states for BLG QD cases, as can be clearly seen in Figs.~\ref{Fig.8}(d) for $\theta = 60^{\circ}$ tBLG QD that has very close energies but being nondegenerate, \cite{da2014analytical, PhysRevB.93.165410, DiegomagQD, PhysRevB.94.035415} and in addition to that, the QR structures [Figs.~\ref{Fig.8}(e)-\ref{Fig.8}(h)] exhibit an extra quantum Hall state with increasing energy \cite{PhysRevB.95.235427, PhysRevB.105.115430}. Also, energy states of tBLG QRs, shown in Figs.~\ref{Fig.8}(e)-\ref{Fig.8}(h), exhibit oscillations as a function of the magnetic flux $\Phi$ through the ring, as it is ubiquitous to QRs under magnetic field. Surprisingly, though, the lowest energy states of QDs with small angles ($\theta = 0^{\circ}$ - $5.09^{\circ}$) also exhibit periodic oscillations at high magnetic flux - let us discuss the origin of these oscillations.

The length of the moiré pattern in tBLG can be inferred from its twist angle by Eq.~\eqref{eq.Lm}. The area of the moiré pattern hexagon of side $L_M$ is then used to obtain the magnetic flux through it, which is then compared to the magnetic flux quantum $\Phi_0$ so that an Aharonov-Bohm minimum is observed whenever the flux through that area is an integer multiple of $\Phi_0$. Indeed, for a QD in tBLG with $\theta = 5.09^\circ$, the period of the minima of the lowest energy states shown in Fig.~\ref{Fig.8}(c) matches very well the expected values for an Aharonov-Bohm oscillation due to confinement in its hexagonal moiré pattern [see sketches and preferential localization of the probability densities in the BA/AB-staked regions in tBLG QR cases in Figs.~\ref{Fig.6} and \ref{Fig.7}]. Therefore, even though the crystal structure of this system looks like a \textit{dot}, the moiré pattern confinement leads effectively to a hexagonal \textit{ring} for the lowest energy states, thus explaining the Aharonov-Bohm oscillation observed here for tBLG QDs.

On the other hand, for $\theta = 1.08^\circ$, the moiré period $L_M$ is larger than the QD radius considered here [$L_M(1.08^\circ)=13.05~\mbox{nm}>R=4.81~\mbox{nm}$]. Therefore, the lowest energy states' oscillations observed for such tBLG QD in Fig.~\ref{Fig.8}(b), especially at high $\Phi$, cannot be due to an effective ring-like confinement resulting from a moiré pattern. Indeed, the period of these oscillations does not match the one expected for its moiré pattern but rather the one for \textit{edge localization states} of the QD. Furthermore, since the $\theta = 1.08^\circ$ tBLG QR considered in Fig.~\ref{Fig.8}(f) has the same outer radius as the tBLG QD in Fig.~\ref{Fig.8}(b), it should also exhibit such states confined around its outer edge. In fact, the Aharonov-Bohm oscillations observed in the results magnified by the inset of Fig.~\ref{Fig.8}(f) match the same Aharonov-Bohm oscillation period as the tBLG QD in Fig.~\ref{Fig.8}(b), suggesting the same origin for both. This shall be demonstrated here further when discussing the tBLG QD and QR probability densities in Figs.~\ref{Fig.10}, \ref{Fig.11}, and \ref{Fig.12}. 

\begin{figure}[t]
	\centering
	\includegraphics[width=\linewidth]{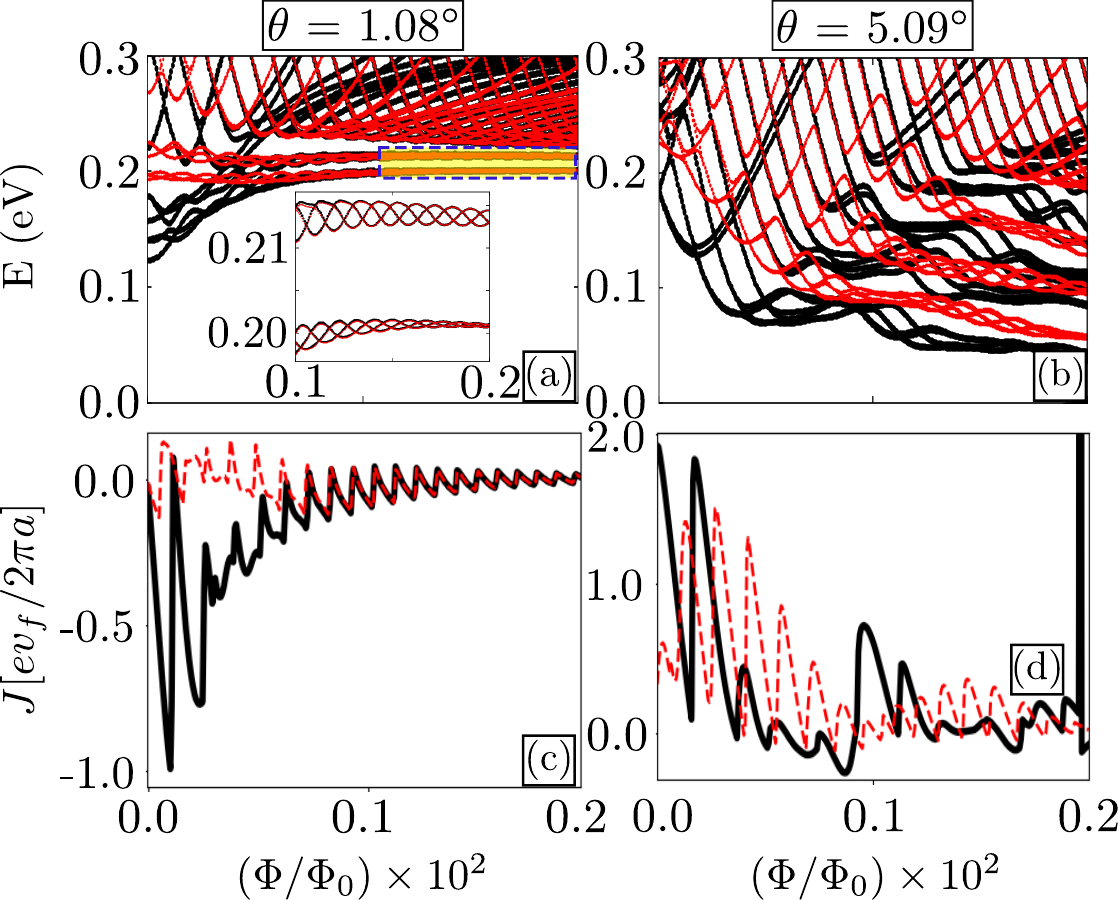}
	\caption{(\textcolor{blue}{Color online}) (a, b) Lowest energy levels of a tBLG QR as a function of magnetic flux $\Phi$ (in units of $\Phi_{0}$) with outer radius $R_e = 4.81$ nm and inner radius $R_i = 1$ nm (black curves) [$R_i = 2.0$ nm (red curves)]. (c, d) Current densities for the ground state of panels (a) and (b). Left and right panels correspond to results for $\theta=1.08^{\circ}$ and $\theta=5.09^{\circ}$, respectively. The inset in panel (a) shows a magnification of the results highlighted by the shaded area in the spectrum.}
	\label{Fig.9}
\end{figure}

To help support these arguments about the nature of the lowest energy states involved in the Aharonov-Bohm oscillations of tBLG QRs, a comparison between lowest energy states for QRs with inner ring radii $R_i =1$ nm (black) and 2 nm (red) as a function of magnetic flux is shown in Figs.~\ref{Fig.9}(a) and \ref{Fig.9}(b) for $\theta=1.08^\circ$ and $\theta=5.09^\circ$, respectively. Results for the probability density current calculated with the ground state of the same ring structures as in Figs.~\ref{Fig.9}(a) and \ref{Fig.9}(b) are shown in Figs.~\ref{Fig.9}(c) and \ref{Fig.9}(d), respectively, to improve the visualization of the oscillation period. The probability current density is computed here using the method developed in Ref.~\cite{de1992probability} and widely employed in graphene-based nanostructured systems, as demonstrated in Refs.~\cite{da2014geometry, PhysRevB.86.115434, lavor2020magnetic, araujo2022modulation}, that is based on the finite difference scheme for the probability density that, in turn, obeys the continuity relation and is calculated using tight-binding eigenfunctions.

Figures~\ref{Fig.9}(a) and \ref{Fig.9}(c) show that the Aharonov-Bohm oscillations observed in the lowest energy states of the $\theta = 1.08^\circ$ tBLG QR case at high magnetic flux do not depend on the inner radius, which indicates that the states involved in them are localized along the outer ring edge. This is the reason why we shall name these states as \textit{edge localization states}. The results for different inner radii differ only for low fields, such that the results of the ring with a smaller inner radius, $R_i= 1$ nm (black), are reminiscent of that of the QD shown in Fig.~\ref{Fig.8}(b), as expected. Conversely, although the lowest energy states for QRs with different inner radii for $\theta = 5.09^\circ$ tBLG QR case present similar energetic aspects with the formation of energy subbands that oscillate with the magnetic flux, Figs.~\ref{Fig.9}(b) and \ref{Fig.9}(d) demonstrate that these energies and their oscillation periods for the $\theta = 5.09^\circ$ case strongly depend on the inner radius of the QR in the whole range of magnetic flux considered here, which implies that the states involved in these Aharonov-Bohm oscillations are not localized in its outer edge.   

\begin{figure}[t]
    \centering
    \includegraphics[width=\linewidth]{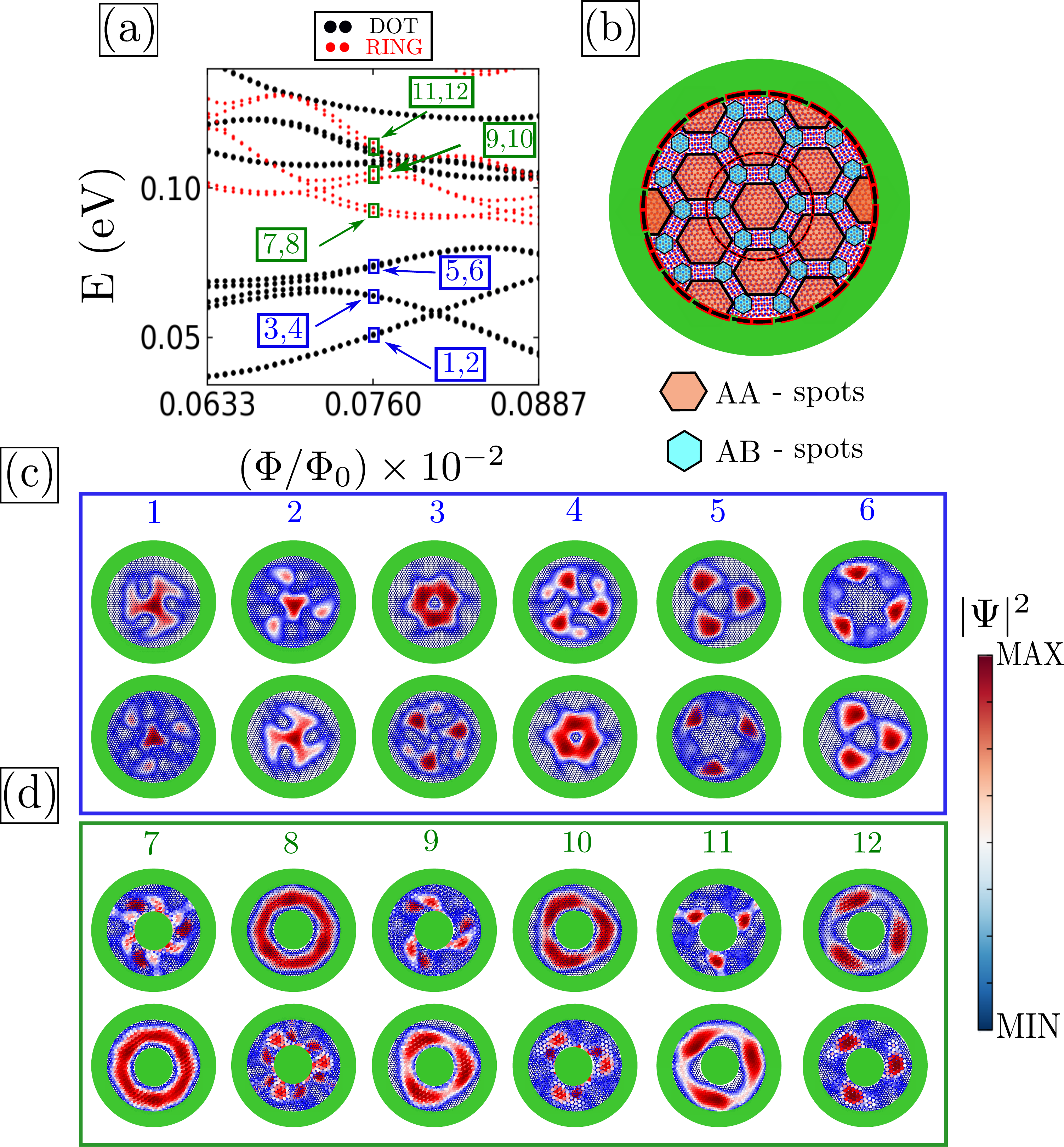}
    \caption{(\textcolor{blue}{Color online}) (a) Lowest energy levels of tBLG QD (black symbols) and QR (red symbols) as a function of the magnetic flux $\Phi$ (in units of $\Phi_{0}$) for a twist angle $\theta = 5.09^{\circ}$. The dot and ring outer radius is $R_e = 4.81$ nm, and the inner radius for the ring is $R_i = 2$ nm. (b) The lattice structure corresponding to the computed nanostructures, where the AA (orange hexagons) and AB (blue hexagons) stacking regions in the moiré pattern are highlighted. The inner red dashed circle denotes the inner radius of the QR. Probability densities for the states labeled as (1-6) are illustrated in (c) for the QD case, while those labeled as (7-12) for the QR case are shown in (d). States (1) and (2) correspond to the energy $E \approx 0.051$ eV, (3) and (4) to $E \approx 0.064$ eV, (5) and (6) to $E \approx 0.074$ eV, (7) and (8) to $E \approx 0.092$ eV, (9) and (10) to $E \approx 0.105$ eV, and (11) and (12) to $E \approx 0.115$ eV. The top and bottom rows of color plots represent contributions from the top and bottom layers of the tBLG.}
	\label{Fig.10}
\end{figure}

A deeper investigation of the lowest energy states involved in subbands' formation composed of (anti)crossings that lead to the Aharonov-Bohm oscillations for both tBLG QD [Figs.~\ref{Fig.8}(b) and \ref{Fig.8}(c)] and QR [Figs.~\ref{Fig.8}(f) and \ref{Fig.8}(g)] at non-zero low twist angles is performed in Figs.~\ref{Fig.10} and \ref{Fig.11} for the $\theta = 5.09^\circ$ and $\theta = 1.08^\circ$ cases, respectively. Figure~\ref{Fig.10} shows a magnification of the results of Figs.~\ref{Fig.8}(c) (for QDs - black symbols) and \ref{Fig.8}(g) (for QRs - red symbols) for values of magnetic flux in the vicinity of $\Phi/\Phi_0 \approx 7.5 \times 10^{-4}$. The probability densities of the eigenstates labeled by (1)-(6) and (7)-(12) indicated in Fig.~\ref{Fig.10}(a) are plotted in Figs.~\ref{Fig.10}(c) and \ref{Fig.10}(d) for tBLG QDs and QRs, respectively. The electron densities for the bottom and upper layers are shown separately in the bottom and top rows of plots, respectively. Generally speaking, the probability densities in Figs.~\ref{Fig.10}(c) and \ref{Fig.10}(d) for both tBLG QDs and QRs exhibit a three-fold symmetry, \textit{i.e.}, a ring-like structure with high-intensity peaks in the vertices of a triangle, as a consequence of the moiré pattern and the preferential localization of the charged particles in such nanostructures to be localized in the AB/BA spots, as discussed in Sec.~\ref{sec.B.eq.0}. This can be verified by analyzing the different AA and AB registers in Figs.~\ref{Fig.10}(b) (where the inner and outer red dashed circles indicate the two boundaries of the ring-like structure, the outer black dashed circle denotes the boundary of the QD, and orange and cyan hexagons correspond to AA and AB spots) and the probability densities in Figs.~\ref{Fig.10}(c) and \ref{Fig.10}(d). In addition, note that the pair of states (1)-(2), (3)-(4), and (5)-(6) for tBLG QD [(7)-(8), (9)-(10), and (11)-(12) for tBLG QR] will present nearly similar total probability density (by adding $|\Psi_{up}|^2 + |\Psi_{bottom}|^2$), since they have very close energies [see Fig.~\ref{Fig.10}(a)] and their wave functions are almost layer symmetric, such that the electron densities for the upper and lower layers for one of the states is nearly transformed into the lower and upper contributions for the other state by a $\pi/3$-reflection, respectively.


The lowest energy levels for the QRs in all cases shown in Figs.~\ref{Fig.8}(e)-\ref{Fig.8}(h) exhibit Aharonov-Bohm oscillations, but a qualitative difference is observed with a non-zero twist angle: the eigenstates are arranged in three-fold groups of oscillating energies, in which each of the oscillating energy states that form the three-fold subband is, in fact, quasi-double degenerate, \textit{i.e.} it corresponds to a total of six states, as one can see more clearly in Figs.~\ref{Fig.8}(f) and \ref{Fig.8}(g) for $\theta = 1.08^{\circ}$ and $\theta = 5.09^{\circ}$, respectively, and in the zoom of the lowest-energy subband of the latter case counting the six states as depicted in Fig.~\ref{Fig.10}(a). The probability densities in Fig.~\ref{Fig.10}(d) help to understand the origin of this phenomenon: the QR eigenstates in $\theta = 5.09^{\circ}$ tBLG capture the moiré pattern structure and are thus confined in a three-fold symmetric configuration, thus leading to such three-fold groups of oscillating energies, which are reminiscent of the Aharonov-Bohm oscillation of triangular graphene QR. \cite{da2014geometry, TGQDs2B, PhysRevB.83.174441} 

\begin{figure}[t!]
	\centering
	\includegraphics[width=0.9\linewidth]{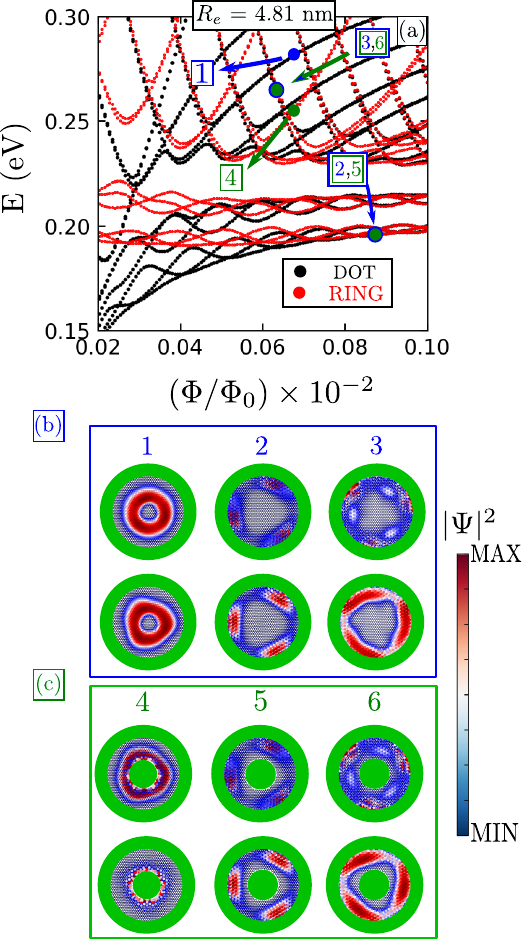}
	\caption{(\textcolor{blue}{Color online}) (a) Lowest energy levels of the tBLG QD (black symbols) and QR (red symbols) as a function of the magnetic flux $\Phi$ (in units of $\Phi_{0}$) for a twist angle $\theta = 1.08^{\circ}$. The dot and ring outer radius is $R_e = 4.81$ nm, and the inner radius for the ring is $R_i = 2$ nm. Panels (b) and (c) display the probability densities for the QD [(1-3)] and QR [(4-6)] cases, respectively, for the states labeled as (1), where $E \approx 0.270$ eV and $(\Phi/\Phi_0)\times 10^{-2} \approx 0.076$, (2), where $E \approx 0.198$ eV and $(\Phi/\Phi_0)\times 10^{-2} \approx 0.088$, (3), where $E \approx 0.265$ eV and $(\Phi/\Phi_0)\times 10^{-2} \approx 0.063$, (4), where $E \approx 0.282$ eV and $(\Phi/\Phi_0)\times 10^{-2} \approx 0.065$, (5), where $E \approx 0.198$ eV and $(\Phi/\Phi_0)\times 10^{-2} \approx 0.088$, and (6), where $E \approx 0.256$ eV and $(\Phi/\Phi_0)\times 10^{-2} \approx 0.063$. The top and bottom rows of color plots represent contributions from the top and bottom layers of the tBLG.}
	\label{Fig.11}
\end{figure}

\begin{figure}[t!]
	\centering
	\includegraphics[width=0.9\linewidth]{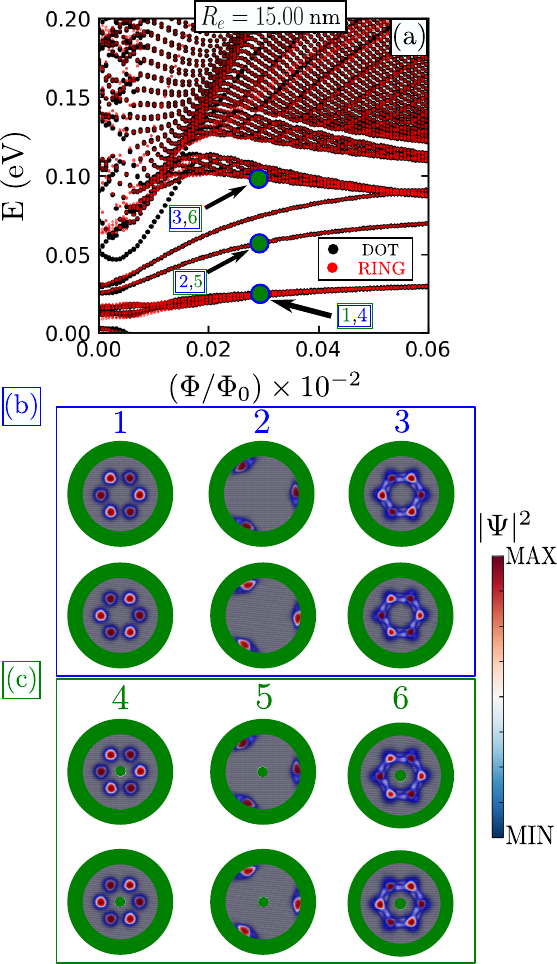}
	\caption{(\textcolor{blue}{Color online}) (a) Lowest energy levels of the tBLG QD (black symbols) and QR (red symbols) as a function of the magnetic flux $\Phi$ (in units of $\Phi_{0}$) for a twist angle $\theta = 1.08^{\circ}$. The dot and ring outer radius is $R_e = 15.0$ nm, and the inner radius for the ring is $R_i = 2$ nm. Panels (b) and (c) display the probability densities for the QD [(1-3)] and QR [(4-6)] cases, respectively, taking $(\Phi/\Phi_0)\times 10^{-2}\approx 0.038$ for the states labeled by (1) and (4) for $E \approx 0.026$ eV, (2) and (5) for $E \approx 0.062$ eV, and (3) and (6) for $E \approx 0.095$ eV. The top and bottom rows of color plots represent contributions from the top and bottom layers of the tBLG.}
	\label{Fig.12}
\end{figure}

\begin{figure*}[t!]
	\centering
	\includegraphics[width=0.7\linewidth]{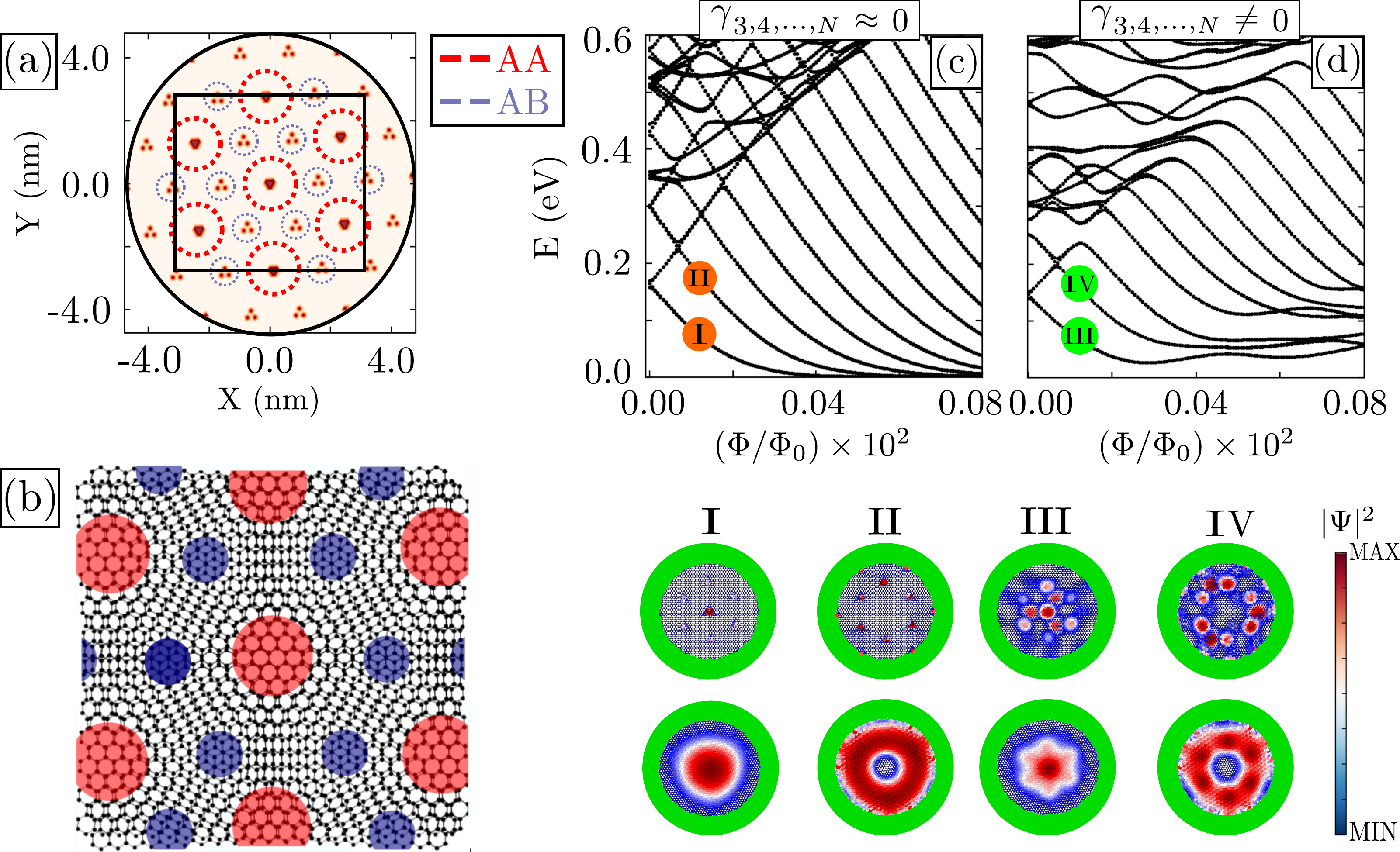}
	\caption{(\textcolor{blue}{Color online}) Contour plot of the spatial distribution of interlayer hopping energies in the lattice structure of a $\theta = 5.09^\circ$ tBLG QD with radius of $R = 4.81$ nm (black circle), emphasizing the AA (dashed red circles) and AB (dashed blue circles) spots. White (low) to red (high) color range is associated with the intensity of hopping between layers. The crystal structure in the square defined by black lines in (a) is illustrated in (b), where AA (red) and AB (blue) regions are also emphasized. Lowest energy levels of a tBLG QD  considering (c) only the nearly perpendicular interlayer hoppings (\textit{i.e.}, assuming $\gamma_{3,4,...N} = 0$) and (d) all interlayer hoppings within a cutoff radius of $d \leq 4a_{cc}$ ($\gamma_{3,4,...N}\neq 0$). Panels (I)-(IV) correspond to the probability densities for the states labeled in the spectra (c) and (d). The top and bottom rows of color plots represent contributions from the top and bottom layers of the tBLG.}
	\label{Fig.13}
\end{figure*}

Figure~\ref{Fig.11}(a) magnifies and combines the results of Figs.~\ref{Fig.8}(b) (for QDs - black symbols) and \ref{Fig.8}(f) (for QRs - red symbols) within the range of energies $E \in [0.15, 0.3]$ eV and magnetic flux $\Phi/\Phi_0 \in [2, 10]\times 10^{-4}$. The probability densities for eigenstates of these tBLG QDs and QRs labeled by (1)-(3) and (4)-(6) are shown in Figs.~\ref{Fig.11}(b) and \ref{Fig.11}(c), respectively, with the layer contributions being plotted separately in the top and bottom rows of plots. As previously discussed, the lowest energy states for QDs and QRs at $\theta = 1.08^{\circ}$ twist angle under high magnetic flux amplitudes are almost the same - this is now explained by the fact that their wavefunctions [states (2) for the QD and (5) for the QR] are both localized in the outer edge of such nanostructures with three-fold symmetry so that cutting a hole in the middle of the QD to form a QR does not affect the results. Similar spatial distribution is observed for the higher energy levels (3) and (6) as well, where eigenenergies of QR and QD also match. Conversely, states (1) and (4) are exclusively in the tBLG QD and QR spectra, respectively. They both exhibit a ring-like shape, with three high-intensity peaks similar to the three-fold symmetry also observed for the other states, but in the QD case, this is simply due to the fact that this is an excited non-zero angular momentum state, for which wavefunctions are naturally zero in origin. Notice, however, that what we call \textit{edge localization state} here is not the same as the so-called \textit{zigzag edge state} due to dangling bonds in graphene \cite{ZareniaQDs, DiegomagQD, BQDsDiego, lavor2020magnetic} - the latter originates from a purely microscopic description of the edges, whereas the system boundaries here are defined by infinite mass terms and consequently pure edge states are absent.

In order to verify the previous assumption that the absence of moiré-induced Aharonov-Bohm oscillations in the $\theta = 1.08^{\circ}$ tBLG QD in Fig.~\ref{Fig.8}(b) results from the small radius of the dot ($R_e = 4.81$ nm), we show in Fig.~\ref{Fig.12} results for $\theta = 1.08^{\circ}$ tBLG QD and QR with a larger outer radius, $R_e = 15$ nm, that is larger than the moiré length [$L_M(1.08^\circ)=13.05~\mbox{nm}<R_e = 15~\mbox{nm}$]. Very shallow Aharonov-Bohm oscillations are observed in the energy levels in this case, as one can verify in Fig.~\ref{Fig.12}(a). Interestingly, though, the results for tBLG QD and QR are almost exactly the same in this case regardless the range of energy and magnetic flux - this is due to the fact that the wavefunctions of the QD are strongly localized either in the moiré pattern or in the outer edge so that cutting a $R_i = 2$ nm circular hole, namely, one with a radius smaller than the moiré hexagon size, in the QD to form a QR does not affect the results. This can be easily verified in the probability densities for the states labeled by (1)-(3) and (4)-(6) in Fig.~\ref{Fig.12}(a), which are depicted in Figs.~\ref{Fig.12}(b) and \ref{Fig.12}(c) for QDs and QRs, respectively, where moiré and edge localization are clearly observed. Note the equivalence between states (1)-(4), (2)-(5), and (3)-(6).

Notice that our theoretical model considers all inter-layer hopping energies for atoms separated by a distance $d_{ij} \leq 4a_{cc}$, as explained in Sec.~\ref{Sec.Model}. This means that once we set an atom $i$ in the top or bottom crystal lattice, we consider all the hopping energies $\gamma_2 = t(d_{i2})$, $\gamma_3 = t(d_{i3})$, $\gamma_4 = t(d_{i4})$, etc., to its first, second, third, etc. nearest neighbor in the adjacent layer, respectively, up to a distance $\approx 4a_{cc}$. The importance of accounting for such wide range in the description of tBLG is emphasized in Fig.~\ref{Fig.13}: considering only the hopping to the closest atom in the other layer, \textit{i.e.}, setting the ``skewed'' hopping parameters $\gamma_{3,4...,N}$ to zero and keeping only the nearly vertical one in each site in the model, the eigenstates of the QD behave similar to those of a non-twisted BLG QD \cite{PhysRevB.93.165410, pereira2007tunable, DiegomagQD, BQDsDiego, da2014analytical}, where lowest energy states - the quantum Hall edge states - converge to zeroth Landau Level at $E = 0$ at high magnetic fields, as shown in Fig.~\ref{Fig.13}(c). The wavefunctions of the eigenstates labeled by (I) and (II) in Fig.~\ref{Fig.13}(c) exhibit only very weak confinement in the AA spots of the moiré pattern, as sketched in Figs.~\ref{Fig.13}(a) (see red dashed circles) and \ref{Fig.13}(b) (red spots), and their probability densities are nearly circular. On the other hand, a proper confinement of the eigenfunctions at the AB/BA regions is only observed in the case where $\gamma_{3,4...,N} \neq 0$, which leads to the previously discussed Aharonov-Bohm oscillations of the lowest levels of this tBLG QD [compare energy spectra Fig.~\ref{Fig.8}(c) and Fig.~\ref{Fig.13}(d)], as indicated in the energy spectra of Fig.~\ref{Fig.13}(d) by states (III) and (IV) and shown in the panel below of Fig.~\ref{Fig.13}(d), exhibiting the previously discussed three-fold symmetry.  

\section{Summary and final remarks}\label{Sec.Conclusion}

In summary, we have systematically investigated the electronic properties of tBLG QDs and QRs for circular samples defined by means of a mass potential profile, which prevents edge states and leads to a gaped energy spectrum. For that, the role of different system parameters, such as interlayer twist angles, dot radii, ring widths, and ring average radii, were explored. In order to understand the results emerging from ring-like confinement in tBLG, we compared them with those for tBLG QDs. In this context, discussions were conducted in view of the already-reported results for twisted \cite{mirzakhani2020circular} and untwisted \cite{da2014analytical, BQDsDiego, pereira2007tunable, DiegomagQD} BLG QDs. 


In the absence of a magnetic field, our findings reveal that (i) the energy levels of tBLG QRs scale differently with the ring width depending on the twist angle, with the lowest-energy levels exhibiting a faster decaying to $E=0$ at the magic angle $\theta=1.08^\circ$ due to the band flattening of the moiré band for the infinite tBLG case; and that (ii) the lowest energy levels oscillate as a function of the QR average radius with a period in the order of half the moiré length $L_{m}(\theta)$. The latter is a consequence of the interplay between the different distributions of AA and AB/BA stacking regions for each value of radius and the strong confinement induced by the moiré superlattice at AB/BA stacking spots. 

In the presence of a magnetic field, a mix of two sets of quantum Hall states, emerging from inner and outer QR edges, was observed, along with a variety of crossings between the energy states as the magnetic flux increases. For $0^\circ<\theta < 60^\circ$, one observes the formation of low energy subbands, and, surprisingly, both QR and QD states exhibit very similar results, with a better agreement between them for high magnetic field amplitudes. We showed that the origin of such unexpected matching of the tBLG QD and QR energy levels lies in the preferential localization of their eigenfunctions, exhibiting a three-fold symmetry for both tBLG QD and QR samples. By manually decoupling the skewed interlayer hoppings, we verified that a circular symmetry aspect for the wavefunctions took place in the otherwise three-fold symmetric wave functions. This led us to conclude that such three-fold symmetry is a consequence of the trigonal warping and moiré confinement effects in tBLG, caused by the inclusion of next-nearest interlayer hoppings in the model and its tuning by twisting the two coupled graphene layers.

Tailoring the physical properties of BLG-based quantum systems via interlayer rotation of the two graphene layers has been one of the main goals of the condensed matter community in the last few years. In this context, our results reveal relevant information for future technological applications based on the physics of quantum confined charge carriers in twisted bilayers, as it demonstrates the role of the interplay between moiré pattern-induced features and the structural limitations imposed by the QR and QD confinement structures on their electronic properties.


\section*{Acknowledgments}

The authors would like to thank the National Council of Scientific and Technological Development (CNPq) through Universal and PQ programs and the Brazilian National Council for the Improvement of Higher Education (CAPES) for their financial support. D.R.C gratefully acknowledges the support from CNPq grants $313211/2021-3$, $437067/2018-1$, $423423/2021-5$, $408144/2022-0$, the Research Foundation—Flanders (FWO), and the Fundação Cearense de Apoio ao Desenvolvimento Científico e Tecnológico (FUNCAP).

\bibliography{refs.bib}

\begin{thebibliography}{94}%
\makeatletter
\providecommand \@ifxundefined [1]{%
 \@ifx{#1\undefined}
}%
\providecommand \@ifnum [1]{%
 \ifnum #1\expandafter \@firstoftwo
 \else \expandafter \@secondoftwo
 \fi
}%
\providecommand \@ifx [1]{%
 \ifx #1\expandafter \@firstoftwo
 \else \expandafter \@secondoftwo
 \fi
}%
\providecommand \natexlab [1]{#1}%
\providecommand \enquote  [1]{``#1''}%
\providecommand \bibnamefont  [1]{#1}%
\providecommand \bibfnamefont [1]{#1}%
\providecommand \citenamefont [1]{#1}%
\providecommand \href@noop [0]{\@secondoftwo}%
\providecommand \href [0]{\begingroup \@sanitize@url \@href}%
\providecommand \@href[1]{\@@startlink{#1}\@@href}%
\providecommand \@@href[1]{\endgroup#1\@@endlink}%
\providecommand \@sanitize@url [0]{\catcode `\\12\catcode `\$12\catcode
  `\&12\catcode `\#12\catcode `\^12\catcode `\_12\catcode `\%12\relax}%
\providecommand \@@startlink[1]{}%
\providecommand \@@endlink[0]{}%
\providecommand \url  [0]{\begingroup\@sanitize@url \@url }%
\providecommand \@url [1]{\endgroup\@href {#1}{\urlprefix }}%
\providecommand \urlprefix  [0]{URL }%
\providecommand \Eprint [0]{\href }%
\providecommand \doibase [0]{https://doi.org/}%
\providecommand \selectlanguage [0]{\@gobble}%
\providecommand \bibinfo  [0]{\@secondoftwo}%
\providecommand \bibfield  [0]{\@secondoftwo}%
\providecommand \translation [1]{[#1]}%
\providecommand \BibitemOpen [0]{}%
\providecommand \bibitemStop [0]{}%
\providecommand \bibitemNoStop [0]{.\EOS\space}%
\providecommand \EOS [0]{\spacefactor3000\relax}%
\providecommand \BibitemShut  [1]{\csname bibitem#1\endcsname}%
\let\auto@bib@innerbib\@empty
\bibitem [{\citenamefont {Cao}\ \emph {et~al.}(2018{\natexlab{a}})\citenamefont
  {Cao}, \citenamefont {Fatemi}, \citenamefont {Fang}, \citenamefont
  {Watanabe}, \citenamefont {Taniguchi}, \citenamefont {Kaxiras},\ and\
  \citenamefont {Jarillo-Herrero}}]{cao2018unconventional}%
  \BibitemOpen
  \bibfield  {author} {\bibinfo {author} {\bibfnamefont {Y.}~\bibnamefont
  {Cao}}, \bibinfo {author} {\bibfnamefont {V.}~\bibnamefont {Fatemi}},
  \bibinfo {author} {\bibfnamefont {S.}~\bibnamefont {Fang}}, \bibinfo {author}
  {\bibfnamefont {K.}~\bibnamefont {Watanabe}}, \bibinfo {author}
  {\bibfnamefont {T.}~\bibnamefont {Taniguchi}}, \bibinfo {author}
  {\bibfnamefont {E.}~\bibnamefont {Kaxiras}},\ and\ \bibinfo {author}
  {\bibfnamefont {P.}~\bibnamefont {Jarillo-Herrero}},\ }\bibfield  {title}
  {\bibinfo {title} {Unconventional superconductivity in magic-angle graphene
  superlattices},\ }\href {https://doi.org/10.1038/nature26160} {\bibfield
  {journal} {\bibinfo  {journal} {Nature}\ }\textbf {\bibinfo {volume} {556}},\
  \bibinfo {pages} {43} (\bibinfo {year} {2018}{\natexlab{a}})}\BibitemShut
  {NoStop}%
\bibitem [{\citenamefont {Yankowitz}\ \emph {et~al.}(2019)\citenamefont
  {Yankowitz}, \citenamefont {Chen}, \citenamefont {Polshyn}, \citenamefont
  {Zhang}, \citenamefont {Watanabe}, \citenamefont {Taniguchi}, \citenamefont
  {Graf}, \citenamefont {Young},\ and\ \citenamefont
  {Dean}}]{yankowitz2019tuning}%
  \BibitemOpen
  \bibfield  {author} {\bibinfo {author} {\bibfnamefont {M.}~\bibnamefont
  {Yankowitz}}, \bibinfo {author} {\bibfnamefont {S.}~\bibnamefont {Chen}},
  \bibinfo {author} {\bibfnamefont {H.}~\bibnamefont {Polshyn}}, \bibinfo
  {author} {\bibfnamefont {Y.}~\bibnamefont {Zhang}}, \bibinfo {author}
  {\bibfnamefont {K.}~\bibnamefont {Watanabe}}, \bibinfo {author}
  {\bibfnamefont {T.}~\bibnamefont {Taniguchi}}, \bibinfo {author}
  {\bibfnamefont {D.}~\bibnamefont {Graf}}, \bibinfo {author} {\bibfnamefont
  {A.~F.}\ \bibnamefont {Young}},\ and\ \bibinfo {author} {\bibfnamefont
  {C.~R.}\ \bibnamefont {Dean}},\ }\bibfield  {title} {\bibinfo {title} {Tuning
  superconductivity in twisted bilayer graphene},\ }\href
  {https://www.science.org/doi/10.1126/science.aav1910} {\bibfield  {journal}
  {\bibinfo  {journal} {Science}\ }\textbf {\bibinfo {volume} {363}},\ \bibinfo
  {pages} {1059} (\bibinfo {year} {2019})}\BibitemShut {NoStop}%
\bibitem [{\citenamefont {Cao}\ \emph {et~al.}(2016)\citenamefont {Cao},
  \citenamefont {Luo}, \citenamefont {Fatemi}, \citenamefont {Fang},
  \citenamefont {Sanchez-Yamagishi}, \citenamefont {Watanabe}, \citenamefont
  {Taniguchi}, \citenamefont {Kaxiras},\ and\ \citenamefont
  {Jarillo-Herrero}}]{cao2016superlattice}%
  \BibitemOpen
  \bibfield  {author} {\bibinfo {author} {\bibfnamefont {Y.}~\bibnamefont
  {Cao}}, \bibinfo {author} {\bibfnamefont {J.~Y.}\ \bibnamefont {Luo}},
  \bibinfo {author} {\bibfnamefont {V.}~\bibnamefont {Fatemi}}, \bibinfo
  {author} {\bibfnamefont {S.}~\bibnamefont {Fang}}, \bibinfo {author}
  {\bibfnamefont {J.~D.}\ \bibnamefont {Sanchez-Yamagishi}}, \bibinfo {author}
  {\bibfnamefont {K.}~\bibnamefont {Watanabe}}, \bibinfo {author}
  {\bibfnamefont {T.}~\bibnamefont {Taniguchi}}, \bibinfo {author}
  {\bibfnamefont {E.}~\bibnamefont {Kaxiras}},\ and\ \bibinfo {author}
  {\bibfnamefont {P.}~\bibnamefont {Jarillo-Herrero}},\ }\bibfield  {title}
  {\bibinfo {title} {Superlattice-induced insulating states and
  valley-protected orbits in twisted bilayer graphene},\ }\href
  {https://doi.org/10.1103/PhysRevLett.117.116804} {\bibfield  {journal}
  {\bibinfo  {journal} {Physical Review Letters}\ }\textbf {\bibinfo {volume}
  {117}},\ \bibinfo {pages} {116804} (\bibinfo {year} {2016})}\BibitemShut
  {NoStop}%
\bibitem [{\citenamefont {Lu}\ \emph {et~al.}(2019)\citenamefont {Lu},
  \citenamefont {Stepanov}, \citenamefont {Yang}, \citenamefont {Xie},
  \citenamefont {Aamir}, \citenamefont {Das}, \citenamefont {Urgell},
  \citenamefont {Watanabe}, \citenamefont {Taniguchi}, \citenamefont {Zhang},
  \citenamefont {Bachtold}, \citenamefont {MacDonald},\ and\ \citenamefont
  {Efetov}}]{lu2019superconductors}%
  \BibitemOpen
  \bibfield  {author} {\bibinfo {author} {\bibfnamefont {X.}~\bibnamefont
  {Lu}}, \bibinfo {author} {\bibfnamefont {P.}~\bibnamefont {Stepanov}},
  \bibinfo {author} {\bibfnamefont {W.}~\bibnamefont {Yang}}, \bibinfo {author}
  {\bibfnamefont {M.}~\bibnamefont {Xie}}, \bibinfo {author} {\bibfnamefont
  {M.~A.}\ \bibnamefont {Aamir}}, \bibinfo {author} {\bibfnamefont
  {I.}~\bibnamefont {Das}}, \bibinfo {author} {\bibfnamefont {C.}~\bibnamefont
  {Urgell}}, \bibinfo {author} {\bibfnamefont {K.}~\bibnamefont {Watanabe}},
  \bibinfo {author} {\bibfnamefont {T.}~\bibnamefont {Taniguchi}}, \bibinfo
  {author} {\bibfnamefont {G.}~\bibnamefont {Zhang}}, \bibinfo {author}
  {\bibfnamefont {A.}~\bibnamefont {Bachtold}}, \bibinfo {author}
  {\bibfnamefont {A.~H.}\ \bibnamefont {MacDonald}},\ and\ \bibinfo {author}
  {\bibfnamefont {D.~K.}\ \bibnamefont {Efetov}},\ }\bibfield  {title}
  {\bibinfo {title} {Superconductors, orbital magnets and correlated states in
  magic-angle bilayer graphene},\ }\href
  {https://doi.org/10.1038/s41586-019-1695-0} {\bibfield  {journal} {\bibinfo
  {journal} {Nature}\ }\textbf {\bibinfo {volume} {574}},\ \bibinfo {pages}
  {653} (\bibinfo {year} {2019})}\BibitemShut {NoStop}%
\bibitem [{\citenamefont {Cao}\ \emph {et~al.}(2018{\natexlab{b}})\citenamefont
  {Cao}, \citenamefont {Fatemi}, \citenamefont {Demir}, \citenamefont {Fang},
  \citenamefont {Tomarken}, \citenamefont {Luo}, \citenamefont
  {Sanchez-Yamagishi}, \citenamefont {Watanabe}, \citenamefont {Taniguchi},
  \citenamefont {Kaxiras}, \citenamefont {Ashoori},\ and\ \citenamefont
  {Jarillo-Herrero}}]{cao2018correlated}%
  \BibitemOpen
  \bibfield  {author} {\bibinfo {author} {\bibfnamefont {Y.}~\bibnamefont
  {Cao}}, \bibinfo {author} {\bibfnamefont {V.}~\bibnamefont {Fatemi}},
  \bibinfo {author} {\bibfnamefont {A.}~\bibnamefont {Demir}}, \bibinfo
  {author} {\bibfnamefont {S.}~\bibnamefont {Fang}}, \bibinfo {author}
  {\bibfnamefont {S.~L.}\ \bibnamefont {Tomarken}}, \bibinfo {author}
  {\bibfnamefont {J.~Y.}\ \bibnamefont {Luo}}, \bibinfo {author} {\bibfnamefont
  {J.~D.}\ \bibnamefont {Sanchez-Yamagishi}}, \bibinfo {author} {\bibfnamefont
  {K.}~\bibnamefont {Watanabe}}, \bibinfo {author} {\bibfnamefont
  {T.}~\bibnamefont {Taniguchi}}, \bibinfo {author} {\bibfnamefont
  {E.}~\bibnamefont {Kaxiras}}, \bibinfo {author} {\bibfnamefont {R.~C.}\
  \bibnamefont {Ashoori}},\ and\ \bibinfo {author} {\bibfnamefont
  {P.}~\bibnamefont {Jarillo-Herrero}},\ }\bibfield  {title} {\bibinfo {title}
  {Correlated insulator behaviour at half-filling in magic-angle graphene
  superlattices},\ }\href {https://doi.org/10.1038/nature26154} {\bibfield
  {journal} {\bibinfo  {journal} {Nature}\ }\textbf {\bibinfo {volume} {556}},\
  \bibinfo {pages} {80} (\bibinfo {year} {2018}{\natexlab{b}})}\BibitemShut
  {NoStop}%
\bibitem [{\citenamefont {Lee}\ \emph {et~al.}(2019)\citenamefont {Lee},
  \citenamefont {Khalaf}, \citenamefont {Liu}, \citenamefont {Liu},
  \citenamefont {Hao}, \citenamefont {Kim},\ and\ \citenamefont
  {Vishwanath}}]{lee2019theory}%
  \BibitemOpen
  \bibfield  {author} {\bibinfo {author} {\bibfnamefont {J.~Y.}\ \bibnamefont
  {Lee}}, \bibinfo {author} {\bibfnamefont {E.}~\bibnamefont {Khalaf}},
  \bibinfo {author} {\bibfnamefont {S.}~\bibnamefont {Liu}}, \bibinfo {author}
  {\bibfnamefont {X.}~\bibnamefont {Liu}}, \bibinfo {author} {\bibfnamefont
  {Z.}~\bibnamefont {Hao}}, \bibinfo {author} {\bibfnamefont {P.}~\bibnamefont
  {Kim}},\ and\ \bibinfo {author} {\bibfnamefont {A.}~\bibnamefont
  {Vishwanath}},\ }\bibfield  {title} {\bibinfo {title} {Theory of correlated
  insulating behaviour and spin-triplet superconductivity in twisted double
  bilayer graphene},\ }\href {https://doi.org/10.1038/s41467-019-12981-1}
  {\bibfield  {journal} {\bibinfo  {journal} {Nature Communications}\ }\textbf
  {\bibinfo {volume} {10}},\ \bibinfo {pages} {5333} (\bibinfo {year}
  {2019})}\BibitemShut {NoStop}%
\bibitem [{\citenamefont {Cao}\ \emph {et~al.}(2020)\citenamefont {Cao},
  \citenamefont {Rodan-Legrain}, \citenamefont {Rubies-Bigorda}, \citenamefont
  {Park}, \citenamefont {Watanabe}, \citenamefont {Taniguchi},\ and\
  \citenamefont {Jarillo-Herrero}}]{cao2020tunable}%
  \BibitemOpen
  \bibfield  {author} {\bibinfo {author} {\bibfnamefont {Y.}~\bibnamefont
  {Cao}}, \bibinfo {author} {\bibfnamefont {D.}~\bibnamefont {Rodan-Legrain}},
  \bibinfo {author} {\bibfnamefont {O.}~\bibnamefont {Rubies-Bigorda}},
  \bibinfo {author} {\bibfnamefont {J.~M.}\ \bibnamefont {Park}}, \bibinfo
  {author} {\bibfnamefont {K.}~\bibnamefont {Watanabe}}, \bibinfo {author}
  {\bibfnamefont {T.}~\bibnamefont {Taniguchi}},\ and\ \bibinfo {author}
  {\bibfnamefont {P.}~\bibnamefont {Jarillo-Herrero}},\ }\bibfield  {title}
  {\bibinfo {title} {Tunable correlated states and spin-polarized phases in
  twisted bilayer--bilayer graphene},\ }\href
  {https://doi.org/10.1038/s41586-020-2260-6} {\bibfield  {journal} {\bibinfo
  {journal} {Nature}\ }\textbf {\bibinfo {volume} {583}},\ \bibinfo {pages}
  {215} (\bibinfo {year} {2020})}\BibitemShut {NoStop}%
\bibitem [{\citenamefont {Wang}\ \emph {et~al.}(2019)\citenamefont {Wang},
  \citenamefont {Mu}, \citenamefont {Wang},\ and\ \citenamefont
  {Sun}}]{wang2019properties}%
  \BibitemOpen
  \bibfield  {author} {\bibinfo {author} {\bibfnamefont {J.}~\bibnamefont
  {Wang}}, \bibinfo {author} {\bibfnamefont {X.}~\bibnamefont {Mu}}, \bibinfo
  {author} {\bibfnamefont {L.}~\bibnamefont {Wang}},\ and\ \bibinfo {author}
  {\bibfnamefont {M.}~\bibnamefont {Sun}},\ }\bibfield  {title} {\bibinfo
  {title} {Properties and applications of new superlattice: twisted bilayer
  graphene},\ }\href
  {https://www.sciencedirect.com/science/article/abs/pii/S2542529319300562}
  {\bibfield  {journal} {\bibinfo  {journal} {Materials Today Physics}\
  }\textbf {\bibinfo {volume} {9}},\ \bibinfo {pages} {100099} (\bibinfo {year}
  {2019})}\BibitemShut {NoStop}%
\bibitem [{\citenamefont {Andrei}\ and\ \citenamefont
  {MacDonald}(2020)}]{andrei2020graphene}%
  \BibitemOpen
  \bibfield  {author} {\bibinfo {author} {\bibfnamefont {E.~Y.}\ \bibnamefont
  {Andrei}}\ and\ \bibinfo {author} {\bibfnamefont {A.~H.}\ \bibnamefont
  {MacDonald}},\ }\bibfield  {title} {\bibinfo {title} {Graphene bilayers with
  a twist},\ }\href {https://www.nature.com/articles/s41563-020-00840-0}
  {\bibfield  {journal} {\bibinfo  {journal} {Nature Materials}\ }\textbf
  {\bibinfo {volume} {19}},\ \bibinfo {pages} {1265} (\bibinfo {year}
  {2020})}\BibitemShut {NoStop}%
\bibitem [{\citenamefont {Lopes~dos Santos}\ \emph {et~al.}(2012)\citenamefont
  {Lopes~dos Santos}, \citenamefont {Peres},\ and\ \citenamefont
  {Castro~Neto}}]{dos2012continuum}%
  \BibitemOpen
  \bibfield  {author} {\bibinfo {author} {\bibfnamefont {J.~M.~B.}\
  \bibnamefont {Lopes~dos Santos}}, \bibinfo {author} {\bibfnamefont
  {N.~M.~R.}\ \bibnamefont {Peres}},\ and\ \bibinfo {author} {\bibfnamefont
  {A.~H.}\ \bibnamefont {Castro~Neto}},\ }\bibfield  {title} {\bibinfo {title}
  {Continuum model of the twisted graphene bilayer},\ }\href
  {https://doi.org/10.1103/PhysRevB.86.155449} {\bibfield  {journal} {\bibinfo
  {journal} {Physical Review B}\ }\textbf {\bibinfo {volume} {86}},\ \bibinfo
  {pages} {155449} (\bibinfo {year} {2012})}\BibitemShut {NoStop}%
\bibitem [{\citenamefont {Moon}\ and\ \citenamefont
  {Koshino}(2013)}]{moon2013optical}%
  \BibitemOpen
  \bibfield  {author} {\bibinfo {author} {\bibfnamefont {P.}~\bibnamefont
  {Moon}}\ and\ \bibinfo {author} {\bibfnamefont {M.}~\bibnamefont {Koshino}},\
  }\bibfield  {title} {\bibinfo {title} {Optical absorption in twisted bilayer
  graphene},\ }\href {https://doi.org/10.1103/PhysRevB.87.205404} {\bibfield
  {journal} {\bibinfo  {journal} {Physical Review B}\ }\textbf {\bibinfo
  {volume} {87}},\ \bibinfo {pages} {205404} (\bibinfo {year}
  {2013})}\BibitemShut {NoStop}%
\bibitem [{\citenamefont {Dindorkar}\ \emph {et~al.}(2023)\citenamefont
  {Dindorkar}, \citenamefont {Kurade},\ and\ \citenamefont
  {Shaikh}}]{dindorkar2023magical}%
  \BibitemOpen
  \bibfield  {author} {\bibinfo {author} {\bibfnamefont {S.~S.}\ \bibnamefont
  {Dindorkar}}, \bibinfo {author} {\bibfnamefont {A.~S.}\ \bibnamefont
  {Kurade}},\ and\ \bibinfo {author} {\bibfnamefont {A.~H.}\ \bibnamefont
  {Shaikh}},\ }\bibfield  {title} {\bibinfo {title} {Magical moir{\'e} patterns
  in twisted bilayer graphene: A review on recent advances in graphene
  twistronics},\ }\href {https://doi.org/10.1016/j.chphi.2023.100325}
  {\bibfield  {journal} {\bibinfo  {journal} {Chemical Physics Impact}\
  }\textbf {\bibinfo {volume} {7}},\ \bibinfo {pages} {100325} (\bibinfo {year}
  {2023})}\BibitemShut {NoStop}%
\bibitem [{\citenamefont {Wang}\ \emph
  {et~al.}(2022{\natexlab{a}})\citenamefont {Wang}, \citenamefont {Ma},
  \citenamefont {Zhang},\ and\ \citenamefont {Lei}}]{wang2022intrinsic}%
  \BibitemOpen
  \bibfield  {author} {\bibinfo {author} {\bibfnamefont {H.}~\bibnamefont
  {Wang}}, \bibinfo {author} {\bibfnamefont {S.}~\bibnamefont {Ma}}, \bibinfo
  {author} {\bibfnamefont {S.}~\bibnamefont {Zhang}},\ and\ \bibinfo {author}
  {\bibfnamefont {D.}~\bibnamefont {Lei}},\ }\bibfield  {title} {\bibinfo
  {title} {Intrinsic superflat bands in general twisted bilayer systems},\
  }\href {https://doi.org/10.1038/s41377-022-00838-0} {\bibfield  {journal}
  {\bibinfo  {journal} {Light: Science \& Applications}\ }\textbf {\bibinfo
  {volume} {11}},\ \bibinfo {pages} {159} (\bibinfo {year}
  {2022}{\natexlab{a}})}\BibitemShut {NoStop}%
\bibitem [{\citenamefont {He}\ \emph {et~al.}(2021)\citenamefont {He},
  \citenamefont {Zhou}, \citenamefont {Ye}, \citenamefont {Cho}, \citenamefont
  {Jeong}, \citenamefont {Meng},\ and\ \citenamefont {Wang}}]{he2021moire}%
  \BibitemOpen
  \bibfield  {author} {\bibinfo {author} {\bibfnamefont {F.}~\bibnamefont
  {He}}, \bibinfo {author} {\bibfnamefont {Y.}~\bibnamefont {Zhou}}, \bibinfo
  {author} {\bibfnamefont {Z.}~\bibnamefont {Ye}}, \bibinfo {author}
  {\bibfnamefont {S.-H.}\ \bibnamefont {Cho}}, \bibinfo {author} {\bibfnamefont
  {J.}~\bibnamefont {Jeong}}, \bibinfo {author} {\bibfnamefont
  {X.}~\bibnamefont {Meng}},\ and\ \bibinfo {author} {\bibfnamefont
  {Y.}~\bibnamefont {Wang}},\ }\bibfield  {title} {\bibinfo {title} {Moir{\'e}
  patterns in 2d materials: a review},\ }\href
  {https://doi.org/10.1021/acsnano.0c10435} {\bibfield  {journal} {\bibinfo
  {journal} {ACS nano}\ }\textbf {\bibinfo {volume} {15}},\ \bibinfo {pages}
  {5944} (\bibinfo {year} {2021})}\BibitemShut {NoStop}%
\bibitem [{\citenamefont {Lopes~dos Santos}\ \emph {et~al.}(2007)\citenamefont
  {Lopes~dos Santos}, \citenamefont {Peres},\ and\ \citenamefont
  {Castro~Neto}}]{PhysRevLett.99.256802}%
  \BibitemOpen
  \bibfield  {author} {\bibinfo {author} {\bibfnamefont {J.~M.~B.}\
  \bibnamefont {Lopes~dos Santos}}, \bibinfo {author} {\bibfnamefont
  {N.~M.~R.}\ \bibnamefont {Peres}},\ and\ \bibinfo {author} {\bibfnamefont
  {A.~H.}\ \bibnamefont {Castro~Neto}},\ }\bibfield  {title} {\bibinfo {title}
  {Graphene bilayer with a twist: Electronic structure},\ }\href
  {https://doi.org/10.1103/PhysRevLett.99.256802} {\bibfield  {journal}
  {\bibinfo  {journal} {Physical Review Letters}\ }\textbf {\bibinfo {volume}
  {99}},\ \bibinfo {pages} {256802} (\bibinfo {year} {2007})}\BibitemShut
  {NoStop}%
\bibitem [{\citenamefont {Bistritzer}\ and\ \citenamefont
  {MacDonald}(2011)}]{bistritzer2011moire}%
  \BibitemOpen
  \bibfield  {author} {\bibinfo {author} {\bibfnamefont {R.}~\bibnamefont
  {Bistritzer}}\ and\ \bibinfo {author} {\bibfnamefont {A.~H.}\ \bibnamefont
  {MacDonald}},\ }\bibfield  {title} {\bibinfo {title} {Moir{\'e} bands in
  twisted double-layer graphene},\ }\href
  {https://doi.org/10.1073/pnas.1108174108} {\bibfield  {journal} {\bibinfo
  {journal} {Proceedings of the National Academy of Sciences}\ }\textbf
  {\bibinfo {volume} {108}},\ \bibinfo {pages} {12233} (\bibinfo {year}
  {2011})}\BibitemShut {NoStop}%
\bibitem [{\citenamefont {de~Andres}\ \emph {et~al.}(2008)\citenamefont
  {de~Andres}, \citenamefont {Ram\'{\i}rez},\ and\ \citenamefont
  {Verg\'es}}]{PhysRevB.77.045403}%
  \BibitemOpen
  \bibfield  {author} {\bibinfo {author} {\bibfnamefont {P.~L.}\ \bibnamefont
  {de~Andres}}, \bibinfo {author} {\bibfnamefont {R.}~\bibnamefont
  {Ram\'{\i}rez}},\ and\ \bibinfo {author} {\bibfnamefont {J.~A.}\ \bibnamefont
  {Verg\'es}},\ }\bibfield  {title} {\bibinfo {title} {Strong covalent bonding
  between two graphene layers},\ }\href
  {https://doi.org/10.1103/PhysRevB.77.045403} {\bibfield  {journal} {\bibinfo
  {journal} {Physical Review B}\ }\textbf {\bibinfo {volume} {77}},\ \bibinfo
  {pages} {045403} (\bibinfo {year} {2008})}\BibitemShut {NoStop}%
\bibitem [{\citenamefont {Koshino}\ and\ \citenamefont
  {Moon}(2015)}]{koshino2015electronic}%
  \BibitemOpen
  \bibfield  {author} {\bibinfo {author} {\bibfnamefont {M.}~\bibnamefont
  {Koshino}}\ and\ \bibinfo {author} {\bibfnamefont {P.}~\bibnamefont {Moon}},\
  }\bibfield  {title} {\bibinfo {title} {Electronic properties of
  incommensurate atomic layers},\ }\href
  {https://doi.org/10.7566/JPSJ.84.121001} {\bibfield  {journal} {\bibinfo
  {journal} {Journal of the Physical Society of Japan}\ }\textbf {\bibinfo
  {volume} {84}},\ \bibinfo {pages} {121001} (\bibinfo {year}
  {2015})}\BibitemShut {NoStop}%
\bibitem [{\citenamefont {Su\'arez~Morell}\ \emph {et~al.}(2010)\citenamefont
  {Su\'arez~Morell}, \citenamefont {Correa}, \citenamefont {Vargas},
  \citenamefont {Pacheco},\ and\ \citenamefont
  {Barticevic}}]{PhysRevB.82.121407}%
  \BibitemOpen
  \bibfield  {author} {\bibinfo {author} {\bibfnamefont {E.}~\bibnamefont
  {Su\'arez~Morell}}, \bibinfo {author} {\bibfnamefont {J.~D.}\ \bibnamefont
  {Correa}}, \bibinfo {author} {\bibfnamefont {P.}~\bibnamefont {Vargas}},
  \bibinfo {author} {\bibfnamefont {M.}~\bibnamefont {Pacheco}},\ and\ \bibinfo
  {author} {\bibfnamefont {Z.}~\bibnamefont {Barticevic}},\ }\bibfield  {title}
  {\bibinfo {title} {Flat bands in slightly twisted bilayer graphene:
  Tight-binding calculations},\ }\href
  {https://doi.org/10.1103/PhysRevB.82.121407} {\bibfield  {journal} {\bibinfo
  {journal} {Physical Review B}\ }\textbf {\bibinfo {volume} {82}},\ \bibinfo
  {pages} {121407(R)} (\bibinfo {year} {2010})}\BibitemShut {NoStop}%
\bibitem [{\citenamefont {Shallcross}\ \emph {et~al.}(2010)\citenamefont
  {Shallcross}, \citenamefont {Sharma}, \citenamefont {Kandelaki},\ and\
  \citenamefont {Pankratov}}]{PhysRevB.81.165105}%
  \BibitemOpen
  \bibfield  {author} {\bibinfo {author} {\bibfnamefont {S.}~\bibnamefont
  {Shallcross}}, \bibinfo {author} {\bibfnamefont {S.}~\bibnamefont {Sharma}},
  \bibinfo {author} {\bibfnamefont {E.}~\bibnamefont {Kandelaki}},\ and\
  \bibinfo {author} {\bibfnamefont {O.~A.}\ \bibnamefont {Pankratov}},\
  }\bibfield  {title} {\bibinfo {title} {Electronic structure of turbostratic
  graphene},\ }\href {https://doi.org/10.1103/PhysRevB.81.165105} {\bibfield
  {journal} {\bibinfo  {journal} {Physical Review B}\ }\textbf {\bibinfo
  {volume} {81}},\ \bibinfo {pages} {165105} (\bibinfo {year}
  {2010})}\BibitemShut {NoStop}%
\bibitem [{\citenamefont {Ni}\ \emph {et~al.}(2008)\citenamefont {Ni},
  \citenamefont {Wang}, \citenamefont {Yu}, \citenamefont {You},\ and\
  \citenamefont {Shen}}]{PhysRevB.77.235403}%
  \BibitemOpen
  \bibfield  {author} {\bibinfo {author} {\bibfnamefont {Z.}~\bibnamefont
  {Ni}}, \bibinfo {author} {\bibfnamefont {Y.}~\bibnamefont {Wang}}, \bibinfo
  {author} {\bibfnamefont {T.}~\bibnamefont {Yu}}, \bibinfo {author}
  {\bibfnamefont {Y.}~\bibnamefont {You}},\ and\ \bibinfo {author}
  {\bibfnamefont {Z.}~\bibnamefont {Shen}},\ }\bibfield  {title} {\bibinfo
  {title} {Reduction of fermi velocity in folded graphene observed by resonance
  raman spectroscopy},\ }\href {https://doi.org/10.1103/PhysRevB.77.235403}
  {\bibfield  {journal} {\bibinfo  {journal} {Physical Review B}\ }\textbf
  {\bibinfo {volume} {77}},\ \bibinfo {pages} {235403} (\bibinfo {year}
  {2008})}\BibitemShut {NoStop}%
\bibitem [{\citenamefont {Trambly~de Laissardi{\`e}re}\ \emph
  {et~al.}(2010)\citenamefont {Trambly~de Laissardi{\`e}re}, \citenamefont
  {Mayou},\ and\ \citenamefont {Magaud}}]{trambly2010localization}%
  \BibitemOpen
  \bibfield  {author} {\bibinfo {author} {\bibfnamefont {G.}~\bibnamefont
  {Trambly~de Laissardi{\`e}re}}, \bibinfo {author} {\bibfnamefont
  {D.}~\bibnamefont {Mayou}},\ and\ \bibinfo {author} {\bibfnamefont
  {L.}~\bibnamefont {Magaud}},\ }\bibfield  {title} {\bibinfo {title}
  {Localization of dirac electrons in rotated graphene bilayers},\ }\href
  {https://doi.org/10.1021/nl902948m} {\bibfield  {journal} {\bibinfo
  {journal} {Nano letters}\ }\textbf {\bibinfo {volume} {10}},\ \bibinfo
  {pages} {804} (\bibinfo {year} {2010})}\BibitemShut {NoStop}%
\bibitem [{\citenamefont {da~Costa}\ \emph {et~al.}(2015)\citenamefont
  {da~Costa}, \citenamefont {Zarenia}, \citenamefont {Chaves}, \citenamefont
  {Farias},\ and\ \citenamefont {Peeters}}]{BQDsDiego}%
  \BibitemOpen
  \bibfield  {author} {\bibinfo {author} {\bibfnamefont {D.~R.}\ \bibnamefont
  {da~Costa}}, \bibinfo {author} {\bibfnamefont {M.}~\bibnamefont {Zarenia}},
  \bibinfo {author} {\bibfnamefont {A.}~\bibnamefont {Chaves}}, \bibinfo
  {author} {\bibfnamefont {G.~A.}\ \bibnamefont {Farias}},\ and\ \bibinfo
  {author} {\bibfnamefont {F.~M.}\ \bibnamefont {Peeters}},\ }\bibfield
  {title} {\bibinfo {title} {Energy levels of bilayer graphene quantum dots},\
  }\href {https://doi.org/10.1103/PhysRevB.92.115437} {\bibfield  {journal}
  {\bibinfo  {journal} {Physical Review B}\ }\textbf {\bibinfo {volume} {92}},\
  \bibinfo {pages} {115437} (\bibinfo {year} {2015})}\BibitemShut {NoStop}%
\bibitem [{\citenamefont {da~Costa}\ \emph
  {et~al.}(2016{\natexlab{a}})\citenamefont {da~Costa}, \citenamefont
  {Zarenia}, \citenamefont {Chaves}, \citenamefont {Farias},\ and\
  \citenamefont {Peeters}}]{DiegomagQD}%
  \BibitemOpen
  \bibfield  {author} {\bibinfo {author} {\bibfnamefont {D.~R.}\ \bibnamefont
  {da~Costa}}, \bibinfo {author} {\bibfnamefont {M.}~\bibnamefont {Zarenia}},
  \bibinfo {author} {\bibfnamefont {A.}~\bibnamefont {Chaves}}, \bibinfo
  {author} {\bibfnamefont {G.~A.}\ \bibnamefont {Farias}},\ and\ \bibinfo
  {author} {\bibfnamefont {F.~M.}\ \bibnamefont {Peeters}},\ }\bibfield
  {title} {\bibinfo {title} {Magnetic field dependence of energy levels in
  biased bilayer graphene quantum dots},\ }\href
  {https://doi.org/10.1103/PhysRevB.93.085401} {\bibfield  {journal} {\bibinfo
  {journal} {Physical Review B}\ }\textbf {\bibinfo {volume} {93}},\ \bibinfo
  {pages} {085401} (\bibinfo {year} {2016}{\natexlab{a}})}\BibitemShut
  {NoStop}%
\bibitem [{\citenamefont {da~Costa}\ \emph
  {et~al.}(2014{\natexlab{a}})\citenamefont {da~Costa}, \citenamefont
  {Zarenia}, \citenamefont {Chaves}, \citenamefont {Farias},\ and\
  \citenamefont {Peeters}}]{da2014analytical}%
  \BibitemOpen
  \bibfield  {author} {\bibinfo {author} {\bibfnamefont {D.~R.}\ \bibnamefont
  {da~Costa}}, \bibinfo {author} {\bibfnamefont {M.}~\bibnamefont {Zarenia}},
  \bibinfo {author} {\bibfnamefont {A.}~\bibnamefont {Chaves}}, \bibinfo
  {author} {\bibfnamefont {G.~A.}\ \bibnamefont {Farias}},\ and\ \bibinfo
  {author} {\bibfnamefont {F.~M.}\ \bibnamefont {Peeters}},\ }\bibfield
  {title} {\bibinfo {title} {Analytical study of the energy levels in bilayer
  graphene quantum dots},\ }\href
  {https://doi.org/10.1016/j.carbon.2014.06.078} {\bibfield  {journal}
  {\bibinfo  {journal} {Carbon}\ }\textbf {\bibinfo {volume} {78}},\ \bibinfo
  {pages} {392} (\bibinfo {year} {2014}{\natexlab{a}})}\BibitemShut {NoStop}%
\bibitem [{\citenamefont {Mirzakhani}\ \emph {et~al.}(2016)\citenamefont
  {Mirzakhani}, \citenamefont {Zarenia}, \citenamefont {Ketabi}, \citenamefont
  {da~Costa},\ and\ \citenamefont {Peeters}}]{PhysRevB.93.165410}%
  \BibitemOpen
  \bibfield  {author} {\bibinfo {author} {\bibfnamefont {M.}~\bibnamefont
  {Mirzakhani}}, \bibinfo {author} {\bibfnamefont {M.}~\bibnamefont {Zarenia}},
  \bibinfo {author} {\bibfnamefont {S.~A.}\ \bibnamefont {Ketabi}}, \bibinfo
  {author} {\bibfnamefont {D.~R.}\ \bibnamefont {da~Costa}},\ and\ \bibinfo
  {author} {\bibfnamefont {F.~M.}\ \bibnamefont {Peeters}},\ }\bibfield
  {title} {\bibinfo {title} {Energy levels of hybrid monolayer-bilayer graphene
  quantum dots},\ }\href {https://doi.org/10.1103/PhysRevB.93.165410}
  {\bibfield  {journal} {\bibinfo  {journal} {Physical Review B}\ }\textbf
  {\bibinfo {volume} {93}},\ \bibinfo {pages} {165410} (\bibinfo {year}
  {2016})}\BibitemShut {NoStop}%
\bibitem [{\citenamefont {da~Costa}\ \emph
  {et~al.}(2016{\natexlab{b}})\citenamefont {da~Costa}, \citenamefont
  {Zarenia}, \citenamefont {Chaves}, \citenamefont {Pereira}, \citenamefont
  {Farias},\ and\ \citenamefont {Peeters}}]{PhysRevB.94.035415}%
  \BibitemOpen
  \bibfield  {author} {\bibinfo {author} {\bibfnamefont {D.~R.}\ \bibnamefont
  {da~Costa}}, \bibinfo {author} {\bibfnamefont {M.}~\bibnamefont {Zarenia}},
  \bibinfo {author} {\bibfnamefont {A.}~\bibnamefont {Chaves}}, \bibinfo
  {author} {\bibfnamefont {J.~M.}\ \bibnamefont {Pereira}}, \bibinfo {author}
  {\bibfnamefont {G.~A.}\ \bibnamefont {Farias}},\ and\ \bibinfo {author}
  {\bibfnamefont {F.~M.}\ \bibnamefont {Peeters}},\ }\bibfield  {title}
  {\bibinfo {title} {Hexagonal-shaped monolayer-bilayer quantum disks in
  graphene: A tight-binding approach},\ }\href
  {https://doi.org/10.1103/PhysRevB.94.035415} {\bibfield  {journal} {\bibinfo
  {journal} {Physical Review B}\ }\textbf {\bibinfo {volume} {94}},\ \bibinfo
  {pages} {035415} (\bibinfo {year} {2016}{\natexlab{b}})}\BibitemShut
  {NoStop}%
\bibitem [{\citenamefont {Pereira}\ \emph {et~al.}(2007)\citenamefont
  {Pereira}, \citenamefont {Vasilopoulos},\ and\ \citenamefont
  {Peeters}}]{pereira2007tunable}%
  \BibitemOpen
  \bibfield  {author} {\bibinfo {author} {\bibfnamefont {J.~M.}\ \bibnamefont
  {Pereira}}, \bibinfo {author} {\bibfnamefont {P.}~\bibnamefont
  {Vasilopoulos}},\ and\ \bibinfo {author} {\bibfnamefont {F.~M.}\ \bibnamefont
  {Peeters}},\ }\bibfield  {title} {\bibinfo {title} {Tunable quantum dots in
  bilayer graphene},\ }\href {https://doi.org/10.1021/nl062967s} {\bibfield
  {journal} {\bibinfo  {journal} {Nano letters}\ }\textbf {\bibinfo {volume}
  {7}},\ \bibinfo {pages} {946} (\bibinfo {year} {2007})}\BibitemShut {NoStop}%
\bibitem [{\citenamefont {Velasco~Jr}\ \emph {et~al.}(2018)\citenamefont
  {Velasco~Jr}, \citenamefont {Lee}, \citenamefont {Wong}, \citenamefont
  {Kahn}, \citenamefont {Tsai}, \citenamefont {Costello}, \citenamefont
  {Umeda}, \citenamefont {Taniguchi}, \citenamefont {Watanabe}, \citenamefont
  {Zettl}, \citenamefont {Wang},\ and\ \citenamefont
  {Crommie}}]{velasco2018visualization}%
  \BibitemOpen
  \bibfield  {author} {\bibinfo {author} {\bibfnamefont {J.}~\bibnamefont
  {Velasco~Jr}}, \bibinfo {author} {\bibfnamefont {J.}~\bibnamefont {Lee}},
  \bibinfo {author} {\bibfnamefont {D.}~\bibnamefont {Wong}}, \bibinfo {author}
  {\bibfnamefont {S.}~\bibnamefont {Kahn}}, \bibinfo {author} {\bibfnamefont
  {H.-Z.}\ \bibnamefont {Tsai}}, \bibinfo {author} {\bibfnamefont
  {J.}~\bibnamefont {Costello}}, \bibinfo {author} {\bibfnamefont
  {T.}~\bibnamefont {Umeda}}, \bibinfo {author} {\bibfnamefont
  {T.}~\bibnamefont {Taniguchi}}, \bibinfo {author} {\bibfnamefont
  {K.}~\bibnamefont {Watanabe}}, \bibinfo {author} {\bibfnamefont
  {A.}~\bibnamefont {Zettl}}, \bibinfo {author} {\bibfnamefont
  {F.}~\bibnamefont {Wang}},\ and\ \bibinfo {author} {\bibfnamefont {M.~F.}\
  \bibnamefont {Crommie}},\ }\bibfield  {title} {\bibinfo {title}
  {Visualization and control of single-electron charging in bilayer graphene
  quantum dots},\ }\href {https://doi.org/10.1021/acs.nanolett.8b01972}
  {\bibfield  {journal} {\bibinfo  {journal} {Nano letters}\ }\textbf {\bibinfo
  {volume} {18}},\ \bibinfo {pages} {5104} (\bibinfo {year}
  {2018})}\BibitemShut {NoStop}%
\bibitem [{\citenamefont {Ge}\ \emph {et~al.}(2020)\citenamefont {Ge},
  \citenamefont {Joucken}, \citenamefont {Quezada}, \citenamefont {da~Costa},
  \citenamefont {Davenport}, \citenamefont {Giraldo}, \citenamefont
  {Taniguchi}, \citenamefont {Watanabe}, \citenamefont {Kobayashi},
  \citenamefont {Low},\ and\ \citenamefont {Velasco~Jr}}]{ge2020visualization}%
  \BibitemOpen
  \bibfield  {author} {\bibinfo {author} {\bibfnamefont {Z.}~\bibnamefont
  {Ge}}, \bibinfo {author} {\bibfnamefont {F.}~\bibnamefont {Joucken}},
  \bibinfo {author} {\bibfnamefont {E.}~\bibnamefont {Quezada}}, \bibinfo
  {author} {\bibfnamefont {D.~R.}\ \bibnamefont {da~Costa}}, \bibinfo {author}
  {\bibfnamefont {J.}~\bibnamefont {Davenport}}, \bibinfo {author}
  {\bibfnamefont {B.}~\bibnamefont {Giraldo}}, \bibinfo {author} {\bibfnamefont
  {T.}~\bibnamefont {Taniguchi}}, \bibinfo {author} {\bibfnamefont
  {K.}~\bibnamefont {Watanabe}}, \bibinfo {author} {\bibfnamefont {N.~P.}\
  \bibnamefont {Kobayashi}}, \bibinfo {author} {\bibfnamefont {T.}~\bibnamefont
  {Low}},\ and\ \bibinfo {author} {\bibfnamefont {J.}~\bibnamefont
  {Velasco~Jr}},\ }\bibfield  {title} {\bibinfo {title} {Visualization and
  manipulation of bilayer graphene quantum dots with broken rotational symmetry
  and nontrivial topology},\ }\href
  {https://doi.org/10.1021/acs.nanolett.0c03453} {\bibfield  {journal}
  {\bibinfo  {journal} {Nano letters}\ }\textbf {\bibinfo {volume} {20}},\
  \bibinfo {pages} {8682} (\bibinfo {year} {2020})}\BibitemShut {NoStop}%
\bibitem [{\citenamefont {Ge}\ \emph {et~al.}(2021)\citenamefont {Ge},
  \citenamefont {Wong}, \citenamefont {Lee}, \citenamefont {Joucken},
  \citenamefont {Quezada-Lopez}, \citenamefont {Kahn}, \citenamefont {Tsai},
  \citenamefont {Taniguchi}, \citenamefont {Watanabe}, \citenamefont {Wang},
  \citenamefont {Zettl}, \citenamefont {Crommie},\ and\ \citenamefont
  {Velasco~Jr.}}]{ge2021imaging}%
  \BibitemOpen
  \bibfield  {author} {\bibinfo {author} {\bibfnamefont {Z.}~\bibnamefont
  {Ge}}, \bibinfo {author} {\bibfnamefont {D.}~\bibnamefont {Wong}}, \bibinfo
  {author} {\bibfnamefont {J.}~\bibnamefont {Lee}}, \bibinfo {author}
  {\bibfnamefont {F.}~\bibnamefont {Joucken}}, \bibinfo {author} {\bibfnamefont
  {E.~A.}\ \bibnamefont {Quezada-Lopez}}, \bibinfo {author} {\bibfnamefont
  {S.}~\bibnamefont {Kahn}}, \bibinfo {author} {\bibfnamefont {H.-Z.}\
  \bibnamefont {Tsai}}, \bibinfo {author} {\bibfnamefont {T.}~\bibnamefont
  {Taniguchi}}, \bibinfo {author} {\bibfnamefont {K.}~\bibnamefont {Watanabe}},
  \bibinfo {author} {\bibfnamefont {F.}~\bibnamefont {Wang}}, \bibinfo {author}
  {\bibfnamefont {A.}~\bibnamefont {Zettl}}, \bibinfo {author} {\bibfnamefont
  {M.~F.}\ \bibnamefont {Crommie}},\ and\ \bibinfo {author} {\bibfnamefont
  {J.}~\bibnamefont {Velasco~Jr.}},\ }\bibfield  {title} {\bibinfo {title}
  {Imaging quantum interference in stadium-shaped monolayer and bilayer
  graphene quantum dots},\ }\href
  {https://doi.org/10.1021/acs.nanolett.1c02271} {\bibfield  {journal}
  {\bibinfo  {journal} {Nano Letters}\ }\textbf {\bibinfo {volume} {21}},\
  \bibinfo {pages} {8993} (\bibinfo {year} {2021})}\BibitemShut {NoStop}%
\bibitem [{\citenamefont {Nascimento}\ \emph {et~al.}(2017)\citenamefont
  {Nascimento}, \citenamefont {da~Costa}, \citenamefont {Zarenia},
  \citenamefont {Chaves},\ and\ \citenamefont {Pereira}}]{PhysRevB.96.115428}%
  \BibitemOpen
  \bibfield  {author} {\bibinfo {author} {\bibfnamefont {J.~S.}\ \bibnamefont
  {Nascimento}}, \bibinfo {author} {\bibfnamefont {D.~R.}\ \bibnamefont
  {da~Costa}}, \bibinfo {author} {\bibfnamefont {M.}~\bibnamefont {Zarenia}},
  \bibinfo {author} {\bibfnamefont {A.}~\bibnamefont {Chaves}},\ and\ \bibinfo
  {author} {\bibfnamefont {J.~M.}\ \bibnamefont {Pereira}},\ }\bibfield
  {title} {\bibinfo {title} {Magnetic properties of bilayer graphene quantum
  dots in the presence of uniaxial strain},\ }\href
  {https://doi.org/10.1103/PhysRevB.96.115428} {\bibfield  {journal} {\bibinfo
  {journal} {Physical Review B}\ }\textbf {\bibinfo {volume} {96}},\ \bibinfo
  {pages} {115428} (\bibinfo {year} {2017})}\BibitemShut {NoStop}%
\bibitem [{\citenamefont {Zarenia}\ \emph {et~al.}(2010)\citenamefont
  {Zarenia}, \citenamefont {Pereira}, \citenamefont {Chaves}, \citenamefont
  {Peeters},\ and\ \citenamefont {Farias}}]{zarenia2010S}%
  \BibitemOpen
  \bibfield  {author} {\bibinfo {author} {\bibfnamefont {M.}~\bibnamefont
  {Zarenia}}, \bibinfo {author} {\bibfnamefont {J.~M.}\ \bibnamefont
  {Pereira}}, \bibinfo {author} {\bibfnamefont {A.}~\bibnamefont {Chaves}},
  \bibinfo {author} {\bibfnamefont {F.~M.}\ \bibnamefont {Peeters}},\ and\
  \bibinfo {author} {\bibfnamefont {G.~A.}\ \bibnamefont {Farias}},\ }\bibfield
   {title} {\bibinfo {title} {Simplified model for the energy levels of quantum
  rings in single layer and bilayer graphene},\ }\href
  {https://doi.org/10.1103/PhysRevB.81.045431} {\bibfield  {journal} {\bibinfo
  {journal} {Physical Review B}\ }\textbf {\bibinfo {volume} {81}},\ \bibinfo
  {pages} {045431} (\bibinfo {year} {2010})}\BibitemShut {NoStop}%
\bibitem [{\citenamefont {Zarenia}\ \emph {et~al.}(2009)\citenamefont
  {Zarenia}, \citenamefont {Pereira~Jr}, \citenamefont {Peeters},\ and\
  \citenamefont {Farias}}]{zarenia2009electrostatically}%
  \BibitemOpen
  \bibfield  {author} {\bibinfo {author} {\bibfnamefont {M.}~\bibnamefont
  {Zarenia}}, \bibinfo {author} {\bibfnamefont {J.~M.}\ \bibnamefont
  {Pereira~Jr}}, \bibinfo {author} {\bibfnamefont {F.~M.}\ \bibnamefont
  {Peeters}},\ and\ \bibinfo {author} {\bibfnamefont {G.~A.}\ \bibnamefont
  {Farias}},\ }\bibfield  {title} {\bibinfo {title} {Electrostatically confined
  quantum rings in bilayer graphene},\ }\href
  {https://doi.org/10.1021/nl902302m} {\bibfield  {journal} {\bibinfo
  {journal} {Nano letters}\ }\textbf {\bibinfo {volume} {9}},\ \bibinfo {pages}
  {4088} (\bibinfo {year} {2009})}\BibitemShut {NoStop}%
\bibitem [{\citenamefont {Mirzakhani}\ \emph {et~al.}(2022)\citenamefont
  {Mirzakhani}, \citenamefont {da~Costa},\ and\ \citenamefont
  {Peeters}}]{PhysRevB.105.115430}%
  \BibitemOpen
  \bibfield  {author} {\bibinfo {author} {\bibfnamefont {M.}~\bibnamefont
  {Mirzakhani}}, \bibinfo {author} {\bibfnamefont {D.~R.}\ \bibnamefont
  {da~Costa}},\ and\ \bibinfo {author} {\bibfnamefont {F.~M.}\ \bibnamefont
  {Peeters}},\ }\bibfield  {title} {\bibinfo {title} {Isolated and hybrid
  bilayer graphene quantum rings},\ }\href
  {https://doi.org/10.1103/PhysRevB.105.115430} {\bibfield  {journal} {\bibinfo
   {journal} {Physical Review B}\ }\textbf {\bibinfo {volume} {105}},\ \bibinfo
  {pages} {115430} (\bibinfo {year} {2022})}\BibitemShut {NoStop}%
\bibitem [{\citenamefont {{Rastegar Sedehi}}\ \emph {et~al.}(2022)\citenamefont
  {{Rastegar Sedehi}}, \citenamefont {Bazrafshan},\ and\ \citenamefont
  {Khordad}}]{RASTEGARSEDEHI2022114853}%
  \BibitemOpen
  \bibfield  {author} {\bibinfo {author} {\bibfnamefont {H.}~\bibnamefont
  {{Rastegar Sedehi}}}, \bibinfo {author} {\bibfnamefont {A.}~\bibnamefont
  {Bazrafshan}},\ and\ \bibinfo {author} {\bibfnamefont {R.}~\bibnamefont
  {Khordad}},\ }\bibfield  {title} {\bibinfo {title} {Thermal properties of
  quantum rings in monolayer and bilayer graphene},\ }\href
  {https://doi.org/https://doi.org/10.1016/j.ssc.2022.114853} {\bibfield
  {journal} {\bibinfo  {journal} {Solid State Communications}\ }\textbf
  {\bibinfo {volume} {353}},\ \bibinfo {pages} {114853} (\bibinfo {year}
  {2022})}\BibitemShut {NoStop}%
\bibitem [{\citenamefont {Zahidi}\ \emph {et~al.}(2017)\citenamefont {Zahidi},
  \citenamefont {Belouad},\ and\ \citenamefont {Jellal}}]{zahidi2017energy}%
  \BibitemOpen
  \bibfield  {author} {\bibinfo {author} {\bibfnamefont {Y.}~\bibnamefont
  {Zahidi}}, \bibinfo {author} {\bibfnamefont {A.}~\bibnamefont {Belouad}},\
  and\ \bibinfo {author} {\bibfnamefont {A.}~\bibnamefont {Jellal}},\
  }\bibfield  {title} {\bibinfo {title} {Energy levels of an ideal quantum ring
  in aa-stacked bilayer graphene},\ }\href
  {https://iopscience.iop.org/article/10.1088/2053-1591/aa69b8} {\bibfield
  {journal} {\bibinfo  {journal} {Materials Research Express}\ }\textbf
  {\bibinfo {volume} {4}},\ \bibinfo {pages} {055603} (\bibinfo {year}
  {2017})}\BibitemShut {NoStop}%
\bibitem [{\citenamefont {Xavier}\ \emph {et~al.}(2010)\citenamefont {Xavier},
  \citenamefont {Pereira}, \citenamefont {Chaves}, \citenamefont {Farias},\
  and\ \citenamefont {Peeters}}]{xavier2010topological}%
  \BibitemOpen
  \bibfield  {author} {\bibinfo {author} {\bibfnamefont {L.~J.~P.}\
  \bibnamefont {Xavier}}, \bibinfo {author} {\bibfnamefont {J.~M.}\
  \bibnamefont {Pereira}}, \bibinfo {author} {\bibfnamefont {A.}~\bibnamefont
  {Chaves}}, \bibinfo {author} {\bibfnamefont {G.~A.}\ \bibnamefont {Farias}},\
  and\ \bibinfo {author} {\bibfnamefont {F.~M.}\ \bibnamefont {Peeters}},\
  }\bibfield  {title} {\bibinfo {title} {Topological confinement in graphene
  bilayer quantum rings},\ }\href {https://doi.org/10.1063/1.3431618}
  {\bibfield  {journal} {\bibinfo  {journal} {Applied Physics Letters}\
  }\textbf {\bibinfo {volume} {96}} (\bibinfo {year} {2010})}\BibitemShut
  {NoStop}%
\bibitem [{\citenamefont {G{\"u}{\c{c}}l{\"u}}\ \emph
  {et~al.}(2016)\citenamefont {G{\"u}{\c{c}}l{\"u}}, \citenamefont {Potasz},
  \citenamefont {Korkusinski},\ and\ \citenamefont {Hawrylak}}]{GQDbook}%
  \BibitemOpen
  \bibfield  {author} {\bibinfo {author} {\bibfnamefont {A.~D.}\ \bibnamefont
  {G{\"u}{\c{c}}l{\"u}}}, \bibinfo {author} {\bibfnamefont {P.}~\bibnamefont
  {Potasz}}, \bibinfo {author} {\bibfnamefont {M.}~\bibnamefont
  {Korkusinski}},\ and\ \bibinfo {author} {\bibfnamefont {P.}~\bibnamefont
  {Hawrylak}},\ }\href@noop {} {\emph {\bibinfo {title} {Graphene quantum
  dots}}}\ (\bibinfo  {publisher} {Springer},\ \bibinfo {year}
  {2016})\BibitemShut {NoStop}%
\bibitem [{\citenamefont {Harrison}\ and\ \citenamefont
  {Valavanis}(2016)}]{harrison2016quantum}%
  \BibitemOpen
  \bibfield  {author} {\bibinfo {author} {\bibfnamefont {P.}~\bibnamefont
  {Harrison}}\ and\ \bibinfo {author} {\bibfnamefont {A.}~\bibnamefont
  {Valavanis}},\ }\href@noop {} {\emph {\bibinfo {title} {Quantum wells, wires
  and dots: theoretical and computational physics of semiconductor
  nanostructures}}}\ (\bibinfo  {publisher} {John Wiley \& Sons},\ \bibinfo
  {year} {2016})\BibitemShut {NoStop}%
\bibitem [{\citenamefont {Li}\ and\ \citenamefont {He}(2022)}]{li2022recent}%
  \BibitemOpen
  \bibfield  {author} {\bibinfo {author} {\bibfnamefont {S.-Y.}\ \bibnamefont
  {Li}}\ and\ \bibinfo {author} {\bibfnamefont {L.}~\bibnamefont {He}},\
  }\bibfield  {title} {\bibinfo {title} {Recent progresses of quantum
  confinement in graphene quantum dots},\ }\href
  {https://link.springer.com/article/10.1007/s11467-021-1125-2} {\bibfield
  {journal} {\bibinfo  {journal} {Frontiers of Physics}\ }\textbf {\bibinfo
  {volume} {17}},\ \bibinfo {pages} {1} (\bibinfo {year} {2022})}\BibitemShut
  {NoStop}%
\bibitem [{\citenamefont {Aharonov}\ and\ \citenamefont
  {Bohm}(1959)}]{PhysRev.115.485}%
  \BibitemOpen
  \bibfield  {author} {\bibinfo {author} {\bibfnamefont {Y.}~\bibnamefont
  {Aharonov}}\ and\ \bibinfo {author} {\bibfnamefont {D.}~\bibnamefont
  {Bohm}},\ }\bibfield  {title} {\bibinfo {title} {Significance of
  electromagnetic potentials in the quantum theory},\ }\href
  {https://doi.org/10.1103/PhysRev.115.485} {\bibfield  {journal} {\bibinfo
  {journal} {Physical Review}\ }\textbf {\bibinfo {volume} {115}},\ \bibinfo
  {pages} {485} (\bibinfo {year} {1959})}\BibitemShut {NoStop}%
\bibitem [{\citenamefont {Timp}\ \emph {et~al.}(1987)\citenamefont {Timp},
  \citenamefont {Chang}, \citenamefont {Cunningham}, \citenamefont {Chang},
  \citenamefont {Mankiewich}, \citenamefont {Behringer},\ and\ \citenamefont
  {Howard}}]{PhysRevLett.58.2814}%
  \BibitemOpen
  \bibfield  {author} {\bibinfo {author} {\bibfnamefont {G.}~\bibnamefont
  {Timp}}, \bibinfo {author} {\bibfnamefont {A.~M.}\ \bibnamefont {Chang}},
  \bibinfo {author} {\bibfnamefont {J.~E.}\ \bibnamefont {Cunningham}},
  \bibinfo {author} {\bibfnamefont {T.~Y.}\ \bibnamefont {Chang}}, \bibinfo
  {author} {\bibfnamefont {P.}~\bibnamefont {Mankiewich}}, \bibinfo {author}
  {\bibfnamefont {R.}~\bibnamefont {Behringer}},\ and\ \bibinfo {author}
  {\bibfnamefont {R.~E.}\ \bibnamefont {Howard}},\ }\bibfield  {title}
  {\bibinfo {title} {Observation of the aharonov-bohm effect for $\omega_c \geq
  1$},\ }\href {https://doi.org/10.1103/PhysRevLett.58.2814} {\bibfield
  {journal} {\bibinfo  {journal} {Physical Review Letters}\ }\textbf {\bibinfo
  {volume} {58}},\ \bibinfo {pages} {2814} (\bibinfo {year}
  {1987})}\BibitemShut {NoStop}%
\bibitem [{\citenamefont {Ford}\ \emph {et~al.}(1988)\citenamefont {Ford},
  \citenamefont {Thornton}, \citenamefont {Newbury}, \citenamefont {Pepper},
  \citenamefont {Ahmed}, \citenamefont {Foxon}, \citenamefont {Harris},\ and\
  \citenamefont {Roberts}}]{ford1988aharonov}%
  \BibitemOpen
  \bibfield  {author} {\bibinfo {author} {\bibfnamefont {C.~J.~B.}\
  \bibnamefont {Ford}}, \bibinfo {author} {\bibfnamefont {T.~J.}\ \bibnamefont
  {Thornton}}, \bibinfo {author} {\bibfnamefont {R.}~\bibnamefont {Newbury}},
  \bibinfo {author} {\bibfnamefont {M.}~\bibnamefont {Pepper}}, \bibinfo
  {author} {\bibfnamefont {H.}~\bibnamefont {Ahmed}}, \bibinfo {author}
  {\bibfnamefont {C.~T.}\ \bibnamefont {Foxon}}, \bibinfo {author}
  {\bibfnamefont {J.~J.}\ \bibnamefont {Harris}},\ and\ \bibinfo {author}
  {\bibfnamefont {C.}~\bibnamefont {Roberts}},\ }\bibfield  {title} {\bibinfo
  {title} {The aharonov-bohm effect in electrostatically defined heterojunction
  rings},\ }\href
  {https://iopscience.iop.org/article/10.1088/0022-3719/21/10/005} {\bibfield
  {journal} {\bibinfo  {journal} {Journal of Physics C: Solid State Physics}\
  }\textbf {\bibinfo {volume} {21}},\ \bibinfo {pages} {L325} (\bibinfo {year}
  {1988})}\BibitemShut {NoStop}%
\bibitem [{\citenamefont {Fuhrer}\ \emph {et~al.}(2001)\citenamefont {Fuhrer},
  \citenamefont {L{\"u}scher}, \citenamefont {Ihn}, \citenamefont {Heinzel},
  \citenamefont {Ensslin}, \citenamefont {Wegscheider},\ and\ \citenamefont
  {Bichler}}]{fuhrer2001energy}%
  \BibitemOpen
  \bibfield  {author} {\bibinfo {author} {\bibfnamefont {A.}~\bibnamefont
  {Fuhrer}}, \bibinfo {author} {\bibfnamefont {S.}~\bibnamefont {L{\"u}scher}},
  \bibinfo {author} {\bibfnamefont {T.}~\bibnamefont {Ihn}}, \bibinfo {author}
  {\bibfnamefont {T.}~\bibnamefont {Heinzel}}, \bibinfo {author} {\bibfnamefont
  {K.}~\bibnamefont {Ensslin}}, \bibinfo {author} {\bibfnamefont
  {W.}~\bibnamefont {Wegscheider}},\ and\ \bibinfo {author} {\bibfnamefont
  {M.}~\bibnamefont {Bichler}},\ }\bibfield  {title} {\bibinfo {title} {Energy
  spectra of quantum rings},\ }\href {https://doi.org/10.1038/35101552}
  {\bibfield  {journal} {\bibinfo  {journal} {Nature}\ }\textbf {\bibinfo
  {volume} {413}},\ \bibinfo {pages} {822} (\bibinfo {year}
  {2001})}\BibitemShut {NoStop}%
\bibitem [{\citenamefont {Ford}\ \emph {et~al.}(1989)\citenamefont {Ford},
  \citenamefont {Thornton}, \citenamefont {Newbury}, \citenamefont {Pepper},
  \citenamefont {Ahmed}, \citenamefont {Peacock}, \citenamefont {Ritchie},
  \citenamefont {Frost},\ and\ \citenamefont
  {Jones}}]{ford1989electrostatically}%
  \BibitemOpen
  \bibfield  {author} {\bibinfo {author} {\bibfnamefont {C.~J.~B.}\
  \bibnamefont {Ford}}, \bibinfo {author} {\bibfnamefont {T.~J.}\ \bibnamefont
  {Thornton}}, \bibinfo {author} {\bibfnamefont {R.}~\bibnamefont {Newbury}},
  \bibinfo {author} {\bibfnamefont {M.}~\bibnamefont {Pepper}}, \bibinfo
  {author} {\bibfnamefont {H.}~\bibnamefont {Ahmed}}, \bibinfo {author}
  {\bibfnamefont {D.~C.}\ \bibnamefont {Peacock}}, \bibinfo {author}
  {\bibfnamefont {D.~A.}\ \bibnamefont {Ritchie}}, \bibinfo {author}
  {\bibfnamefont {J.~E.~F.}\ \bibnamefont {Frost}},\ and\ \bibinfo {author}
  {\bibfnamefont {G.~A.~C.}\ \bibnamefont {Jones}},\ }\bibfield  {title}
  {\bibinfo {title} {Electrostatically defined heterojunction rings and the
  aharonov--bohm effect},\ }\href {https://doi.org/10.1063/1.100818} {\bibfield
   {journal} {\bibinfo  {journal} {Applied Physics Letters}\ }\textbf {\bibinfo
  {volume} {54}},\ \bibinfo {pages} {21} (\bibinfo {year} {1989})}\BibitemShut
  {NoStop}%
\bibitem [{\citenamefont {Viefers}\ \emph {et~al.}(2004)\citenamefont
  {Viefers}, \citenamefont {Koskinen}, \citenamefont {{Singha Deo}},\ and\
  \citenamefont {Manninen}}]{VIEFERS20041}%
  \BibitemOpen
  \bibfield  {author} {\bibinfo {author} {\bibfnamefont {S.}~\bibnamefont
  {Viefers}}, \bibinfo {author} {\bibfnamefont {P.}~\bibnamefont {Koskinen}},
  \bibinfo {author} {\bibfnamefont {P.}~\bibnamefont {{Singha Deo}}},\ and\
  \bibinfo {author} {\bibfnamefont {M.}~\bibnamefont {Manninen}},\ }\bibfield
  {title} {\bibinfo {title} {Quantum rings for beginners: energy spectra and
  persistent currents},\ }\href
  {https://doi.org/https://doi.org/10.1016/j.physe.2003.08.076} {\bibfield
  {journal} {\bibinfo  {journal} {Physica E: Low-dimensional Systems and
  Nanostructures}\ }\textbf {\bibinfo {volume} {21}},\ \bibinfo {pages} {1}
  (\bibinfo {year} {2004})}\BibitemShut {NoStop}%
\bibitem [{\citenamefont {Fuhrer}\ \emph {et~al.}(2002)\citenamefont {Fuhrer},
  \citenamefont {L{\"u}scher}, \citenamefont {Ihn}, \citenamefont {Heinzel},
  \citenamefont {Ensslin}, \citenamefont {Wegscheider},\ and\ \citenamefont
  {Bichler}}]{fuhrer2002energy}%
  \BibitemOpen
  \bibfield  {author} {\bibinfo {author} {\bibfnamefont {A.}~\bibnamefont
  {Fuhrer}}, \bibinfo {author} {\bibfnamefont {S.}~\bibnamefont {L{\"u}scher}},
  \bibinfo {author} {\bibfnamefont {T.}~\bibnamefont {Ihn}}, \bibinfo {author}
  {\bibfnamefont {T.}~\bibnamefont {Heinzel}}, \bibinfo {author} {\bibfnamefont
  {K.}~\bibnamefont {Ensslin}}, \bibinfo {author} {\bibfnamefont
  {W.}~\bibnamefont {Wegscheider}},\ and\ \bibinfo {author} {\bibfnamefont
  {M.}~\bibnamefont {Bichler}},\ }\bibfield  {title} {\bibinfo {title} {Energy
  spectra of quantum rings},\ }\href
  {https://doi.org/10.1016/S0167-9317(02)00598-1} {\bibfield  {journal}
  {\bibinfo  {journal} {Microelectronic Engineering}\ }\textbf {\bibinfo
  {volume} {63}},\ \bibinfo {pages} {47} (\bibinfo {year} {2002})}\BibitemShut
  {NoStop}%
\bibitem [{\citenamefont {Manninen}\ \emph {et~al.}(2012)\citenamefont
  {Manninen}, \citenamefont {Viefers},\ and\ \citenamefont
  {Reimann}}]{MANNINEN2012119}%
  \BibitemOpen
  \bibfield  {author} {\bibinfo {author} {\bibfnamefont {M.}~\bibnamefont
  {Manninen}}, \bibinfo {author} {\bibfnamefont {S.}~\bibnamefont {Viefers}},\
  and\ \bibinfo {author} {\bibfnamefont {S.}~\bibnamefont {Reimann}},\
  }\bibfield  {title} {\bibinfo {title} {Quantum rings for beginners ii: Bosons
  versus fermions},\ }\href
  {https://doi.org/https://doi.org/10.1016/j.physe.2012.09.013} {\bibfield
  {journal} {\bibinfo  {journal} {Physica E: Low-dimensional Systems and
  Nanostructures}\ }\textbf {\bibinfo {volume} {46}},\ \bibinfo {pages} {119}
  (\bibinfo {year} {2012})}\BibitemShut {NoStop}%
\bibitem [{\citenamefont {da~Costa}\ \emph {et~al.}(2017)\citenamefont
  {da~Costa}, \citenamefont {Chaves}, \citenamefont {Ferreira}, \citenamefont
  {Farias},\ and\ \citenamefont {Ferreira}}]{da2017electronic}%
  \BibitemOpen
  \bibfield  {author} {\bibinfo {author} {\bibfnamefont {D.~R.}\ \bibnamefont
  {da~Costa}}, \bibinfo {author} {\bibfnamefont {A.}~\bibnamefont {Chaves}},
  \bibinfo {author} {\bibfnamefont {W.~P.}\ \bibnamefont {Ferreira}}, \bibinfo
  {author} {\bibfnamefont {G.~A.}\ \bibnamefont {Farias}},\ and\ \bibinfo
  {author} {\bibfnamefont {R.}~\bibnamefont {Ferreira}},\ }\bibfield  {title}
  {\bibinfo {title} {Electronic properties of superlattices on quantum rings},\
  }\href {http://iopscience.iop.org/0953-8984/29/16/165501} {\bibfield
  {journal} {\bibinfo  {journal} {Journal of Physics: Condensed Matter}\
  }\textbf {\bibinfo {volume} {29}},\ \bibinfo {pages} {165501} (\bibinfo
  {year} {2017})}\BibitemShut {NoStop}%
\bibitem [{\citenamefont {da~Costa}\ \emph
  {et~al.}(2014{\natexlab{b}})\citenamefont {da~Costa}, \citenamefont {Chaves},
  \citenamefont {Zarenia}, \citenamefont {Pereira}, \citenamefont {Farias},\
  and\ \citenamefont {Peeters}}]{da2014geometry}%
  \BibitemOpen
  \bibfield  {author} {\bibinfo {author} {\bibfnamefont {D.~R.}\ \bibnamefont
  {da~Costa}}, \bibinfo {author} {\bibfnamefont {A.}~\bibnamefont {Chaves}},
  \bibinfo {author} {\bibfnamefont {M.}~\bibnamefont {Zarenia}}, \bibinfo
  {author} {\bibfnamefont {J.~M.}\ \bibnamefont {Pereira}}, \bibinfo {author}
  {\bibfnamefont {G.~A.}\ \bibnamefont {Farias}},\ and\ \bibinfo {author}
  {\bibfnamefont {F.~M.}\ \bibnamefont {Peeters}},\ }\bibfield  {title}
  {\bibinfo {title} {Geometry and edge effects on the energy levels of graphene
  quantum rings: A comparison between tight-binding and simplified dirac
  models},\ }\href {https://doi.org/10.1103/PhysRevB.89.075418} {\bibfield
  {journal} {\bibinfo  {journal} {Physical Review B}\ }\textbf {\bibinfo
  {volume} {89}},\ \bibinfo {pages} {075418} (\bibinfo {year}
  {2014}{\natexlab{b}})}\BibitemShut {NoStop}%
\bibitem [{\citenamefont {Xavier}\ \emph {et~al.}(2016)\citenamefont {Xavier},
  \citenamefont {da~Costa}, \citenamefont {Chaves}, \citenamefont
  {Pereira~Jr},\ and\ \citenamefont {Farias}}]{xavier2016electronic}%
  \BibitemOpen
  \bibfield  {author} {\bibinfo {author} {\bibfnamefont {L.~J.~P.}\
  \bibnamefont {Xavier}}, \bibinfo {author} {\bibfnamefont {D.~R.}\
  \bibnamefont {da~Costa}}, \bibinfo {author} {\bibfnamefont {A.}~\bibnamefont
  {Chaves}}, \bibinfo {author} {\bibfnamefont {J.~M.}\ \bibnamefont
  {Pereira~Jr}},\ and\ \bibinfo {author} {\bibfnamefont {G.~A.}\ \bibnamefont
  {Farias}},\ }\bibfield  {title} {\bibinfo {title} {Electronic confinement in
  graphene quantum rings due to substrate-induced mass radial kink},\ }\href
  {http://iopscience.iop.org/0953-8984/28/50/505501} {\bibfield  {journal}
  {\bibinfo  {journal} {Journal of Physics: Condensed Matter}\ }\textbf
  {\bibinfo {volume} {28}},\ \bibinfo {pages} {505501} (\bibinfo {year}
  {2016})}\BibitemShut {NoStop}%
\bibitem [{\citenamefont {de~Sousa}\ \emph {et~al.}(2017)\citenamefont
  {de~Sousa}, \citenamefont {da~Costa}, \citenamefont {Chaves}, \citenamefont
  {Farias},\ and\ \citenamefont {Peeters}}]{PhysRevB.95.205414}%
  \BibitemOpen
  \bibfield  {author} {\bibinfo {author} {\bibfnamefont {G.~O.}\ \bibnamefont
  {de~Sousa}}, \bibinfo {author} {\bibfnamefont {D.~R.}\ \bibnamefont
  {da~Costa}}, \bibinfo {author} {\bibfnamefont {A.}~\bibnamefont {Chaves}},
  \bibinfo {author} {\bibfnamefont {G.~A.}\ \bibnamefont {Farias}},\ and\
  \bibinfo {author} {\bibfnamefont {F.~M.}\ \bibnamefont {Peeters}},\
  }\bibfield  {title} {\bibinfo {title} {Unusual quantum confined stark effect
  and aharonov-bohm oscillations in semiconductor quantum rings with
  anisotropic effective masses},\ }\href
  {https://doi.org/10.1103/PhysRevB.95.205414} {\bibfield  {journal} {\bibinfo
  {journal} {Physical Review B}\ }\textbf {\bibinfo {volume} {95}},\ \bibinfo
  {pages} {205414} (\bibinfo {year} {2017})}\BibitemShut {NoStop}%
\bibitem [{\citenamefont {Ara{\'u}jo}\ \emph {et~al.}(2022)\citenamefont
  {Ara{\'u}jo}, \citenamefont {da~Costa}, \citenamefont {Chaves}, \citenamefont
  {de~Sousa},\ and\ \citenamefont {Pereira}}]{araujo2022modulation}%
  \BibitemOpen
  \bibfield  {author} {\bibinfo {author} {\bibfnamefont {F.~R.~V.}\
  \bibnamefont {Ara{\'u}jo}}, \bibinfo {author} {\bibfnamefont {D.~R.}\
  \bibnamefont {da~Costa}}, \bibinfo {author} {\bibfnamefont {A.~J.~C.}\
  \bibnamefont {Chaves}}, \bibinfo {author} {\bibfnamefont {F.~E.~B.}\
  \bibnamefont {de~Sousa}},\ and\ \bibinfo {author} {\bibfnamefont {J.~M.}\
  \bibnamefont {Pereira}},\ }\bibfield  {title} {\bibinfo {title} {Modulation
  of persistent current in graphene quantum rings},\ }\href
  {https://doi.org/10.1088/1361-648X/ac452} {\bibfield  {journal} {\bibinfo
  {journal} {Journal of Physics: Condensed Matter}\ }\textbf {\bibinfo {volume}
  {34}},\ \bibinfo {pages} {125503} (\bibinfo {year} {2022})}\BibitemShut
  {NoStop}%
\bibitem [{\citenamefont {Bahamon}\ \emph {et~al.}(2009)\citenamefont
  {Bahamon}, \citenamefont {Pereira},\ and\ \citenamefont {Schulz}}]{bahamon}%
  \BibitemOpen
  \bibfield  {author} {\bibinfo {author} {\bibfnamefont {D.~A.}\ \bibnamefont
  {Bahamon}}, \bibinfo {author} {\bibfnamefont {A.~L.~C.}\ \bibnamefont
  {Pereira}},\ and\ \bibinfo {author} {\bibfnamefont {P.~A.}\ \bibnamefont
  {Schulz}},\ }\bibfield  {title} {\bibinfo {title} {Inner and outer edge
  states in graphene rings: A numerical investigation},\ }\href
  {https://doi.org/10.1103/PhysRevB.79.125414} {\bibfield  {journal} {\bibinfo
  {journal} {Physical Review B}\ }\textbf {\bibinfo {volume} {79}},\ \bibinfo
  {pages} {125414} (\bibinfo {year} {2009})}\BibitemShut {NoStop}%
\bibitem [{\citenamefont {Landgraf}\ \emph {et~al.}(2013)\citenamefont
  {Landgraf}, \citenamefont {Shallcross}, \citenamefont {T\"urschmann},
  \citenamefont {Weckbecker},\ and\ \citenamefont
  {Pankratov}}]{PhysRevB.87.075433}%
  \BibitemOpen
  \bibfield  {author} {\bibinfo {author} {\bibfnamefont {W.}~\bibnamefont
  {Landgraf}}, \bibinfo {author} {\bibfnamefont {S.}~\bibnamefont
  {Shallcross}}, \bibinfo {author} {\bibfnamefont {K.}~\bibnamefont
  {T\"urschmann}}, \bibinfo {author} {\bibfnamefont {D.}~\bibnamefont
  {Weckbecker}},\ and\ \bibinfo {author} {\bibfnamefont {O.}~\bibnamefont
  {Pankratov}},\ }\bibfield  {title} {\bibinfo {title} {Electronic structure of
  twisted graphene flakes},\ }\href
  {https://doi.org/10.1103/PhysRevB.87.075433} {\bibfield  {journal} {\bibinfo
  {journal} {Physical Review B}\ }\textbf {\bibinfo {volume} {87}},\ \bibinfo
  {pages} {075433} (\bibinfo {year} {2013})}\BibitemShut {NoStop}%
\bibitem [{\citenamefont {Mirzakhani}\ \emph {et~al.}(2020)\citenamefont
  {Mirzakhani}, \citenamefont {Peeters},\ and\ \citenamefont
  {Zarenia}}]{mirzakhani2020circular}%
  \BibitemOpen
  \bibfield  {author} {\bibinfo {author} {\bibfnamefont {M.}~\bibnamefont
  {Mirzakhani}}, \bibinfo {author} {\bibfnamefont {F.~M.}\ \bibnamefont
  {Peeters}},\ and\ \bibinfo {author} {\bibfnamefont {M.}~\bibnamefont
  {Zarenia}},\ }\bibfield  {title} {\bibinfo {title} {Circular quantum dots in
  twisted bilayer graphene},\ }\href
  {https://doi.org/10.1103/PhysRevB.101.075413} {\bibfield  {journal} {\bibinfo
   {journal} {Physical Review B}\ }\textbf {\bibinfo {volume} {101}},\ \bibinfo
  {pages} {075413} (\bibinfo {year} {2020})}\BibitemShut {NoStop}%
\bibitem [{\citenamefont {Wang}\ \emph {et~al.}(2021)\citenamefont {Wang},
  \citenamefont {Cui}, \citenamefont {Zhang},\ and\ \citenamefont
  {Yang}}]{wang2021enhanced}%
  \BibitemOpen
  \bibfield  {author} {\bibinfo {author} {\bibfnamefont {X.}~\bibnamefont
  {Wang}}, \bibinfo {author} {\bibfnamefont {Y.}~\bibnamefont {Cui}}, \bibinfo
  {author} {\bibfnamefont {L.}~\bibnamefont {Zhang}},\ and\ \bibinfo {author}
  {\bibfnamefont {M.}~\bibnamefont {Yang}},\ }\bibfield  {title} {\bibinfo
  {title} {Enhanced second-order stark effect in twisted bilayer graphene
  quantum dots},\ }\href {https://doi.org/10.1007/s12274-021-3318-y} {\bibfield
   {journal} {\bibinfo  {journal} {Nano Research}\ }\textbf {\bibinfo {volume}
  {14}},\ \bibinfo {pages} {1} (\bibinfo {year} {2021})}\BibitemShut {NoStop}%
\bibitem [{\citenamefont {Wang}\ and\ \citenamefont
  {Yang}(2022)}]{wang2022enhanced}%
  \BibitemOpen
  \bibfield  {author} {\bibinfo {author} {\bibfnamefont {X.}~\bibnamefont
  {Wang}}\ and\ \bibinfo {author} {\bibfnamefont {M.}~\bibnamefont {Yang}},\
  }\bibfield  {title} {\bibinfo {title} {Enhanced interlayer coupling in
  twisted bilayer graphene quantum dots},\ }\href
  {https://doi.org/10.1016/j.apsusc.2022.154148} {\bibfield  {journal}
  {\bibinfo  {journal} {Applied Surface Science}\ }\textbf {\bibinfo {volume}
  {600}},\ \bibinfo {pages} {154148} (\bibinfo {year} {2022})}\BibitemShut
  {NoStop}%
\bibitem [{\citenamefont {Tilak}\ \emph {et~al.}(2021)\citenamefont {Tilak},
  \citenamefont {Lai}, \citenamefont {Wu}, \citenamefont {Zhang}, \citenamefont
  {Xu}, \citenamefont {Ribeiro}, \citenamefont {Canfield},\ and\ \citenamefont
  {Andrei}}]{tilak2021flat}%
  \BibitemOpen
  \bibfield  {author} {\bibinfo {author} {\bibfnamefont {N.}~\bibnamefont
  {Tilak}}, \bibinfo {author} {\bibfnamefont {X.}~\bibnamefont {Lai}}, \bibinfo
  {author} {\bibfnamefont {S.}~\bibnamefont {Wu}}, \bibinfo {author}
  {\bibfnamefont {Z.}~\bibnamefont {Zhang}}, \bibinfo {author} {\bibfnamefont
  {M.}~\bibnamefont {Xu}}, \bibinfo {author} {\bibfnamefont {R.~d.~A.}\
  \bibnamefont {Ribeiro}}, \bibinfo {author} {\bibfnamefont {P.~C.}\
  \bibnamefont {Canfield}},\ and\ \bibinfo {author} {\bibfnamefont {E.~Y.}\
  \bibnamefont {Andrei}},\ }\bibfield  {title} {\bibinfo {title} {Flat band
  carrier confinement in magic-angle twisted bilayer graphene},\ }\href
  {https://doi.org/10.1038/s41467-021-24480-3} {\bibfield  {journal} {\bibinfo
  {journal} {Nature Communications}\ }\textbf {\bibinfo {volume} {12}},\
  \bibinfo {pages} {4180} (\bibinfo {year} {2021})}\BibitemShut {NoStop}%
\bibitem [{\citenamefont {Zhou}\ \emph {et~al.}(2021)\citenamefont {Zhou},
  \citenamefont {Liu}, \citenamefont {Yan}, \citenamefont {Fu}, \citenamefont
  {Liu},\ and\ \citenamefont {He}}]{PhysRevB.104.235417}%
  \BibitemOpen
  \bibfield  {author} {\bibinfo {author} {\bibfnamefont {X.-F.}\ \bibnamefont
  {Zhou}}, \bibinfo {author} {\bibfnamefont {Y.-W.}\ \bibnamefont {Liu}},
  \bibinfo {author} {\bibfnamefont {H.-Y.}\ \bibnamefont {Yan}}, \bibinfo
  {author} {\bibfnamefont {Z.-Q.}\ \bibnamefont {Fu}}, \bibinfo {author}
  {\bibfnamefont {H.}~\bibnamefont {Liu}},\ and\ \bibinfo {author}
  {\bibfnamefont {L.}~\bibnamefont {He}},\ }\bibfield  {title} {\bibinfo
  {title} {Electronic confinement in quantum dots of twisted bilayer
  graphene},\ }\href {https://doi.org/10.1103/PhysRevB.104.235417} {\bibfield
  {journal} {\bibinfo  {journal} {Physical Review B}\ }\textbf {\bibinfo
  {volume} {104}},\ \bibinfo {pages} {235417} (\bibinfo {year}
  {2021})}\BibitemShut {NoStop}%
\bibitem [{\citenamefont {Wang}\ \emph
  {et~al.}(2022{\natexlab{b}})\citenamefont {Wang}, \citenamefont {Yu},
  \citenamefont {R\"osner}, \citenamefont {Katsnelson}, \citenamefont {Lin},\
  and\ \citenamefont {Yuan}}]{PhysRevX.12.021055}%
  \BibitemOpen
  \bibfield  {author} {\bibinfo {author} {\bibfnamefont {Y.}~\bibnamefont
  {Wang}}, \bibinfo {author} {\bibfnamefont {G.}~\bibnamefont {Yu}}, \bibinfo
  {author} {\bibfnamefont {M.}~\bibnamefont {R\"osner}}, \bibinfo {author}
  {\bibfnamefont {M.~I.}\ \bibnamefont {Katsnelson}}, \bibinfo {author}
  {\bibfnamefont {H.-Q.}\ \bibnamefont {Lin}},\ and\ \bibinfo {author}
  {\bibfnamefont {S.}~\bibnamefont {Yuan}},\ }\bibfield  {title} {\bibinfo
  {title} {Polarization-dependent selection rules and optical spectrum atlas of
  twisted bilayer graphene quantum dots},\ }\href
  {https://doi.org/10.1103/PhysRevX.12.021055} {\bibfield  {journal} {\bibinfo
  {journal} {Physical Review X}\ }\textbf {\bibinfo {volume} {12}},\ \bibinfo
  {pages} {021055} (\bibinfo {year} {2022}{\natexlab{b}})}\BibitemShut
  {NoStop}%
\bibitem [{\citenamefont {Tiutiunnyk}\ \emph {et~al.}(2019)\citenamefont
  {Tiutiunnyk}, \citenamefont {Duque}, \citenamefont {Caro-Lopera},
  \citenamefont {Mora-Ramos},\ and\ \citenamefont
  {Correa}}]{tiutiunnyk2019opto}%
  \BibitemOpen
  \bibfield  {author} {\bibinfo {author} {\bibfnamefont {A.}~\bibnamefont
  {Tiutiunnyk}}, \bibinfo {author} {\bibfnamefont {C.~A.}\ \bibnamefont
  {Duque}}, \bibinfo {author} {\bibfnamefont {F.~J.}\ \bibnamefont
  {Caro-Lopera}}, \bibinfo {author} {\bibfnamefont {M.}~\bibnamefont
  {Mora-Ramos}},\ and\ \bibinfo {author} {\bibfnamefont {J.~D.}\ \bibnamefont
  {Correa}},\ }\bibfield  {title} {\bibinfo {title} {Opto-electronic properties
  of twisted bilayer graphene quantum dots},\ }\href
  {https://doi.org/10.1016/j.physe.2019.03.028} {\bibfield  {journal} {\bibinfo
   {journal} {Physica E: Low-dimensional Systems and Nanostructures}\ }\textbf
  {\bibinfo {volume} {112}},\ \bibinfo {pages} {36} (\bibinfo {year}
  {2019})}\BibitemShut {NoStop}%
\bibitem [{\citenamefont {Tepliakov}\ \emph {et~al.}(2020)\citenamefont
  {Tepliakov}, \citenamefont {Orlov}, \citenamefont {Kundelev},\ and\
  \citenamefont {Rukhlenko}}]{tepliakov2020twisted}%
  \BibitemOpen
  \bibfield  {author} {\bibinfo {author} {\bibfnamefont {N.~V.}\ \bibnamefont
  {Tepliakov}}, \bibinfo {author} {\bibfnamefont {A.~V.}\ \bibnamefont
  {Orlov}}, \bibinfo {author} {\bibfnamefont {E.~V.}\ \bibnamefont
  {Kundelev}},\ and\ \bibinfo {author} {\bibfnamefont {I.~D.}\ \bibnamefont
  {Rukhlenko}},\ }\bibfield  {title} {\bibinfo {title} {Twisted bilayer
  graphene quantum dots for chiral nanophotonics},\ }\href
  {https://doi.org/10.1021/acs.jpcc.0c07416} {\bibfield  {journal} {\bibinfo
  {journal} {The Journal of Physical Chemistry C}\ }\textbf {\bibinfo {volume}
  {124}},\ \bibinfo {pages} {22704} (\bibinfo {year} {2020})}\BibitemShut
  {NoStop}%
\bibitem [{\citenamefont {Liu}\ \emph {et~al.}(2023)\citenamefont {Liu},
  \citenamefont {Wang}, \citenamefont {Yu}, \citenamefont {Wang}, \citenamefont
  {Li}, \citenamefont {Cui}, \citenamefont {Lou},\ and\ \citenamefont
  {Ge}}]{liu2023polarizability}%
  \BibitemOpen
  \bibfield  {author} {\bibinfo {author} {\bibfnamefont {X.}~\bibnamefont
  {Liu}}, \bibinfo {author} {\bibfnamefont {X.}~\bibnamefont {Wang}}, \bibinfo
  {author} {\bibfnamefont {S.}~\bibnamefont {Yu}}, \bibinfo {author}
  {\bibfnamefont {G.}~\bibnamefont {Wang}}, \bibinfo {author} {\bibfnamefont
  {B.}~\bibnamefont {Li}}, \bibinfo {author} {\bibfnamefont {T.}~\bibnamefont
  {Cui}}, \bibinfo {author} {\bibfnamefont {Z.}~\bibnamefont {Lou}},\ and\
  \bibinfo {author} {\bibfnamefont {H.}~\bibnamefont {Ge}},\ }\bibfield
  {title} {\bibinfo {title} {Polarizability characteristics of twisted bilayer
  graphene quantum dots in the absence of periodic moir{\'e} potential},\
  }\href {https://doi.org/10.1039/D3RA03444E} {\bibfield  {journal} {\bibinfo
  {journal} {RSC Advances}\ }\textbf {\bibinfo {volume} {13}},\ \bibinfo
  {pages} {23590} (\bibinfo {year} {2023})}\BibitemShut {NoStop}%
\bibitem [{\citenamefont {Luo}(2024)}]{luo2024tuning}%
  \BibitemOpen
  \bibfield  {author} {\bibinfo {author} {\bibfnamefont {M.}~\bibnamefont
  {Luo}},\ }\bibfield  {title} {\bibinfo {title} {Tuning the magnetic
  configuration of bilayer graphene quantum dot by twisting},\ }\href
  {https://doi.org/10.1557/s43577-023-00608-2} {\bibfield  {journal} {\bibinfo
  {journal} {MRS Bulletin}\ }\textbf {\bibinfo {volume} {49}},\ \bibinfo
  {pages} {194} (\bibinfo {year} {2024})}\BibitemShut {NoStop}%
\bibitem [{\citenamefont {Mirzakhani}\ \emph {et~al.}(2023)\citenamefont
  {Mirzakhani}, \citenamefont {Park}, \citenamefont {Peeters},\ and\
  \citenamefont {da~Costa}}]{mirzakhani2023magnetism}%
  \BibitemOpen
  \bibfield  {author} {\bibinfo {author} {\bibfnamefont {M.}~\bibnamefont
  {Mirzakhani}}, \bibinfo {author} {\bibfnamefont {H.~C.}\ \bibnamefont
  {Park}}, \bibinfo {author} {\bibfnamefont {F.~M.}\ \bibnamefont {Peeters}},\
  and\ \bibinfo {author} {\bibfnamefont {D.~R.}\ \bibnamefont {da~Costa}},\
  }\bibfield  {title} {\bibinfo {title} {Magnetism in twisted triangular
  bilayer graphene quantum dots},\ }\href
  {https://doi.org/10.48550/arXiv.2304.06228} {\bibfield  {journal} {\bibinfo
  {journal} {arXiv preprint arXiv:2304.06228}\ } (\bibinfo {year}
  {2023})}\BibitemShut {NoStop}%
\bibitem [{\citenamefont {Han}\ \emph {et~al.}(2020)\citenamefont {Han},
  \citenamefont {Zeng}, \citenamefont {Ren}, \citenamefont {Dong},
  \citenamefont {Ren},\ and\ \citenamefont {Qiao}}]{PhysRevB.101.235432}%
  \BibitemOpen
  \bibfield  {author} {\bibinfo {author} {\bibfnamefont {Y.}~\bibnamefont
  {Han}}, \bibinfo {author} {\bibfnamefont {J.}~\bibnamefont {Zeng}}, \bibinfo
  {author} {\bibfnamefont {Y.}~\bibnamefont {Ren}}, \bibinfo {author}
  {\bibfnamefont {X.}~\bibnamefont {Dong}}, \bibinfo {author} {\bibfnamefont
  {W.}~\bibnamefont {Ren}},\ and\ \bibinfo {author} {\bibfnamefont
  {Z.}~\bibnamefont {Qiao}},\ }\bibfield  {title} {\bibinfo {title} {Mesoscopic
  electronic transport in twisted bilayer graphene},\ }\href
  {https://doi.org/10.1103/PhysRevB.101.235432} {\bibfield  {journal} {\bibinfo
   {journal} {Physical Review B}\ }\textbf {\bibinfo {volume} {101}},\ \bibinfo
  {pages} {235432} (\bibinfo {year} {2020})}\BibitemShut {NoStop}%
\bibitem [{\citenamefont {Thomsen}\ and\ \citenamefont
  {Pedersen}(2017)}]{PhysRevB.95.235427}%
  \BibitemOpen
  \bibfield  {author} {\bibinfo {author} {\bibfnamefont {M.~R.}\ \bibnamefont
  {Thomsen}}\ and\ \bibinfo {author} {\bibfnamefont {T.~G.}\ \bibnamefont
  {Pedersen}},\ }\bibfield  {title} {\bibinfo {title} {Analytical dirac model
  of graphene rings, dots, and antidots in magnetic fields},\ }\href
  {https://doi.org/10.1103/PhysRevB.95.235427} {\bibfield  {journal} {\bibinfo
  {journal} {Physical Review B}\ }\textbf {\bibinfo {volume} {95}},\ \bibinfo
  {pages} {235427} (\bibinfo {year} {2017})}\BibitemShut {NoStop}%
\bibitem [{\citenamefont {McCann}\ and\ \citenamefont
  {Koshino}(2013)}]{mccann2013electronic}%
  \BibitemOpen
  \bibfield  {author} {\bibinfo {author} {\bibfnamefont {E.}~\bibnamefont
  {McCann}}\ and\ \bibinfo {author} {\bibfnamefont {M.}~\bibnamefont
  {Koshino}},\ }\bibfield  {title} {\bibinfo {title} {The electronic properties
  of bilayer graphene},\ }\href
  {https://iopscience.iop.org/article/10.1088/0034-4885/76/5/056503/meta}
  {\bibfield  {journal} {\bibinfo  {journal} {Reports on Progress in Physics}\
  }\textbf {\bibinfo {volume} {76}},\ \bibinfo {pages} {056503} (\bibinfo
  {year} {2013})}\BibitemShut {NoStop}%
\bibitem [{\citenamefont {Mele}(2010)}]{PhysRevB.81.161405}%
  \BibitemOpen
  \bibfield  {author} {\bibinfo {author} {\bibfnamefont {E.~J.}\ \bibnamefont
  {Mele}},\ }\bibfield  {title} {\bibinfo {title} {Commensuration and
  interlayer coherence in twisted bilayer graphene},\ }\href
  {https://doi.org/10.1103/PhysRevB.81.161405} {\bibfield  {journal} {\bibinfo
  {journal} {Physical Review B}\ }\textbf {\bibinfo {volume} {81}},\ \bibinfo
  {pages} {161405(R)} (\bibinfo {year} {2010})}\BibitemShut {NoStop}%
\bibitem [{\citenamefont {Berry}\ and\ \citenamefont
  {Mondragon}(1987)}]{berry1987neutrino}%
  \BibitemOpen
  \bibfield  {author} {\bibinfo {author} {\bibfnamefont {M.~V.}\ \bibnamefont
  {Berry}}\ and\ \bibinfo {author} {\bibfnamefont {R.~J.}\ \bibnamefont
  {Mondragon}},\ }\bibfield  {title} {\bibinfo {title} {Neutrino billiards:
  time-reversal symmetry-breaking without magnetic fields},\ }\href
  {https://doi.org/10.1098/rspa.1987.0080} {\bibfield  {journal} {\bibinfo
  {journal} {Proceedings of the Royal Society of London. A. Mathematical and
  Physical Sciences}\ }\textbf {\bibinfo {volume} {412}},\ \bibinfo {pages}
  {53} (\bibinfo {year} {1987})}\BibitemShut {NoStop}%
\bibitem [{\citenamefont {Schnez}\ \emph {et~al.}(2008)\citenamefont {Schnez},
  \citenamefont {Ensslin}, \citenamefont {Sigrist},\ and\ \citenamefont
  {Ihn}}]{schnez2008analytic}%
  \BibitemOpen
  \bibfield  {author} {\bibinfo {author} {\bibfnamefont {S.}~\bibnamefont
  {Schnez}}, \bibinfo {author} {\bibfnamefont {K.}~\bibnamefont {Ensslin}},
  \bibinfo {author} {\bibfnamefont {M.}~\bibnamefont {Sigrist}},\ and\ \bibinfo
  {author} {\bibfnamefont {T.}~\bibnamefont {Ihn}},\ }\bibfield  {title}
  {\bibinfo {title} {Analytic model of the energy spectrum of a graphene
  quantum dot in a perpendicular magnetic field},\ }\href
  {https://doi.org/10.1103/PhysRevB.78.195427} {\bibfield  {journal} {\bibinfo
  {journal} {Physical Review B}\ }\textbf {\bibinfo {volume} {78}},\ \bibinfo
  {pages} {195427} (\bibinfo {year} {2008})}\BibitemShut {NoStop}%
\bibitem [{\citenamefont {Zhou}\ \emph {et~al.}(2007)\citenamefont {Zhou},
  \citenamefont {Gweon}, \citenamefont {Fedorov}, \citenamefont {First},
  \citenamefont {De~Heer}, \citenamefont {Lee}, \citenamefont {Guinea},
  \citenamefont {Castro~Neto},\ and\ \citenamefont
  {Lanzara}}]{zhou2007substrate}%
  \BibitemOpen
  \bibfield  {author} {\bibinfo {author} {\bibfnamefont {S.~Y.}\ \bibnamefont
  {Zhou}}, \bibinfo {author} {\bibfnamefont {G.-H.}\ \bibnamefont {Gweon}},
  \bibinfo {author} {\bibfnamefont {A.~V.}\ \bibnamefont {Fedorov}}, \bibinfo
  {author} {\bibfnamefont {d.}~\bibnamefont {First}, \bibfnamefont {P.~N.}},
  \bibinfo {author} {\bibfnamefont {W.~A.}\ \bibnamefont {De~Heer}}, \bibinfo
  {author} {\bibfnamefont {D.-H.}\ \bibnamefont {Lee}}, \bibinfo {author}
  {\bibfnamefont {F.}~\bibnamefont {Guinea}}, \bibinfo {author} {\bibfnamefont
  {A.~H.}\ \bibnamefont {Castro~Neto}},\ and\ \bibinfo {author} {\bibfnamefont
  {A.}~\bibnamefont {Lanzara}},\ }\bibfield  {title} {\bibinfo {title}
  {Substrate-induced bandgap opening in epitaxial graphene},\ }\href
  {https://www.nature.com/articles/nmat2003} {\bibfield  {journal} {\bibinfo
  {journal} {Nature Materials}\ }\textbf {\bibinfo {volume} {6}},\ \bibinfo
  {pages} {770} (\bibinfo {year} {2007})}\BibitemShut {NoStop}%
\bibitem [{\citenamefont {Giovannetti}\ \emph
  {et~al.}(2007{\natexlab{a}})\citenamefont {Giovannetti}, \citenamefont
  {Khomyakov}, \citenamefont {Brocks}, \citenamefont {Kelly},\ and\
  \citenamefont {van~den Brink}}]{PhysRevB.76.073103}%
  \BibitemOpen
  \bibfield  {author} {\bibinfo {author} {\bibfnamefont {G.}~\bibnamefont
  {Giovannetti}}, \bibinfo {author} {\bibfnamefont {P.~A.}\ \bibnamefont
  {Khomyakov}}, \bibinfo {author} {\bibfnamefont {G.}~\bibnamefont {Brocks}},
  \bibinfo {author} {\bibfnamefont {P.~J.}\ \bibnamefont {Kelly}},\ and\
  \bibinfo {author} {\bibfnamefont {J.}~\bibnamefont {van~den Brink}},\
  }\bibfield  {title} {\bibinfo {title} {Substrate-induced band gap in graphene
  on hexagonal boron nitride: Ab initio density functional calculations},\
  }\href {https://doi.org/10.1103/PhysRevB.76.073103} {\bibfield  {journal}
  {\bibinfo  {journal} {Physical Review B}\ }\textbf {\bibinfo {volume} {76}},\
  \bibinfo {pages} {073103} (\bibinfo {year} {2007}{\natexlab{a}})}\BibitemShut
  {NoStop}%
\bibitem [{\citenamefont {Nevius}\ \emph {et~al.}(2015)\citenamefont {Nevius},
  \citenamefont {Conrad}, \citenamefont {Wang}, \citenamefont {Celis},
  \citenamefont {Nair}, \citenamefont {Taleb-Ibrahimi}, \citenamefont
  {Tejeda},\ and\ \citenamefont {Conrad}}]{PhysRevLett.115.136802}%
  \BibitemOpen
  \bibfield  {author} {\bibinfo {author} {\bibfnamefont {M.~S.}\ \bibnamefont
  {Nevius}}, \bibinfo {author} {\bibfnamefont {M.}~\bibnamefont {Conrad}},
  \bibinfo {author} {\bibfnamefont {F.}~\bibnamefont {Wang}}, \bibinfo {author}
  {\bibfnamefont {A.}~\bibnamefont {Celis}}, \bibinfo {author} {\bibfnamefont
  {M.~N.}\ \bibnamefont {Nair}}, \bibinfo {author} {\bibfnamefont
  {A.}~\bibnamefont {Taleb-Ibrahimi}}, \bibinfo {author} {\bibfnamefont
  {A.}~\bibnamefont {Tejeda}},\ and\ \bibinfo {author} {\bibfnamefont {E.~H.}\
  \bibnamefont {Conrad}},\ }\bibfield  {title} {\bibinfo {title}
  {Semiconducting graphene from highly ordered substrate interactions},\ }\href
  {https://doi.org/10.1103/PhysRevLett.115.136802} {\bibfield  {journal}
  {\bibinfo  {journal} {Physical Review Letters}\ }\textbf {\bibinfo {volume}
  {115}},\ \bibinfo {pages} {136802} (\bibinfo {year} {2015})}\BibitemShut
  {NoStop}%
\bibitem [{\citenamefont {Giovannetti}\ \emph
  {et~al.}(2007{\natexlab{b}})\citenamefont {Giovannetti}, \citenamefont
  {Khomyakov}, \citenamefont {Brocks}, \citenamefont {Kelly},\ and\
  \citenamefont {van~den Brink}}]{giovannetti2007substrate}%
  \BibitemOpen
  \bibfield  {author} {\bibinfo {author} {\bibfnamefont {G.}~\bibnamefont
  {Giovannetti}}, \bibinfo {author} {\bibfnamefont {P.~A.}\ \bibnamefont
  {Khomyakov}}, \bibinfo {author} {\bibfnamefont {G.}~\bibnamefont {Brocks}},
  \bibinfo {author} {\bibfnamefont {P.~J.}\ \bibnamefont {Kelly}},\ and\
  \bibinfo {author} {\bibfnamefont {J.}~\bibnamefont {van~den Brink}},\
  }\bibfield  {title} {\bibinfo {title} {Substrate-induced band gap in graphene
  on hexagonal boron nitride: Ab initio density functional calculations},\
  }\href {https://doi.org/10.1103/PhysRevB.76.073103} {\bibfield  {journal}
  {\bibinfo  {journal} {Physical Review B}\ }\textbf {\bibinfo {volume} {76}},\
  \bibinfo {pages} {073103} (\bibinfo {year} {2007}{\natexlab{b}})}\BibitemShut
  {NoStop}%
\bibitem [{\citenamefont {Wang}\ \emph {et~al.}(2016)\citenamefont {Wang},
  \citenamefont {Lu}, \citenamefont {Ding}, \citenamefont {Yao}, \citenamefont
  {Yan}, \citenamefont {Wan}, \citenamefont {Deng}, \citenamefont {Wang},
  \citenamefont {Chen}, \citenamefont {Ma}, \citenamefont {Jung}, \citenamefont
  {Fedorov}, \citenamefont {Zhang}, \citenamefont {Zhang},\ and\ \citenamefont
  {Zhou}}]{wang2016gaps}%
  \BibitemOpen
  \bibfield  {author} {\bibinfo {author} {\bibfnamefont {E.}~\bibnamefont
  {Wang}}, \bibinfo {author} {\bibfnamefont {X.}~\bibnamefont {Lu}}, \bibinfo
  {author} {\bibfnamefont {S.}~\bibnamefont {Ding}}, \bibinfo {author}
  {\bibfnamefont {W.}~\bibnamefont {Yao}}, \bibinfo {author} {\bibfnamefont
  {M.}~\bibnamefont {Yan}}, \bibinfo {author} {\bibfnamefont {G.}~\bibnamefont
  {Wan}}, \bibinfo {author} {\bibfnamefont {K.}~\bibnamefont {Deng}}, \bibinfo
  {author} {\bibfnamefont {S.}~\bibnamefont {Wang}}, \bibinfo {author}
  {\bibfnamefont {G.}~\bibnamefont {Chen}}, \bibinfo {author} {\bibfnamefont
  {L.}~\bibnamefont {Ma}}, \bibinfo {author} {\bibfnamefont {J.}~\bibnamefont
  {Jung}}, \bibinfo {author} {\bibfnamefont {A.~V.}\ \bibnamefont {Fedorov}},
  \bibinfo {author} {\bibfnamefont {Y.}~\bibnamefont {Zhang}}, \bibinfo
  {author} {\bibfnamefont {G.}~\bibnamefont {Zhang}},\ and\ \bibinfo {author}
  {\bibfnamefont {S.}~\bibnamefont {Zhou}},\ }\bibfield  {title} {\bibinfo
  {title} {Gaps induced by inversion symmetry breaking and second-generation
  dirac cones in graphene/hexagonal boron nitride},\ }\href
  {https://doi.org/10.1038/nphys3856} {\bibfield  {journal} {\bibinfo
  {journal} {Nature Physics}\ }\textbf {\bibinfo {volume} {12}},\ \bibinfo
  {pages} {1111} (\bibinfo {year} {2016})}\BibitemShut {NoStop}%
\bibitem [{\citenamefont {Duarte}\ \emph {et~al.}(2024)\citenamefont {Duarte},
  \citenamefont {da~Costa}, \citenamefont {Peres}, \citenamefont {Teles},\ and\
  \citenamefont {Chaves}}]{duarte2023moir}%
  \BibitemOpen
  \bibfield  {author} {\bibinfo {author} {\bibfnamefont {V.~G.~M.}\
  \bibnamefont {Duarte}}, \bibinfo {author} {\bibfnamefont {D.~R.}\
  \bibnamefont {da~Costa}}, \bibinfo {author} {\bibfnamefont {N.~M.~R.}\
  \bibnamefont {Peres}}, \bibinfo {author} {\bibfnamefont {L.~K.}\ \bibnamefont
  {Teles}},\ and\ \bibinfo {author} {\bibfnamefont {A.~J.}\ \bibnamefont
  {Chaves}},\ }\bibfield  {title} {\bibinfo {title} {Moir\'e excitons in biased
  twisted bilayer graphene under pressure},\ }\href
  {https://doi.org/10.1103/PhysRevB.110.035405} {\bibfield  {journal} {\bibinfo
   {journal} {Phys. Rev. B}\ }\textbf {\bibinfo {volume} {110}},\ \bibinfo
  {pages} {035405} (\bibinfo {year} {2024})}\BibitemShut {NoStop}%
\bibitem [{\citenamefont {Deng}\ \emph {et~al.}(2020)\citenamefont {Deng},
  \citenamefont {Wang}, \citenamefont {Li}, \citenamefont {Li}, \citenamefont
  {Wang}, \citenamefont {Tang}, \citenamefont {Fu}, \citenamefont {Tian},
  \citenamefont {Gao}, \citenamefont {Xue}, ,\ and\ \citenamefont
  {Peng}}]{deng2020interlayer}%
  \BibitemOpen
  \bibfield  {author} {\bibinfo {author} {\bibfnamefont {B.}~\bibnamefont
  {Deng}}, \bibinfo {author} {\bibfnamefont {B.}~\bibnamefont {Wang}}, \bibinfo
  {author} {\bibfnamefont {N.}~\bibnamefont {Li}}, \bibinfo {author}
  {\bibfnamefont {R.}~\bibnamefont {Li}}, \bibinfo {author} {\bibfnamefont
  {Y.}~\bibnamefont {Wang}}, \bibinfo {author} {\bibfnamefont {J.}~\bibnamefont
  {Tang}}, \bibinfo {author} {\bibfnamefont {Q.}~\bibnamefont {Fu}}, \bibinfo
  {author} {\bibfnamefont {Z.}~\bibnamefont {Tian}}, \bibinfo {author}
  {\bibfnamefont {P.}~\bibnamefont {Gao}}, \bibinfo {author} {\bibfnamefont
  {J.}~\bibnamefont {Xue}}, ,\ and\ \bibinfo {author} {\bibfnamefont
  {H.}~\bibnamefont {Peng}},\ }\bibfield  {title} {\bibinfo {title} {Interlayer
  decoupling in 30 twisted bilayer graphene quasicrystal},\ }\href
  {https://doi.org/10.1021/acsnano.9b07091} {\bibfield  {journal} {\bibinfo
  {journal} {ACS Nano}\ }\textbf {\bibinfo {volume} {14}},\ \bibinfo {pages}
  {1656} (\bibinfo {year} {2020})}\BibitemShut {NoStop}%
\bibitem [{\citenamefont {Joucken}\ \emph {et~al.}(2020)\citenamefont
  {Joucken}, \citenamefont {Ge}, \citenamefont {Quezada-L\'opez}, \citenamefont
  {Davenport}, \citenamefont {Watanabe}, \citenamefont {Taniguchi},\ and\
  \citenamefont {Velasco}}]{PhysRevB.101.161103}%
  \BibitemOpen
  \bibfield  {author} {\bibinfo {author} {\bibfnamefont {F.}~\bibnamefont
  {Joucken}}, \bibinfo {author} {\bibfnamefont {Z.}~\bibnamefont {Ge}},
  \bibinfo {author} {\bibfnamefont {E.~A.}\ \bibnamefont {Quezada-L\'opez}},
  \bibinfo {author} {\bibfnamefont {J.~L.}\ \bibnamefont {Davenport}}, \bibinfo
  {author} {\bibfnamefont {K.}~\bibnamefont {Watanabe}}, \bibinfo {author}
  {\bibfnamefont {T.}~\bibnamefont {Taniguchi}},\ and\ \bibinfo {author}
  {\bibfnamefont {J.}~\bibnamefont {Velasco}},\ }\bibfield  {title} {\bibinfo
  {title} {Determination of the trigonal warping orientation in bernal-stacked
  bilayer graphene via scanning tunneling microscopy},\ }\href
  {https://doi.org/10.1103/PhysRevB.101.161103} {\bibfield  {journal} {\bibinfo
   {journal} {Physical Review B}\ }\textbf {\bibinfo {volume} {101}},\ \bibinfo
  {pages} {161103(R)} (\bibinfo {year} {2020})}\BibitemShut {NoStop}%
\bibitem [{\citenamefont {Chiappini}\ \emph {et~al.}(2015)\citenamefont
  {Chiappini}, \citenamefont {Wiedmann}, \citenamefont {Novoselov},
  \citenamefont {Mishchenko}, \citenamefont {Geim}, \citenamefont {Maan},\ and\
  \citenamefont {Zeitler}}]{PhysRevB.92.201412}%
  \BibitemOpen
  \bibfield  {author} {\bibinfo {author} {\bibfnamefont {F.}~\bibnamefont
  {Chiappini}}, \bibinfo {author} {\bibfnamefont {S.}~\bibnamefont {Wiedmann}},
  \bibinfo {author} {\bibfnamefont {K.}~\bibnamefont {Novoselov}}, \bibinfo
  {author} {\bibfnamefont {A.}~\bibnamefont {Mishchenko}}, \bibinfo {author}
  {\bibfnamefont {A.~K.}\ \bibnamefont {Geim}}, \bibinfo {author}
  {\bibfnamefont {J.~C.}\ \bibnamefont {Maan}},\ and\ \bibinfo {author}
  {\bibfnamefont {U.}~\bibnamefont {Zeitler}},\ }\bibfield  {title} {\bibinfo
  {title} {Lifting of the landau level degeneracy in graphene devices in a
  tilted magnetic field},\ }\href
  {https://link.aps.org/doi/10.1103/PhysRevB.92.201412} {\bibfield  {journal}
  {\bibinfo  {journal} {Physical Review B}\ }\textbf {\bibinfo {volume} {92}},\
  \bibinfo {pages} {201412(R)} (\bibinfo {year} {2015})}\BibitemShut {NoStop}%
\bibitem [{\citenamefont {Xiao}\ \emph {et~al.}(2007)\citenamefont {Xiao},
  \citenamefont {Yao},\ and\ \citenamefont {Niu}}]{PhysRevLett.99.236809}%
  \BibitemOpen
  \bibfield  {author} {\bibinfo {author} {\bibfnamefont {D.}~\bibnamefont
  {Xiao}}, \bibinfo {author} {\bibfnamefont {W.}~\bibnamefont {Yao}},\ and\
  \bibinfo {author} {\bibfnamefont {Q.}~\bibnamefont {Niu}},\ }\bibfield
  {title} {\bibinfo {title} {Valley-contrasting physics in graphene: Magnetic
  moment and topological transport},\ }\href
  {https://link.aps.org/doi/10.1103/PhysRevLett.99.236809} {\bibfield
  {journal} {\bibinfo  {journal} {Physical Review Letters}\ }\textbf {\bibinfo
  {volume} {99}},\ \bibinfo {pages} {236809} (\bibinfo {year}
  {2007})}\BibitemShut {NoStop}%
\bibitem [{\citenamefont {Yao}\ \emph {et~al.}(2008)\citenamefont {Yao},
  \citenamefont {Xiao},\ and\ \citenamefont {Niu}}]{PhysRevB.77.235406}%
  \BibitemOpen
  \bibfield  {author} {\bibinfo {author} {\bibfnamefont {W.}~\bibnamefont
  {Yao}}, \bibinfo {author} {\bibfnamefont {D.}~\bibnamefont {Xiao}},\ and\
  \bibinfo {author} {\bibfnamefont {Q.}~\bibnamefont {Niu}},\ }\bibfield
  {title} {\bibinfo {title} {Valley-dependent optoelectronics from inversion
  symmetry breaking},\ }\href
  {https://link.aps.org/doi/10.1103/PhysRevB.77.235406} {\bibfield  {journal}
  {\bibinfo  {journal} {Physical Review B}\ }\textbf {\bibinfo {volume} {77}},\
  \bibinfo {pages} {235406} (\bibinfo {year} {2008})}\BibitemShut {NoStop}%
\bibitem [{\citenamefont {Zhang}\ \emph {et~al.}(2006)\citenamefont {Zhang},
  \citenamefont {Jiang}, \citenamefont {Small}, \citenamefont {Purewal},
  \citenamefont {Tan}, \citenamefont {Fazlollahi}, \citenamefont {Chudow},
  \citenamefont {Jaszczak}, \citenamefont {Stormer},\ and\ \citenamefont
  {Kim}}]{PhysRevLett.96.136806}%
  \BibitemOpen
  \bibfield  {author} {\bibinfo {author} {\bibfnamefont {Y.}~\bibnamefont
  {Zhang}}, \bibinfo {author} {\bibfnamefont {Z.}~\bibnamefont {Jiang}},
  \bibinfo {author} {\bibfnamefont {J.~P.}\ \bibnamefont {Small}}, \bibinfo
  {author} {\bibfnamefont {M.~S.}\ \bibnamefont {Purewal}}, \bibinfo {author}
  {\bibfnamefont {Y.-W.}\ \bibnamefont {Tan}}, \bibinfo {author} {\bibfnamefont
  {M.}~\bibnamefont {Fazlollahi}}, \bibinfo {author} {\bibfnamefont {J.~D.}\
  \bibnamefont {Chudow}}, \bibinfo {author} {\bibfnamefont {J.~A.}\
  \bibnamefont {Jaszczak}}, \bibinfo {author} {\bibfnamefont {H.~L.}\
  \bibnamefont {Stormer}},\ and\ \bibinfo {author} {\bibfnamefont
  {P.}~\bibnamefont {Kim}},\ }\bibfield  {title} {\bibinfo {title}
  {Landau-level splitting in graphene in high magnetic fields},\ }\href
  {https://link.aps.org/doi/10.1103/PhysRevLett.96.136806} {\bibfield
  {journal} {\bibinfo  {journal} {Physical Review Letters}\ }\textbf {\bibinfo
  {volume} {96}},\ \bibinfo {pages} {136806} (\bibinfo {year}
  {2006})}\BibitemShut {NoStop}%
\bibitem [{\citenamefont {Kandemir}\ and\ \citenamefont
  {Mogulkoc}(2015)}]{KANDEMIR20152120}%
  \BibitemOpen
  \bibfield  {author} {\bibinfo {author} {\bibfnamefont {B.}~\bibnamefont
  {Kandemir}}\ and\ \bibinfo {author} {\bibfnamefont {A.}~\bibnamefont
  {Mogulkoc}},\ }\bibfield  {title} {\bibinfo {title} {Chiral symmetry breaking
  by a magnetic field in graphene},\ }\href
  {https://www.sciencedirect.com/science/article/pii/S0375960115005769}
  {\bibfield  {journal} {\bibinfo  {journal} {Physics Letters A}\ }\textbf
  {\bibinfo {volume} {379}},\ \bibinfo {pages} {2120} (\bibinfo {year}
  {2015})}\BibitemShut {NoStop}%
\bibitem [{\citenamefont {G\"u\ifmmode~\mbox{\c{c}}\else \c{c}\fi{}l\"u}\ \emph
  {et~al.}(2013{\natexlab{a}})\citenamefont {G\"u\ifmmode~\mbox{\c{c}}\else
  \c{c}\fi{}l\"u}, \citenamefont {Potasz},\ and\ \citenamefont
  {Hawrylak}}]{TGQDs2B}%
  \BibitemOpen
  \bibfield  {author} {\bibinfo {author} {\bibfnamefont {A.~D.}\ \bibnamefont
  {G\"u\ifmmode~\mbox{\c{c}}\else \c{c}\fi{}l\"u}}, \bibinfo {author}
  {\bibfnamefont {P.}~\bibnamefont {Potasz}},\ and\ \bibinfo {author}
  {\bibfnamefont {P.}~\bibnamefont {Hawrylak}},\ }\bibfield  {title} {\bibinfo
  {title} {Zero-energy states of graphene triangular quantum dots in a magnetic
  field},\ }\href {https://doi.org/10.1103/PhysRevB.88.155429} {\bibfield
  {journal} {\bibinfo  {journal} {Physical Review B}\ }\textbf {\bibinfo
  {volume} {88}},\ \bibinfo {pages} {155429} (\bibinfo {year}
  {2013}{\natexlab{a}})}\BibitemShut {NoStop}%
\bibitem [{\citenamefont {Lavor}\ \emph {et~al.}(2020)\citenamefont {Lavor},
  \citenamefont {da~Costa}, \citenamefont {Chaves}, \citenamefont {Farias},
  \citenamefont {Mac{\^e}do},\ and\ \citenamefont
  {Peeters}}]{lavor2020magnetic}%
  \BibitemOpen
  \bibfield  {author} {\bibinfo {author} {\bibfnamefont {I.~R.}\ \bibnamefont
  {Lavor}}, \bibinfo {author} {\bibfnamefont {D.~R.}\ \bibnamefont {da~Costa}},
  \bibinfo {author} {\bibfnamefont {A.}~\bibnamefont {Chaves}}, \bibinfo
  {author} {\bibfnamefont {G.~d.~A.}\ \bibnamefont {Farias}}, \bibinfo {author}
  {\bibfnamefont {R.}~\bibnamefont {Mac{\^e}do}},\ and\ \bibinfo {author}
  {\bibfnamefont {F.~M.}\ \bibnamefont {Peeters}},\ }\bibfield  {title}
  {\bibinfo {title} {Magnetic field induced vortices in graphene quantum
  dots},\ }\href {https://iopscience.iop.org/article/10.1088/1361-648X/ab6463}
  {\bibfield  {journal} {\bibinfo  {journal} {Journal of Physics: Condensed
  Matter}\ }\textbf {\bibinfo {volume} {32}},\ \bibinfo {pages} {155501}
  (\bibinfo {year} {2020})}\BibitemShut {NoStop}%
\bibitem [{\citenamefont {Zarenia}\ \emph {et~al.}(2011)\citenamefont
  {Zarenia}, \citenamefont {Chaves}, \citenamefont {Farias},\ and\
  \citenamefont {Peeters}}]{ZareniaQDs}%
  \BibitemOpen
  \bibfield  {author} {\bibinfo {author} {\bibfnamefont {M.}~\bibnamefont
  {Zarenia}}, \bibinfo {author} {\bibfnamefont {A.}~\bibnamefont {Chaves}},
  \bibinfo {author} {\bibfnamefont {G.~A.}\ \bibnamefont {Farias}},\ and\
  \bibinfo {author} {\bibfnamefont {F.~M.}\ \bibnamefont {Peeters}},\
  }\bibfield  {title} {\bibinfo {title} {Energy levels of triangular and
  hexagonal graphene quantum dots: A comparative study between the
  tight-binding and dirac equation approach},\ }\href
  {https://doi.org/10.1103/PhysRevB.84.245403} {\bibfield  {journal} {\bibinfo
  {journal} {Physical Review B}\ }\textbf {\bibinfo {volume} {84}},\ \bibinfo
  {pages} {245403} (\bibinfo {year} {2011})}\BibitemShut {NoStop}%
\bibitem [{\citenamefont {Zhang}\ \emph {et~al.}(2008)\citenamefont {Zhang},
  \citenamefont {Chang},\ and\ \citenamefont {Peeters}}]{PetersGQDsMagnetic}%
  \BibitemOpen
  \bibfield  {author} {\bibinfo {author} {\bibfnamefont {Z.~Z.}\ \bibnamefont
  {Zhang}}, \bibinfo {author} {\bibfnamefont {K.}~\bibnamefont {Chang}},\ and\
  \bibinfo {author} {\bibfnamefont {F.~M.}\ \bibnamefont {Peeters}},\
  }\bibfield  {title} {\bibinfo {title} {Tuning of energy levels and optical
  properties of graphene quantum dots},\ }\href
  {https://link.aps.org/doi/10.1103/PhysRevB.77.235411} {\bibfield  {journal}
  {\bibinfo  {journal} {Physical Review B}\ }\textbf {\bibinfo {volume} {77}},\
  \bibinfo {pages} {235411} (\bibinfo {year} {2008})}\BibitemShut {NoStop}%
\bibitem [{\citenamefont {G\"u\ifmmode~\mbox{\c{c}}\else \c{c}\fi{}l\"u}\ \emph
  {et~al.}(2013{\natexlab{b}})\citenamefont {G\"u\ifmmode~\mbox{\c{c}}\else
  \c{c}\fi{}l\"u}, \citenamefont {Potasz},\ and\ \citenamefont
  {Hawrylak}}]{gucclu2013zero}%
  \BibitemOpen
  \bibfield  {author} {\bibinfo {author} {\bibfnamefont {A.~D.}\ \bibnamefont
  {G\"u\ifmmode~\mbox{\c{c}}\else \c{c}\fi{}l\"u}}, \bibinfo {author}
  {\bibfnamefont {P.}~\bibnamefont {Potasz}},\ and\ \bibinfo {author}
  {\bibfnamefont {P.}~\bibnamefont {Hawrylak}},\ }\bibfield  {title} {\bibinfo
  {title} {Zero-energy states of graphene triangular quantum dots in a magnetic
  field},\ }\href {https://link.aps.org/doi/10.1103/PhysRevB.88.155429}
  {\bibfield  {journal} {\bibinfo  {journal} {Physical Review B}\ }\textbf
  {\bibinfo {volume} {88}},\ \bibinfo {pages} {155429} (\bibinfo {year}
  {2013}{\natexlab{b}})}\BibitemShut {NoStop}%
\bibitem [{\citenamefont {de~Andrada~e Silva}(1992)}]{de1992probability}%
  \BibitemOpen
  \bibfield  {author} {\bibinfo {author} {\bibfnamefont {E.~A.}\ \bibnamefont
  {de~Andrada~e Silva}},\ }\bibfield  {title} {\bibinfo {title} {Probability
  current in the tight-binding model},\ }\href
  {https://doi.org/10.1119/1.17084} {\bibfield  {journal} {\bibinfo  {journal}
  {American Journal of Physics}\ }\textbf {\bibinfo {volume} {60}},\ \bibinfo
  {pages} {753} (\bibinfo {year} {1992})}\BibitemShut {NoStop}%
\bibitem [{\citenamefont {da~Costa}\ \emph {et~al.}(2012)\citenamefont
  {da~Costa}, \citenamefont {Chaves}, \citenamefont {Farias}, \citenamefont
  {Covaci},\ and\ \citenamefont {Peeters}}]{PhysRevB.86.115434}%
  \BibitemOpen
  \bibfield  {author} {\bibinfo {author} {\bibfnamefont {D.~R.}\ \bibnamefont
  {da~Costa}}, \bibinfo {author} {\bibfnamefont {A.}~\bibnamefont {Chaves}},
  \bibinfo {author} {\bibfnamefont {G.~A.}\ \bibnamefont {Farias}}, \bibinfo
  {author} {\bibfnamefont {L.}~\bibnamefont {Covaci}},\ and\ \bibinfo {author}
  {\bibfnamefont {F.~M.}\ \bibnamefont {Peeters}},\ }\bibfield  {title}
  {\bibinfo {title} {Wave-packet scattering on graphene edges in the presence
  of a pseudomagnetic field},\ }\href
  {https://link.aps.org/doi/10.1103/PhysRevB.86.115434} {\bibfield  {journal}
  {\bibinfo  {journal} {Physical Review B}\ }\textbf {\bibinfo {volume} {86}},\
  \bibinfo {pages} {115434} (\bibinfo {year} {2012})}\BibitemShut {NoStop}%
\bibitem [{\citenamefont {Potasz}\ \emph {et~al.}(2011)\citenamefont {Potasz},
  \citenamefont {G\"u\ifmmode~\mbox{\c{c}}\else \c{c}\fi{}l\"u}, \citenamefont
  {Voznyy}, \citenamefont {Folk},\ and\ \citenamefont
  {Hawrylak}}]{PhysRevB.83.174441}%
  \BibitemOpen
  \bibfield  {author} {\bibinfo {author} {\bibfnamefont {P.}~\bibnamefont
  {Potasz}}, \bibinfo {author} {\bibfnamefont {A.~D.}\ \bibnamefont
  {G\"u\ifmmode~\mbox{\c{c}}\else \c{c}\fi{}l\"u}}, \bibinfo {author}
  {\bibfnamefont {O.}~\bibnamefont {Voznyy}}, \bibinfo {author} {\bibfnamefont
  {J.~A.}\ \bibnamefont {Folk}},\ and\ \bibinfo {author} {\bibfnamefont
  {P.}~\bibnamefont {Hawrylak}},\ }\bibfield  {title} {\bibinfo {title}
  {Electronic and magnetic properties of triangular graphene quantum rings},\
  }\href {https://doi.org/10.1103/PhysRevB.83.174441} {\bibfield  {journal}
  {\bibinfo  {journal} {Physical Review B}\ }\textbf {\bibinfo {volume} {83}},\
  \bibinfo {pages} {174441} (\bibinfo {year} {2011})}\BibitemShut {NoStop}%
\end{thebibliography}%

\end{document}